\documentclass[letterpaper]{aa} 
\usepackage{graphicx}
\begin{document}
\title{Direct multi-wavelength limb-darkening measurements of three\\
late-type giants with the Navy Prototype Optical Interferometer}
\titlerunning{Limb-darkening measurements with NPOI}
\author{
M.~Wittkowski\inst{1}\thanks{\emph{Present address:}
European Southern Observatory, Casilla 19001, Santiago 19, Chile, 
mwittkow@eso.org} \and 
C.~A.~Hummel\inst{2} \and
K.~J.~Johnston\inst{2} \and
D.~Mozurkewich\inst{3} \and 
A.~R.~Hajian\inst{2} \and
N.~M.~White\inst{4}}
\institute{
Universities Space Research Association (USRA), 300 D Street SW,
Suite 801, Washington, DC 20024, USA\\
Mailing address: U.S. Naval Observatory, AD5, 3450 Massachusetts Avenue NW, 
Washington, DC 20392, USA \and
U.S. Naval Observatory, 3450 Massachusetts Avenue NW, Washington, DC
20392, USA\\ (cah@,kjj@astro.,hajian@)usno.navy.mil \and
Naval Research Laboratory, Code 7210, Washington, DC 20375, USA, 
mozurk@rsd.nrl.navy.mil \and
Lowell Observatory, 1400 West Mars Hill Road, Flagstaff, AZ 86001, USA, 
nmw@lowell.edu
}
\offprints{M.~Wittkowski}
\mail{M.~Wittkowski}
\date{Received \dots ; accepted \dots} 
\abstract{
We present direct measurements of the limb-darkened intensity profiles
of the late-type giant stars HR\,5299, HR\,7635, and HR\,8621 obtained
with the Navy Prototype Optical Interferometer (NPOI) at the Lowell
Observatory. A triangle 
of baselines with lengths of 18.9\,m, 22.2\,m, and 37.5\,m was used.
We utilized squared visibility amplitudes beyond the first minimum, 
as well as triple amplitudes and phases in up to 10 spectral channels covering 
a wavelength range of $\sim$\,650\,nm to $\sim$\,850\,nm.
We find that our data can best be described by featureless symmetric 
limb-darkened disk models while uniform disk and fully darkened disk
models can be
rejected. We derive high-precision angular limb-darkened 
diameters for the three stars of 7.44\,mas$\,\pm\,$0.11\,mas,
6.18\,mas$\,\pm\,$0.07\,mas, and 6.94\,mas$\,\pm\,$0.12\,mas, respectively.
Using the HIPPARCOS parallaxes, we determine linear limb-darkened
radii of 114\,R$_\odot\,\pm\,$13\,R$_\odot$,
56\,R$_\odot\,\pm\,$4\,R$_\odot$,
and 98\,R$_\odot\,\pm\,$9\,R$_\odot$, respectively.
We compare our data to a grid of Kurucz stellar model atmospheres, with
them derive the effective temperatures and surface gravities without 
additional information, and find agreement with independent estimates 
derived from empirical calibrations and bolometric fluxes. This
confirms the consistency of model predictions and direct observations
of the limb-darkening effect.
}
\maketitle
\section{Introduction}
\label{sec:introduction}
For a detailed understanding of stellar atmospheres and
stellar evolution, it is of essential importance to obtain accurate 
observational estimates of stellar surface structure
parameters on all scales.
These  parameters include diameters, limb-darkening profiles,
photospheric asymmetries, and special features like hot spots. 
Model atmospheres are mainly constrained by observations of stellar spectra.
Direct measurements of the limb-darkening profile can, in principle,
provide an independent estimate of the temperature change with continuum
opacity and, thus, an independent observational verification.
So far, the detailed intensity profile and the whole variety of 
additional surface structure parameters could only be 
observed in the case of the Sun.
Today's interest, however, includes other phases of stellar structure
and evolution,
for instance stars in late evolutionary phases (Manduca et al. \cite{manduca},
Scholz \& Takeda \cite{scholztakeda}, Hofmann \& Scholz \cite{hofmann},
Jacob et al. \cite{jacob}).
These stars might exhibit asymmetric 
(e.g. Wittkowski et al. \cite{wittkowski}) and 
even highly fragmented (e.g. Weigelt et al. \cite{weigelt}) mass-loss 
events, which are believed to be triggered by the conditions on the 
stellar surfaces.
Unfortunately, direct measurements of surface structure parameters 
are rare. While diameters have so far been obtained 
for several hundred stars with interferometric and lunar occultation
techniques, the second-order effect, limb-darkening,
has been directly observed for only a very limited number of stars
(Hanbury Brown et al. \cite{hanbury}; 
Haniff et al. \cite{haniff};
Quirrenbach et al. \cite{quirrenbach};
Hajian et al. \cite{hajian}). Additional surface features 
have been detected on the apparently largest supergiants $\alpha$~Ori,
$\alpha$~Sco, and $\alpha$~Her contributing
with 5\% to 20\% to the total source fluxes at optical wavelengths
(Buscher et al. \cite{buscher};
Wilson et al. \cite{wilson}; Gilliland \& Dupree \cite{gilliland};
Burns et al. \cite{burns}; Tuthill et al. \cite{tuthill};
Young et al. \cite{young}). 

The NPOI, located near Flagstaff, Arizona, is especially designed for 
imaging of stars and their environments and is described in detail by
Armstrong et al. (\cite{armstrong}).
The methods of ''baseline bootstrapping'' and ''wavelength bootstrapping''
(see Roddier \cite{roddier}, Quirrenbach et al. \cite{quirrenbach},
Hajian et al. \cite{hajian}) can be used in order to detect
weak fringe contrasts, i.e. low visibility values, on resolved stars
by detecting the higher-contrast fringes on a chain of shorter effective 
spacings which comprise the long baselines. Hajian et al. (\cite{hajian}) 
demonstrated that bootstrapping with the NPOI astrometric subarray enabled 
the measurement of visibility values of a resolved stellar disk beyond the 
first minimum. By analyzing NPOI triple visibility products, they verified 
that the intensity profiles of $\alpha$\,Ari 
and $\alpha$\,Cas deviate from uniform disks due to the effect 
of limb-darkening.

Here, we use NPOI's bootstrapping ability and apply an improved
bias correction in order to utilize squared visibility amplitudes, 
in addition to triple amplitudes and closure phases, of three much fainter 
but well resolved late-type stars for spatial frequencies on both sides
of the first minimum. Since there are three squared visibility amplitudes for 
each triple amplitude, more information can be used for the analysis of 
stellar intensity profiles. 
In an attempt to check the consistency of direct observations and model
predictions of the limb-darkening effect as a function of continuum
wavelength, effective temperature, and surface gravity,
we compare our multi-wavelength interferometric data to a grid
of Kurucz stellar model atmospheres.
In addition, we determine high-precision limb-darkened diameters.
\section{Observations}
\label{sec:observations}
\begin{table*}
\caption{Names (bright star catalog number HR, FK5 catalog number, common
name) of the program stars, together with their properties 
(spectral type, apparent visual magnitude $m_V$, 
HIPPARCOS parallax $\pi_\mathrm{trig}$, bolometric flux 
$F_\mathrm{Bol}$)
and observational parameters (observing date, number of obtained scans,
names of calibrator stars and their estimated 
diameters $\Theta_\mathrm{Cal.}$).
Spectral Type, $m_V$, and $\pi_\mathrm{trig}$ are taken from the 
HIPPARCOS catalogue (Perryman \& ESA \cite{esa}), references for 
$F_\mathrm{Bol}$ are cited below. 
Note, HR\,7635
was classified as K5-M0 by Morgan \& Keenan (\cite{morgan}), as K5 before.}
\begin{tabular}{rrl|rrrrr|lrlr} 
HR   & FK5  & Name       &Spectral& $m_V$ & $\pi_\mathrm{trig}$ & 
$F_\mathrm{Bol}$ & Ref. & Observing  & \# of & Calibr. & $\Theta_\mathrm{Cal.}$ \\
&      &            &Type    & [mag] & [mas]            & 
[10$^{-17}$\,W/m$^2$] & & Dates      & scans & stars & [mas] \\[1ex]\hline 
5299 & 1368 &BY\,Boo     &M4.2III & 5.13  & 7.01$\pm$0.66    & 2.523$\pm$0.40
& a & 2000-07-07 & 2     & FK5\,527 & 0.5 \\
& & & & & & & & 2000-07-13 & 4 & &      \\[1ex]
7635 & 752  &$\gamma$\,Sge &K5III & 3.51  & 11.90$\pm$0.71   & 2.792$\pm$0.14
& b & 2000-07-21 & 6     & FK5\,768 & 0.3 \\[1ex]
8621 &      &V416\,Lac   &M4III   & 5.11  & 7.59$\pm$0.57    &                
&   & 2000-07-07 & 2     & FK5\,891 & 0.3 \\
& & & & & & & & 2000-07-12 & 2     & HR\,8494 & 0.8 \\
& & & & & & & & 2000-07-13 & 3     &          &  \\
\end{tabular}\\[1ex]
References for bolometric fluxes: (a) Tsuji (\cite{tsuji}), 
(b) Alonso et al. (\cite{alonso}). 
\label{tab:obs}
\end{table*}
\begin{figure*}
\resizebox{0.33\hsize}{!}{\includegraphics{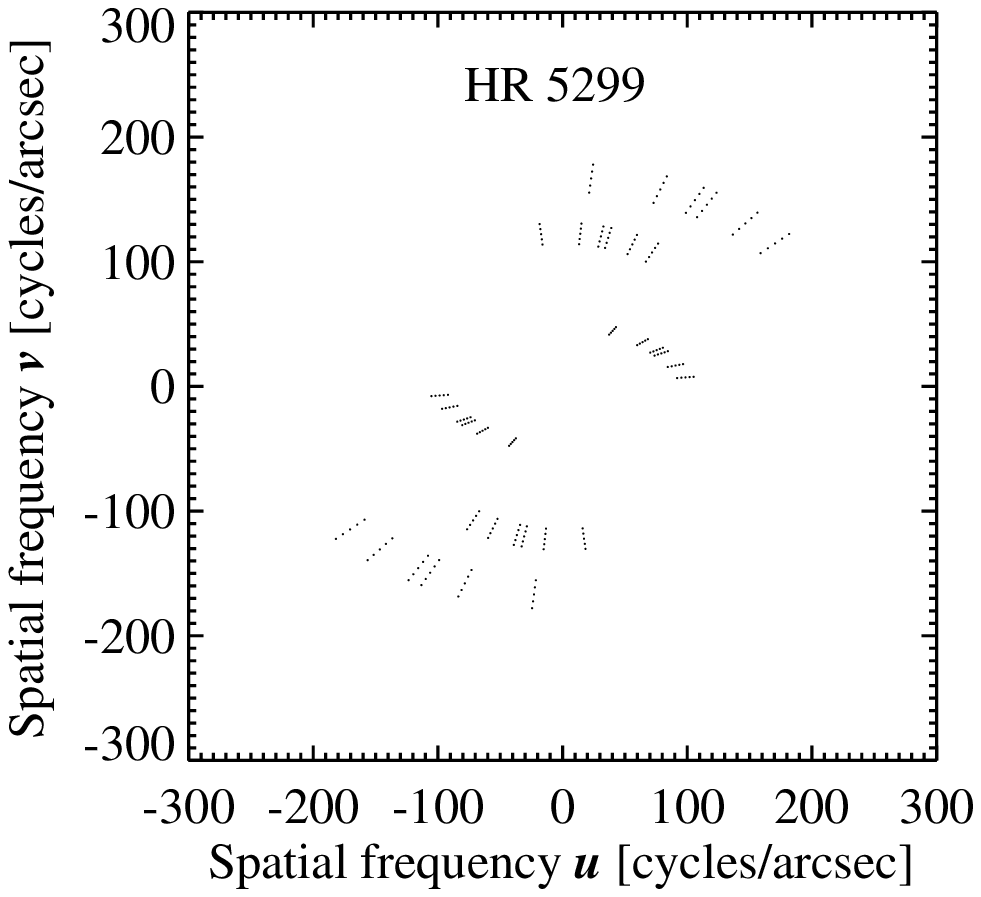}}
\resizebox{0.33\hsize}{!}{\includegraphics{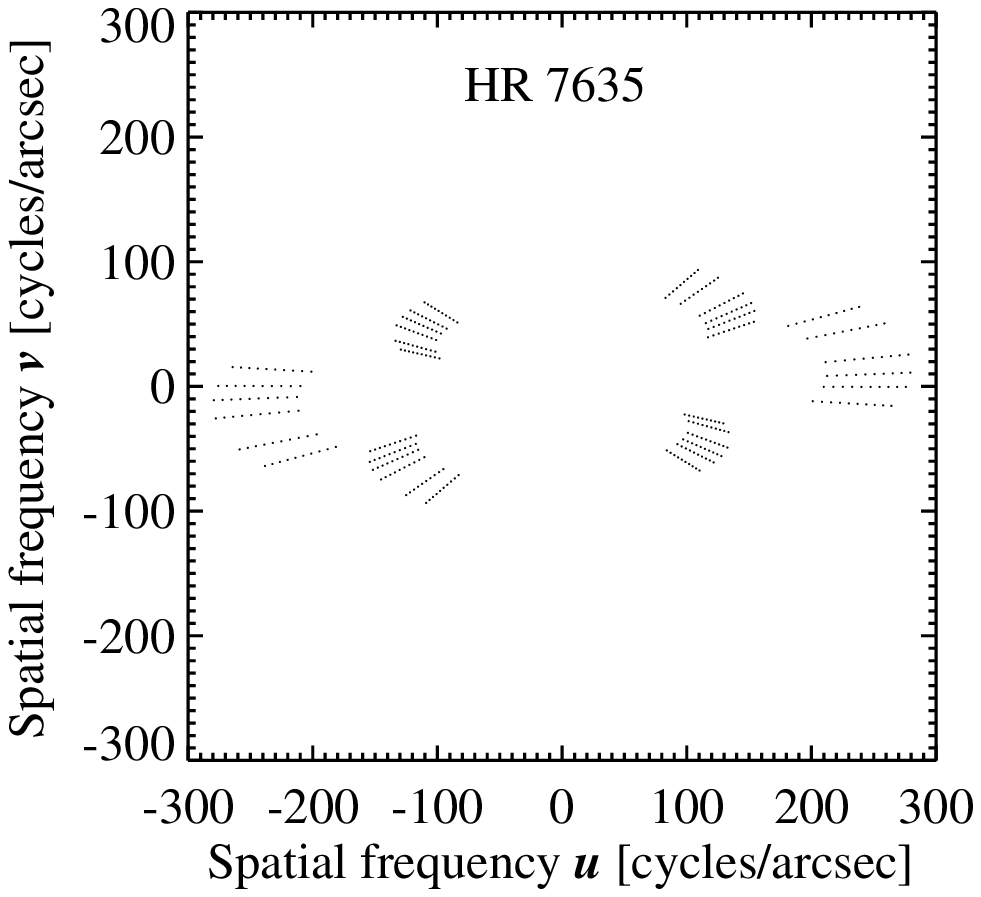}}
\resizebox{0.33\hsize}{!}{\includegraphics{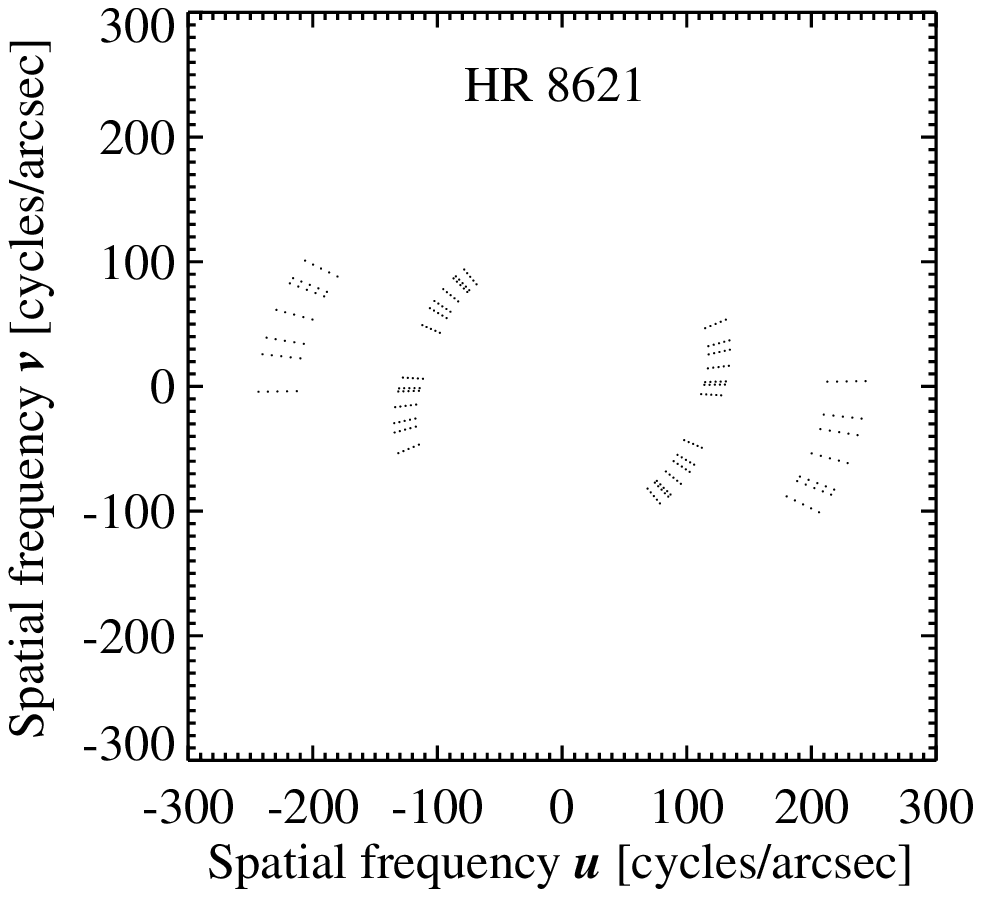}}
\vspace*{-0.8cm}%
\caption{Plot of the obtained coverages of the $uv$-plane of the HR\,5299, 
HR\,7635 and HR\,8621
observations based on all observation dates (see Table~\ref{tab:obs}) and 
on those spectral channels which were used for the data analysis, i.e. the 
10 reddest channels for HR\,7635 and the 5 reddest channels for the fainter 
stars HR\,5299 and HR\,8621.}
\label{fig:uvcov}
\end{figure*}
Interferometric observations of the three late-type giants, 
\object{HR\,5299} (M4), \object{HR\,7635} (K5), and \object{HR\,8621} (M4),
were performed with the NPOI using the configuration described by 
Benson et al. (\cite{benson}).

The center, east, and west siderostats of the astrometric subarray were used 
with effective apertures of 12.5\,cm. They provide baselines with lengths 
of 18.9\,m, 22.2\,m, and 37.5\,m at azimuths (measured east from north) 
of $-$67\fdg5, 63\fdg6, and 86\fdg0, respectively. 
The three afocal beams of light are reduced in diameter and sent into 
vacuum delay-lines for compensation of optical path differences (OPDs) 
before they are combined pairwise using beamsplitters. 
Three of the afocal output beams corresponding to the three baselines
are dispersed by a prism, focused by a lenslet array onto 32 optical 
fibers, and detected by avalanche photodiodes (APDs), covering a spectral
range from 450\,nm to 850\,nm. The fringe packet is detected through 
modulation of the OPDs and is kept centered close to zero
residual delay using the method of ''group delay fringe tracking''
(Armstrong et al. \cite{armstrong}).

Table~\ref{tab:obs} lists names and characteristics of the observed stars 
together with the observing dates, the number of obtained scans, and the 
names and estimated diameters of the calibrator stars. During a scan of 
90\,s, the photon count rate for every channel is determined in eight 
temporal bins (synchronous with the delay line modulation), which sample a 
fringe every 2\,ms. After each scan, a background measurement was taken on 
blank sky near the star. Immediately before or after each scan of a program 
star, a scan on one of the calibrator stars as specified in 
Table~\ref{tab:obs} was recorded. The calibrator stars were chosen to be 
located near the appropriate program stars on the sky. Their diameters were 
estimated using a calibration obtained by Mozurkewich et al. 
(\cite{mozurkewich}) based on the apparent visual magnitude and the 
($R-I$) color index and are small enough so that possible errors in
this estimate do not noticeably affect the calibration of our much larger
resolved program stars. 

In order to compensate for detection noise bias terms 
(see Sect.\,\ref{sec:reduction}), incoherent (i.e. fringeless) data on several 
stars covering a range of apparent visual magnitudes were recorded on 
July 22, 2000, by moving the delay lines off the fringe packet.

The signal-to-noise ratio of the measured visibilities decreases for spectral
channels with shorter wavelengths owing to their narrower bandwidths
(the channels are equally spaced in wavenumber), poorer
seeing, and the red color of the observed stars.
Therefore, only the 10 reddest channels were used for the data analysis of 
HR\,7635, and the 5 reddest channels for that of the fainter stars HR\,5299 
and HR\,8621. The central wavelengths of the 10 reddest spectral channels 
are known to within about 1\% to be (852, 822, 794, 769, 745, 723, 702, 683, 
665, 649)\,nm,
while their bandwidths range from $\sim$31\,nm for the 852\,nm channel to
$\sim$16\,nm for the 649\,nm channel. These passbands mainly provide 
continuum observations and are not dominated by spectral features. 
The absorption band/continuum ratio is 
relatively small for the K5 star and the strong TiO bands at 671\,nm
and 714\,nm are not covered by the spectral channels used for the M4 stars.
However, for the M4 giants, some spectral channels
are affected by molecular absorption bands, which 
will be taken into account.

Figure~\ref{fig:uvcov} shows the obtained coverages of the $uv$-plane for 
HR\,5299, HR\,7635, and HR\,8621 based on all observation dates and those 
spectral channels used for the data analysis. These spatial frequencies 
range up to a radius of $\sim$\,300 cycles/arcsecond, which corresponds to a 
spatial resolution of 3.3\,mas.
\section{Data reduction and calibration}
\label{sec:reduction}
The raw data were processed, reduced, and calibrated as described in detail
by Hummel et al. (\cite{hummel}). 
This process includes (1) 
calculation of the real ($X_i$) and imaginary ($Y_i$) parts of the
complex visibility for each baseline $i$ and each spectral
channel by Fourier transform of the bin counts
as a function of time, (2)
calculation of the squared visibility amplitudes ($|V_i|^2$) and
amplitude ($|V_{123}|$) and phase ($\phi_{123}$, ''closure phase'') 
of the complex triple product, averaged
over the 1\,s intervals, (3) editing of the 1\,s data, (4) further averaging 
over a scan of 90\,s, and (5) calibration of the program star's squared 
visibility amplitudes, triple amplitudes, and closure phases by normalization 
with the corresponding smoothed (time kernel of 20\,min) quantities of the 
calibrator stars. 

In step (2) $|V|^2$ is calculated using the unbiased estimator
\begin{equation}
|V|^2=\underbrace{4\left[\frac{\pi/n}{\sin(\pi/n)}\right]^2}_{=:f}
\frac{<X^2+Y^2-\sigma_I^2(N)>}{<N>^2},
\label{eq:estvis}
\end{equation}
where $n=8$ is the number of bins, $N$ the total photon count rate
in 2\,ms, and $\sigma_I^2$ the variance
of the intensity caused by photon and detection noise.
For stellar observations, $|V|^2$ is compensated for background intensity 
by an additional factor $<N>^2/<N-D>^2$, where $D$ is the background rate.
The use of a
squared quantity requires attention to the bias term $\sigma_I^2$ which
is discussed in detail below.
The triple product in step (2) is calculated using the unbiased estimator
\begin{equation}
|V_{123}|e^{i\phi_{123}}=<(X_1+iY_1)(X_2+iY_2)(X_3+iY_3)>.
\label{eq:esttriple}
\end{equation}
The phase $\phi_{123}$ of the complex triple product, formed by
a triangle of baselines, is not corrupted by the phase noise 
caused by atmospheric turbulence (Jennison \cite{jennison}).
No bias correction is required for the triple product, since the
noise from the three detector arrays receiving the signal from each
baseline is uncorrelated (e.g. Hummel et al. \cite{hummel}).
In a triple with two short baselines and one long baseline resolving
a stellar disk, as in the configuration used here, the triple amplitude
has a higher signal-to-noise ratio than the squared visibility amplitude
on the long baseline. This effect can be seen in our data shown below.
It can be understood due to the fact that the low visibility
on the long baseline is not squared but multiplied with higher
amplitudes from the other two baselines in the triple. This effect was
also confirmed by simulations of visibility data based on Poisson noise.
For the long baseline a visibility amplitude of 0.1 was assumed and for
the two short baselines a visibility amplitude of 0.6.
The signal-to-noise ratio of the squared visibility on the long (east-west) 
baseline was found to be 5 while that of the triple product was 20.

Formal errors for the squared visibility amplitudes, triple amplitudes
and closure phases were calculated based on
the scatter of the 1\,s samples.
Calibration errors of the squared visibility amplitudes were estimated 
to be 7\% for HR\,7635 and 10\% for HR\,5299 and HR\,8621 based on 
comparisons of different scans and on calibrations with other 
calibrator stars at slightly farther distances in time and position. 
The total (formal and calibration) errors for the triple amplitudes 
and closure phases of HR\,7635, HR\,5299, and HR\,8621
were estimated to be  2, 1.5, and 2 times the formal errors,
respectively. 
\paragraph{Noise bias compensation}
The use of the squared quantity $|V|^2$ requires attention to bias
corrections as shown by Hummel et al. (\cite{hummel}) for NPOI data,
by Colavita (\cite{colavita}) for the
case of the Palomar Testbed Interferometer, and by Davis et al. (\cite{davis})
for the Sydney University Stellar Interferometer.
The noise bias term 
$\sigma_I^2$ in Eq.~\ref{eq:estvis} is equal to $N$ in case of 
Poisson statistics (e.g. Shao et al. \cite{shao}).
Since the NPOI detectors exhibit non-Poisson noise, due to
after-pulsing of the APDs, the noise bias is estimated by
$\sigma_I^2=Z^2$ (Hummel et al. \cite{hummel}). Here, $Z^2$ is the fringe 
amplitude floor estimated at four times the modulation frequency
($X$ and $Y$ are calculated at the temporal
frequency $k$=1 to select the component corresponding to the
modulation frequency of the delay lines). 
However, despite the $Z^2$ bias compensation, positive visibility 
amplitudes are observed for fringeless data and this bias,
$B:=|V|^2\left(\mathrm{\,fringeless\,\,data,\,}\sigma_I^2=Z^2,\,D=0\right)$,
was found to be a function of $N$ and to be different for 
each spectral channel $c$ and each spectrometer $i$. It was modeled with 
a power-law
\begin{equation}
B(i,c)=a_0(i,c)\,N^{a_1(i,c)}
\end{equation}
where parameters $a_0$ and $a_1$ were fitted
to the fringeless data. This additional bias was compensated by using 
\begin{equation}
\sigma_I^2(N)=Z^2+B\,N^2/f,
\end{equation}
in Eq.~\ref{eq:estvis}. 
Obtained ranges for parameters $a_0$ and $a_1$ 
are [\,0.25,3.4\,] and [\,-1.48,-0.93\,], respectively.
Thus, $B$ is approximately inversely proportional 
to $N$, i.e. the star's brightness. Consequently, its magnitude relative to 
the squared visibility amplitude is largest for low visibility values of 
faint stars.
Therefore, the compensation of this additional noise bias
is essential
to obtain the very low squared visibility amplitudes of our faint program
stars around and beyond the first minimum with a precision that allows
an analysis of the limb-darkened intensity profiles. For instance, for
our stars HR\,5299, HR\,7635, and HR\,8621, the total biases 
$\sigma_I^2\,f/N^2$ for the bluest used channel and the 37.5\,m baseline 
amount to about 0.056, 0.063, and 0.063, respectively. Residuals of 
$\sim$0.004, $\sim$0.004, and $\sim$0.003 remain after the $Z^2$ correction 
and are compensated by $B$. For comparison, the squared visibility amplitude
of a fully darkened disk at the second maximum has a value of 0.0074.

The triple amplitudes and closure phases are not
affected by this bias, as mentioned above.
\section{Data analysis and results}
\label{sec:results}
\begin{figure*}
\begin{minipage}[b]{12cm}
\resizebox{0.32\hsize}{!}{\includegraphics{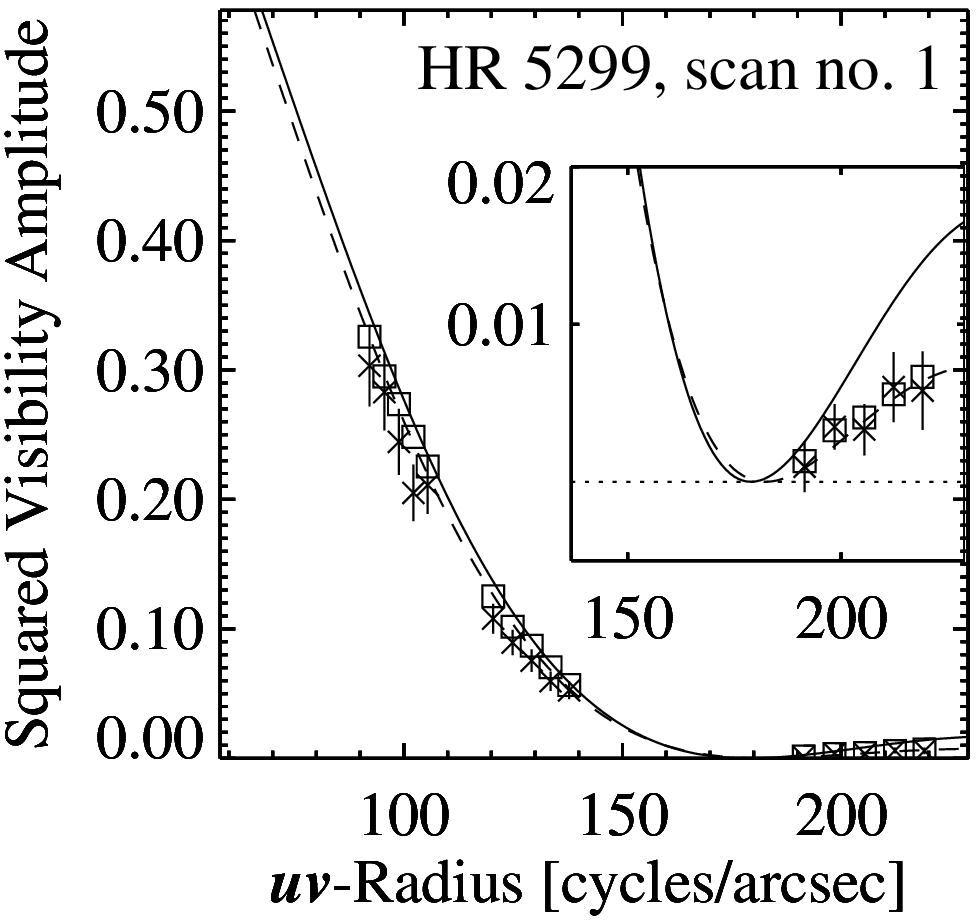}}
\hspace*{2mm}%
\resizebox{0.32\hsize}{!}{\includegraphics{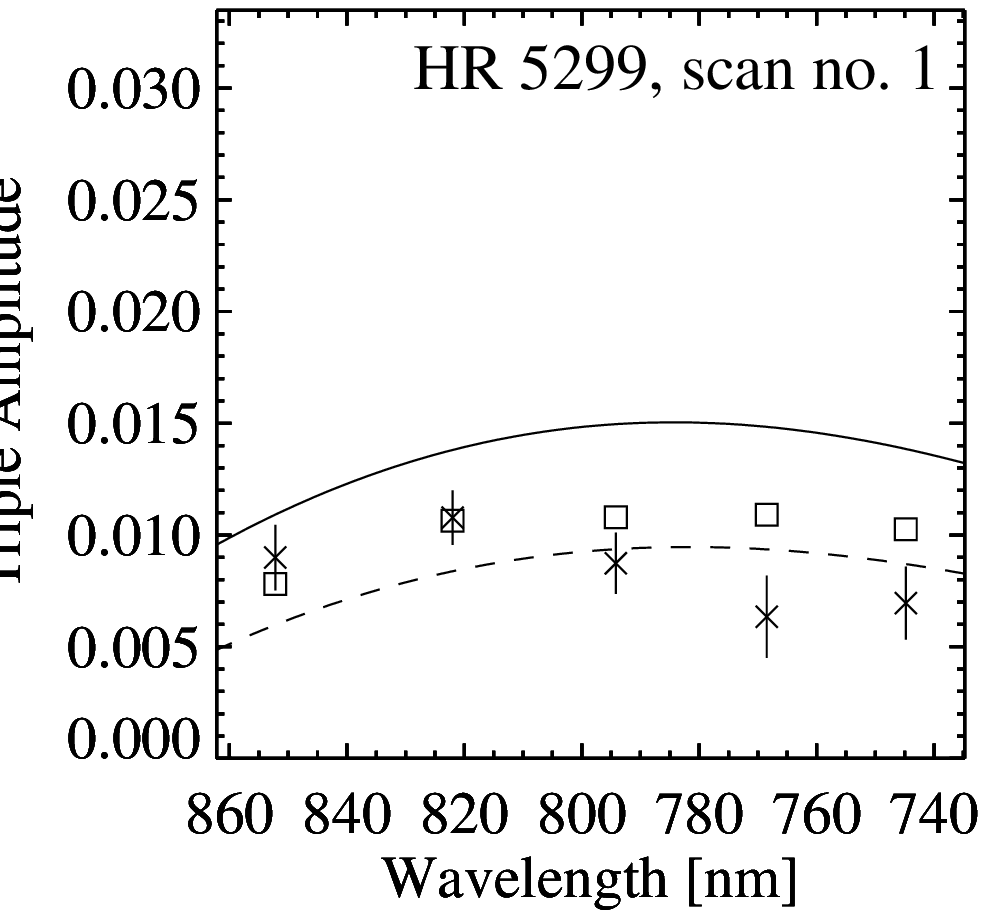}}
\hspace*{-1mm}%
\resizebox{0.32\hsize}{!}{\includegraphics{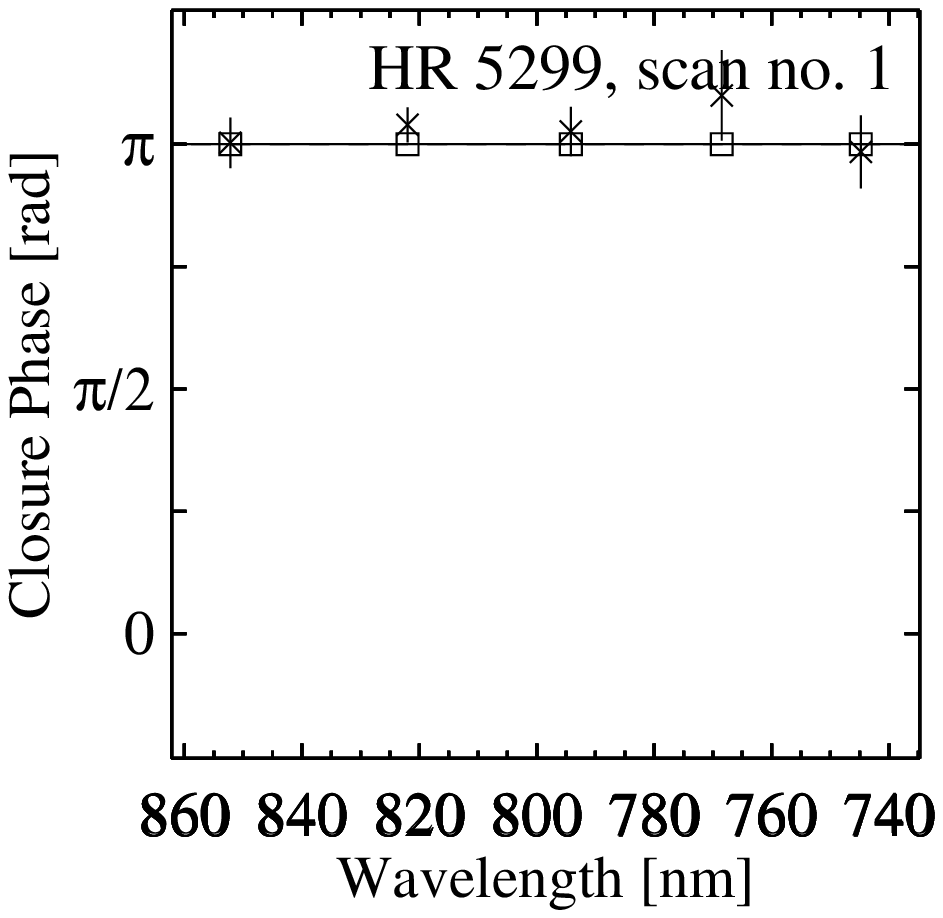}}

\resizebox{0.32\hsize}{!}{\includegraphics{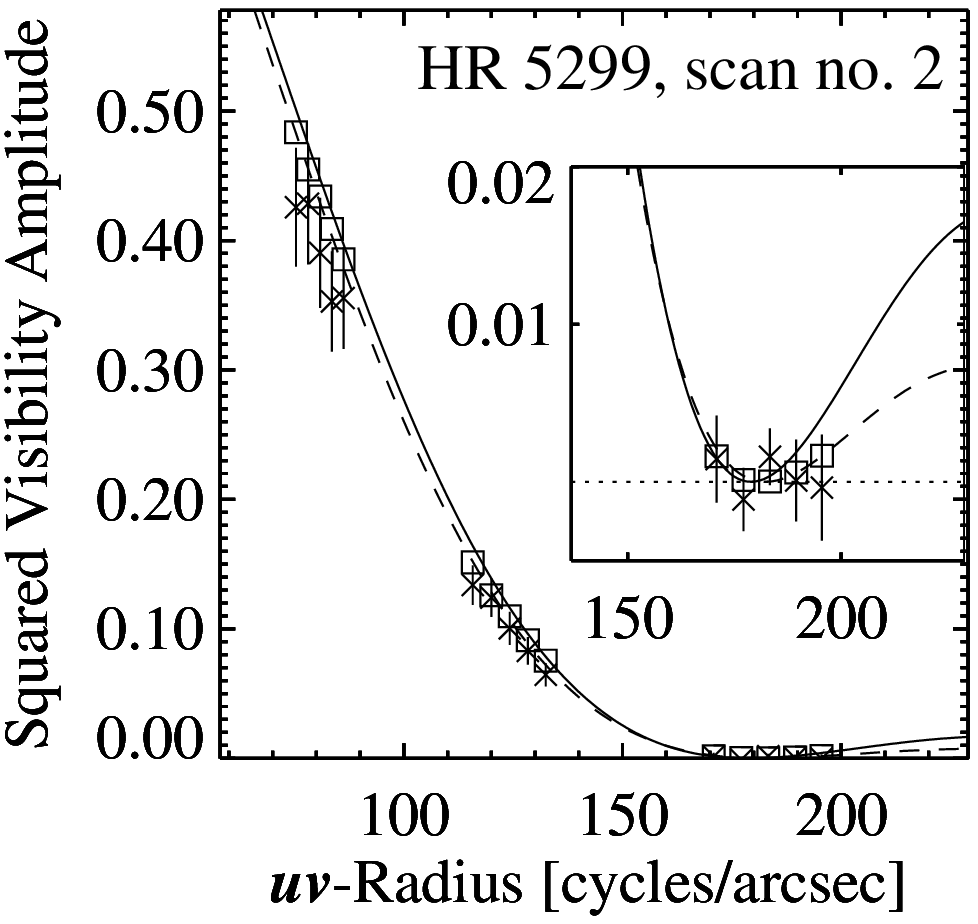}}
\hspace*{2mm}%
\resizebox{0.32\hsize}{!}{\includegraphics{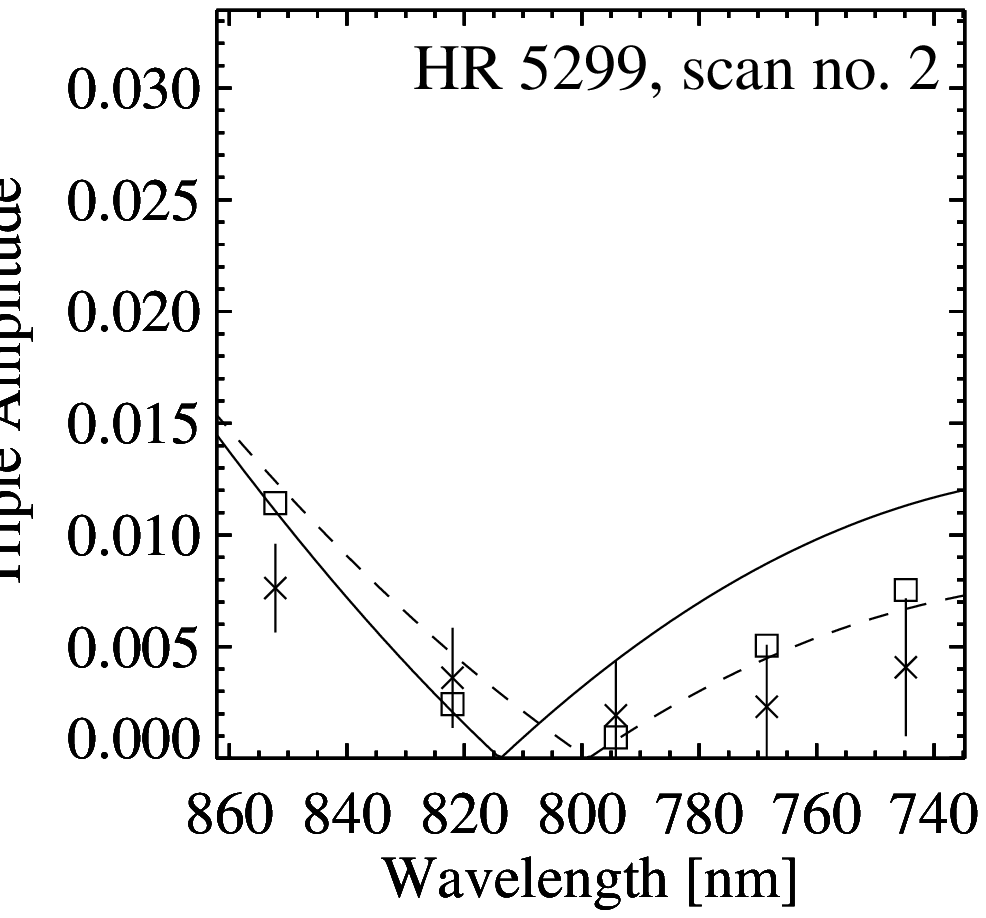}}
\hspace*{-1mm}%
\resizebox{0.32\hsize}{!}{\includegraphics{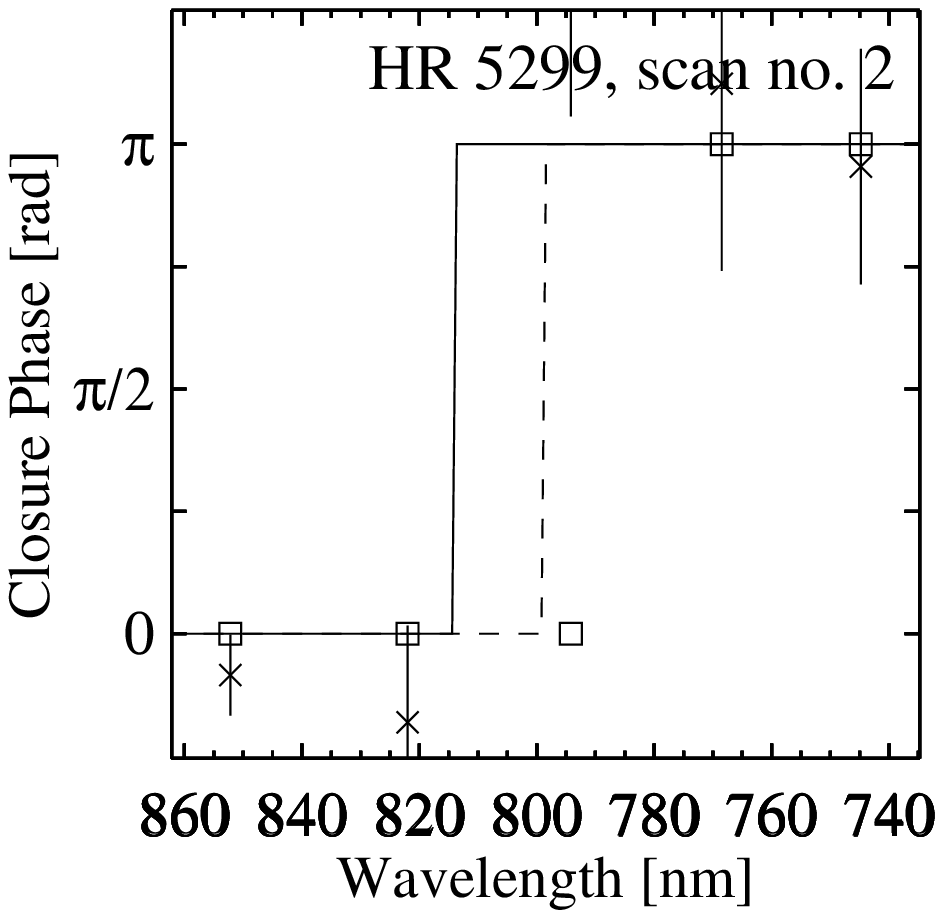}}

\resizebox{0.32\hsize}{!}{\includegraphics{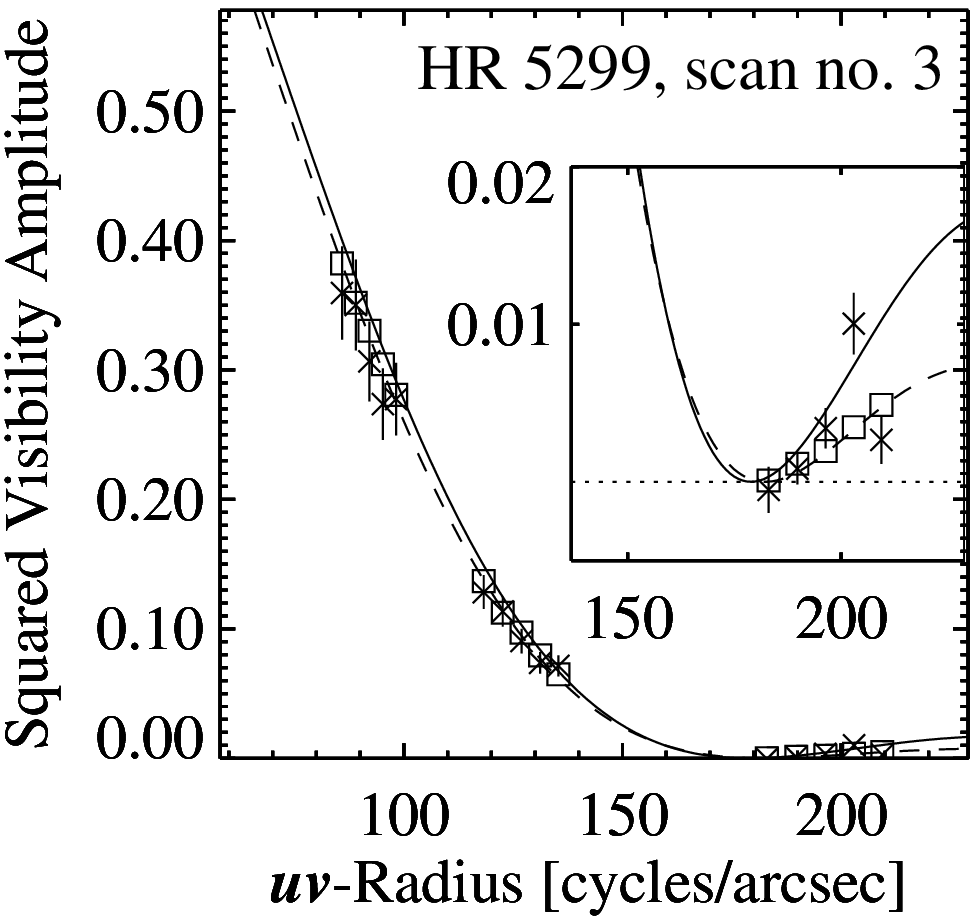}}
\hspace*{2mm}%
\resizebox{0.32\hsize}{!}{\includegraphics{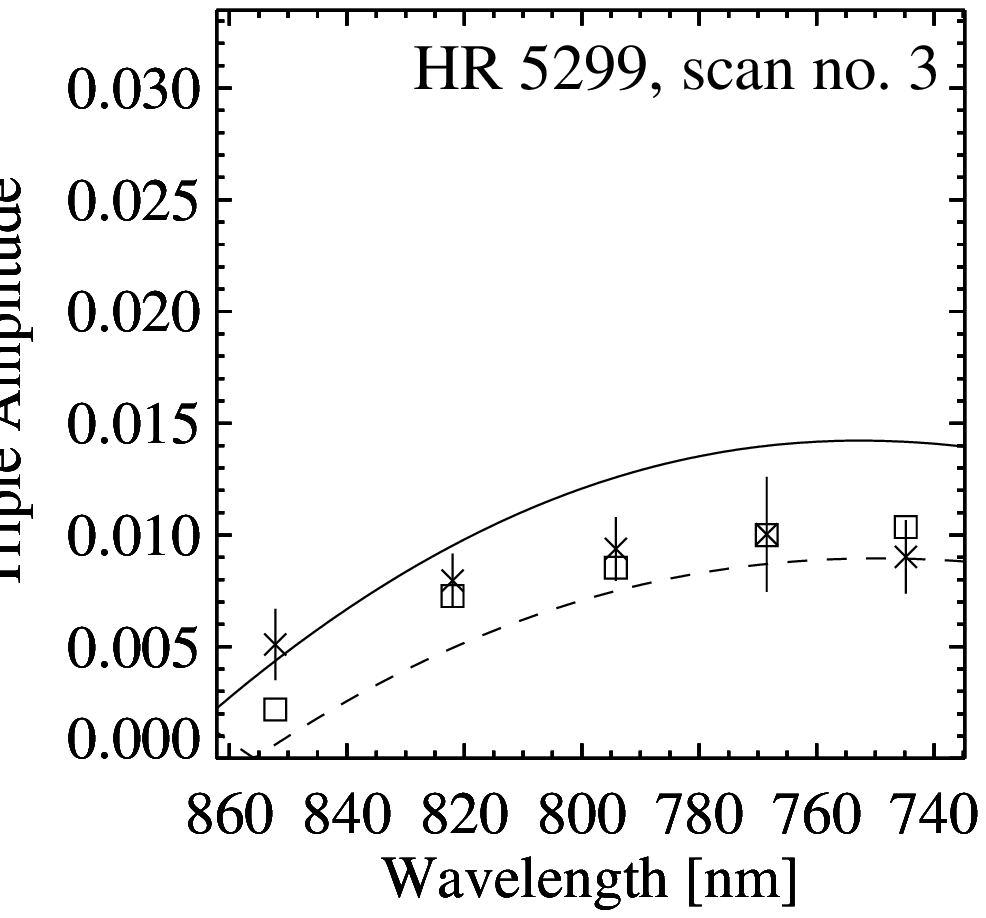}}
\hspace*{-1mm}%
\resizebox{0.32\hsize}{!}{\includegraphics{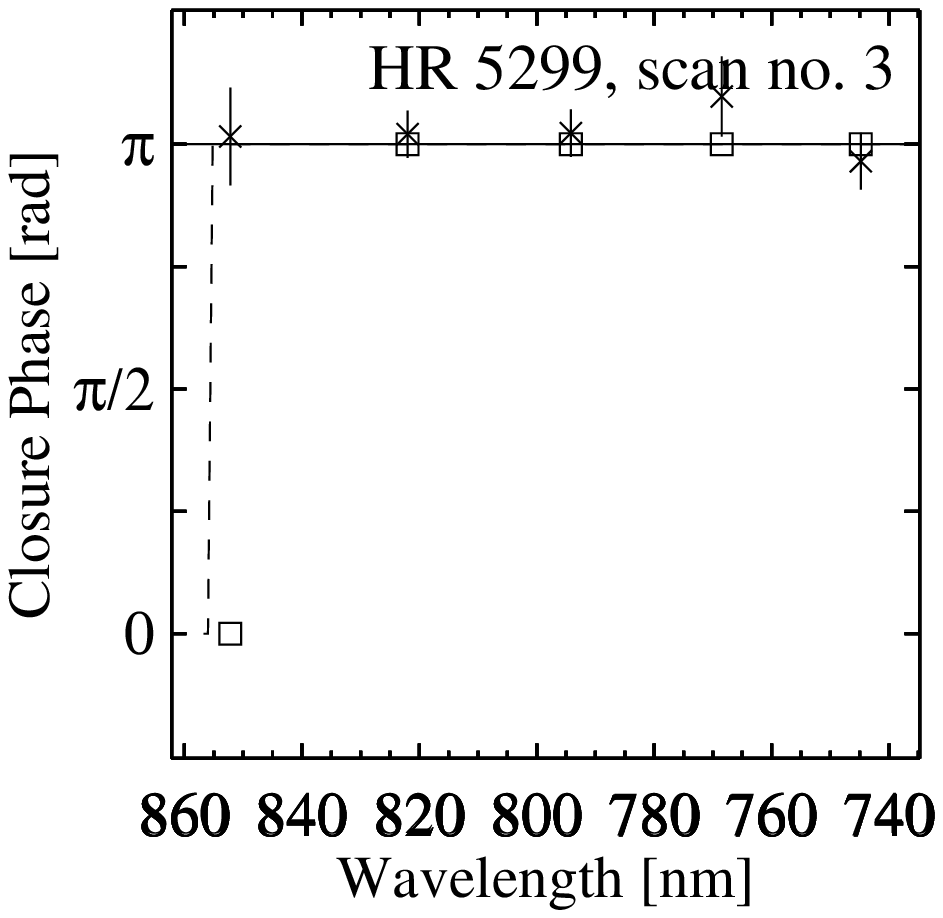}}

\resizebox{0.32\hsize}{!}{\includegraphics{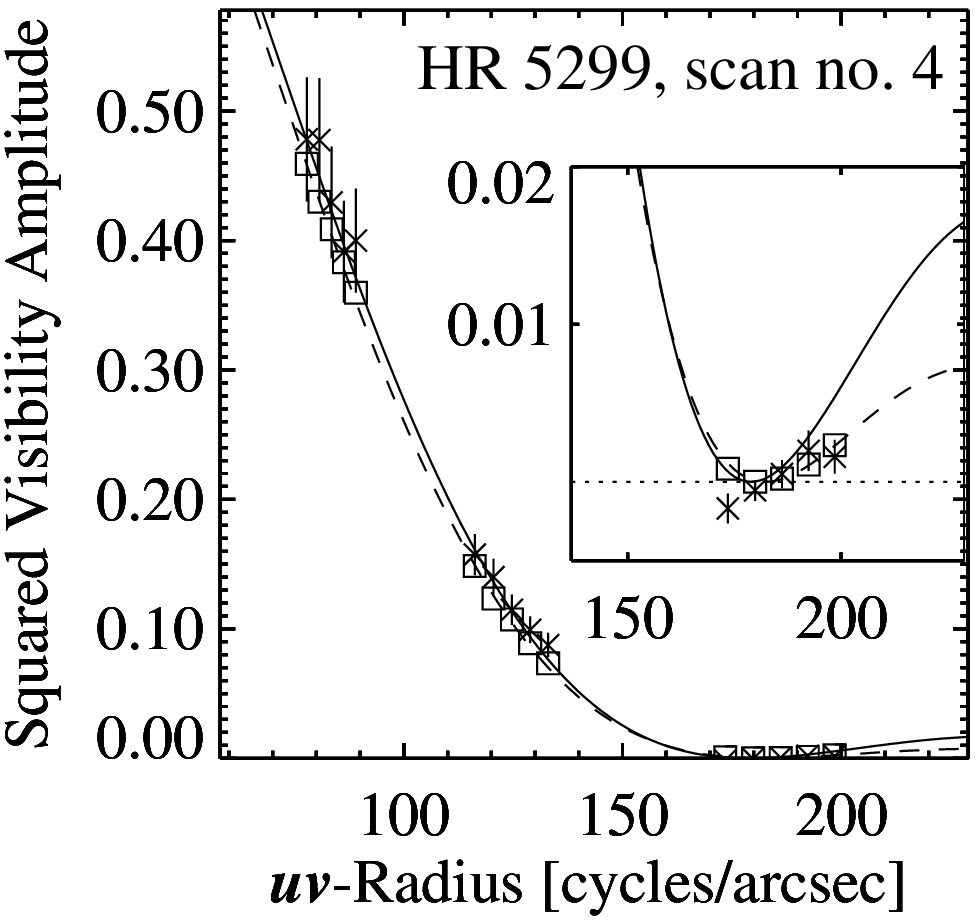}}
\hspace*{2mm}%
\resizebox{0.32\hsize}{!}{\includegraphics{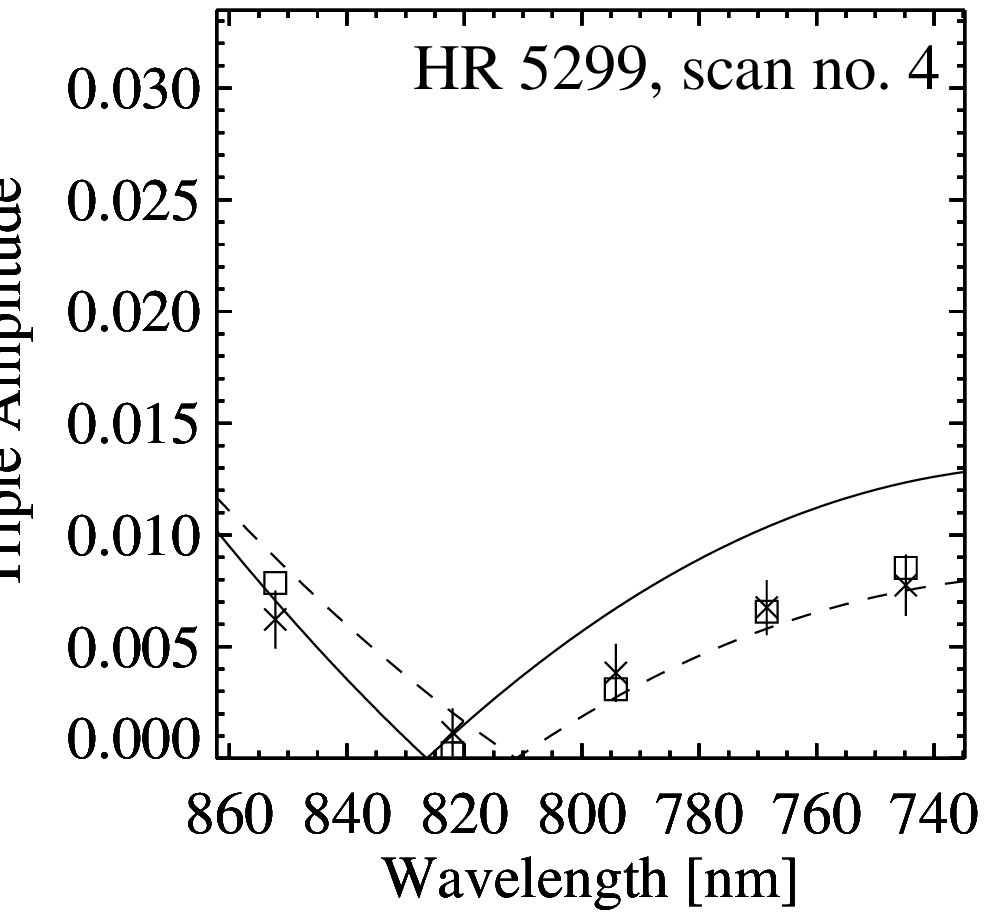}}
\hspace*{-1mm}%
\resizebox{0.32\hsize}{!}{\includegraphics{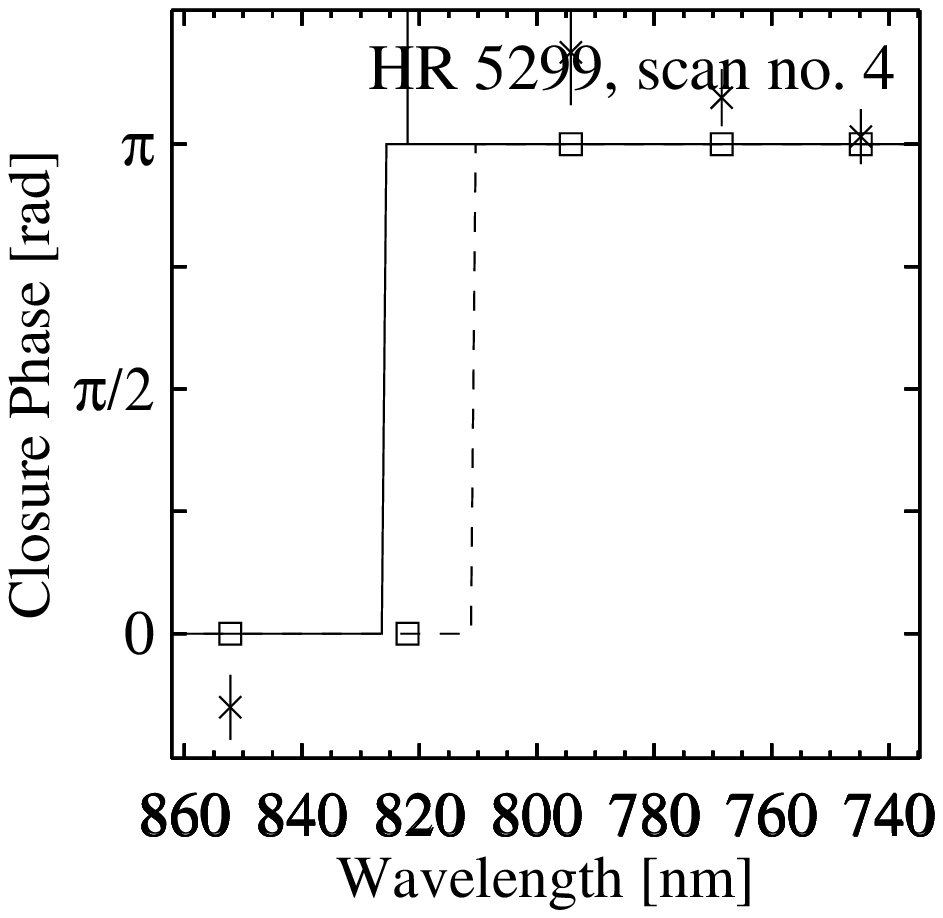}}

\resizebox{0.32\hsize}{!}{\includegraphics{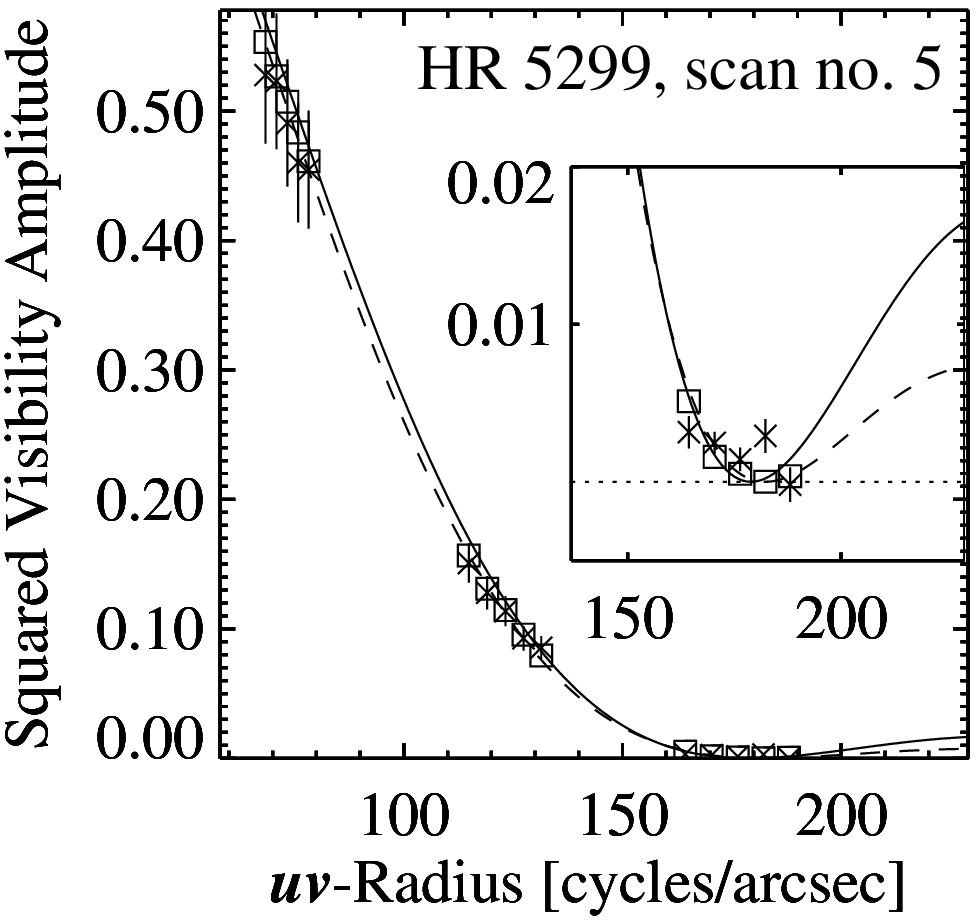}}
\hspace*{2mm}%
\resizebox{0.32\hsize}{!}{\includegraphics{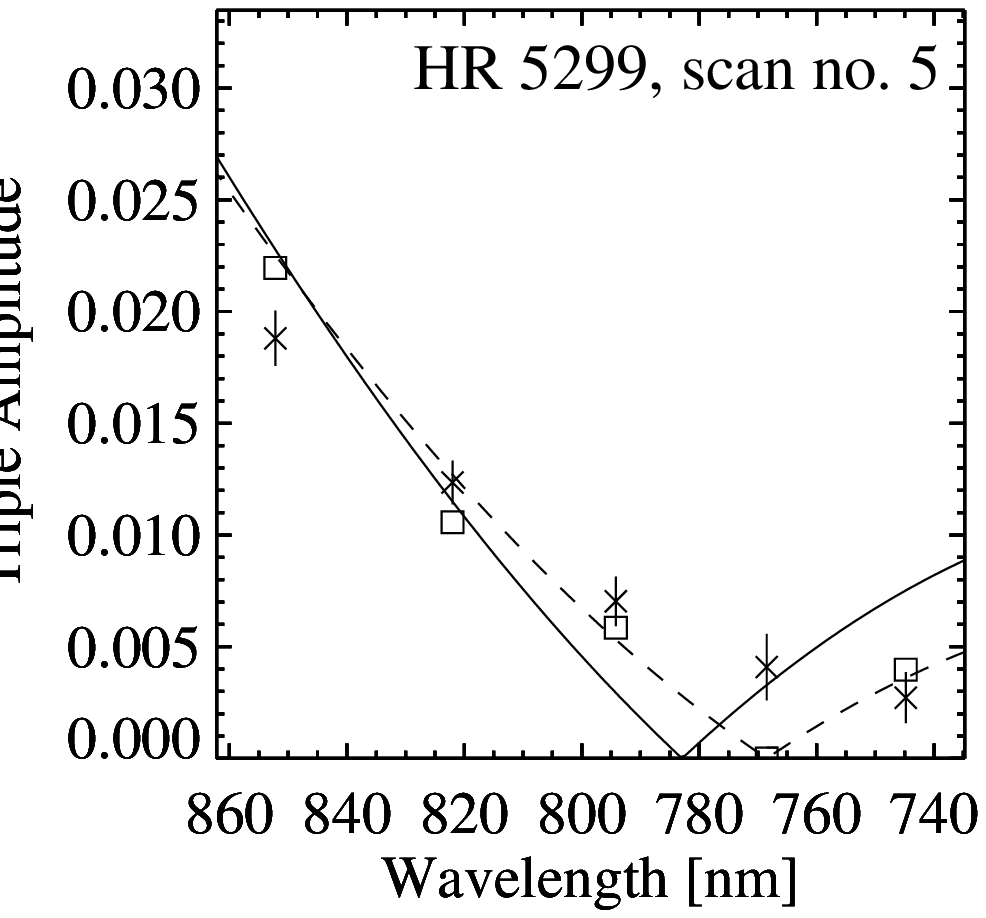}}
\hspace*{-1mm}%
\resizebox{0.32\hsize}{!}{\includegraphics{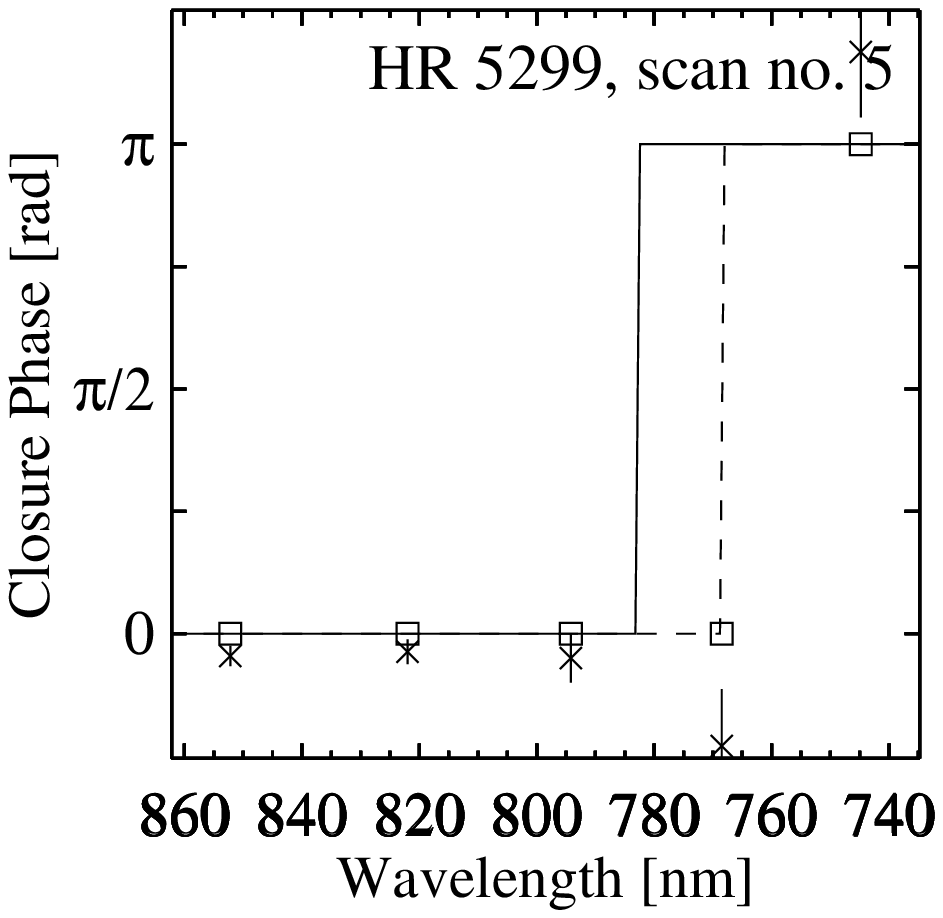}}

\resizebox{0.32\hsize}{!}{\includegraphics{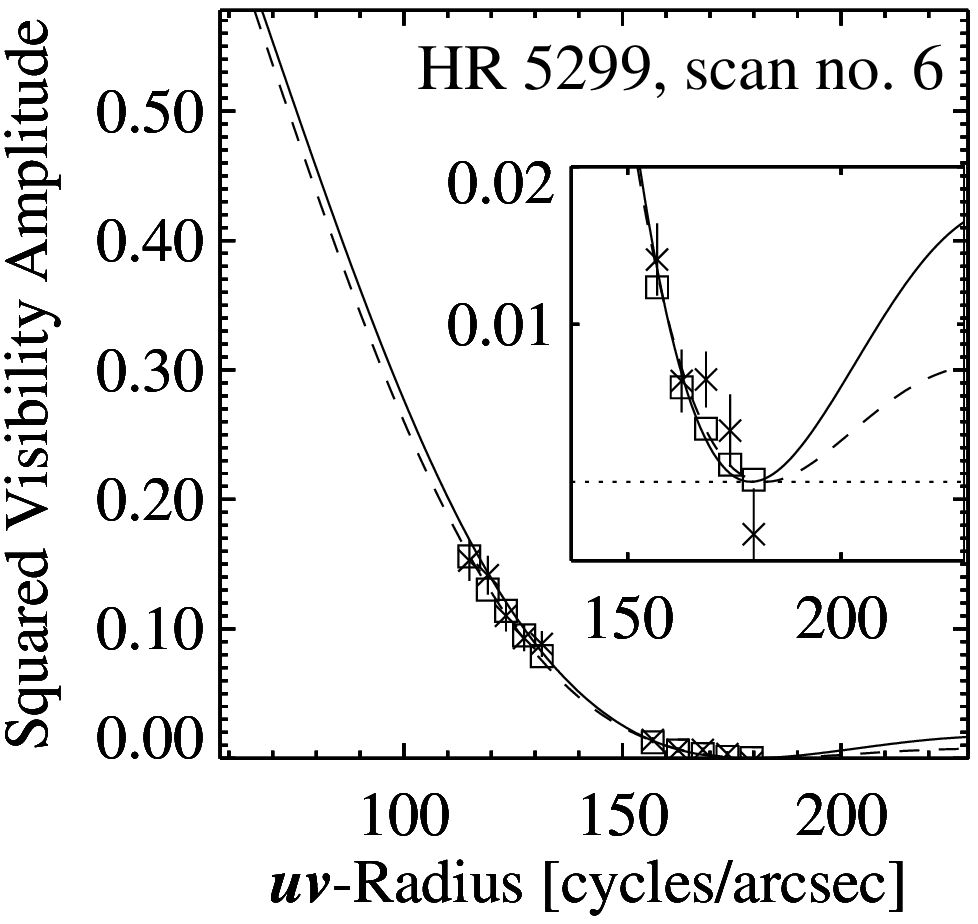}}
\hspace*{2mm}%
\resizebox{0.32\hsize}{!}{\includegraphics{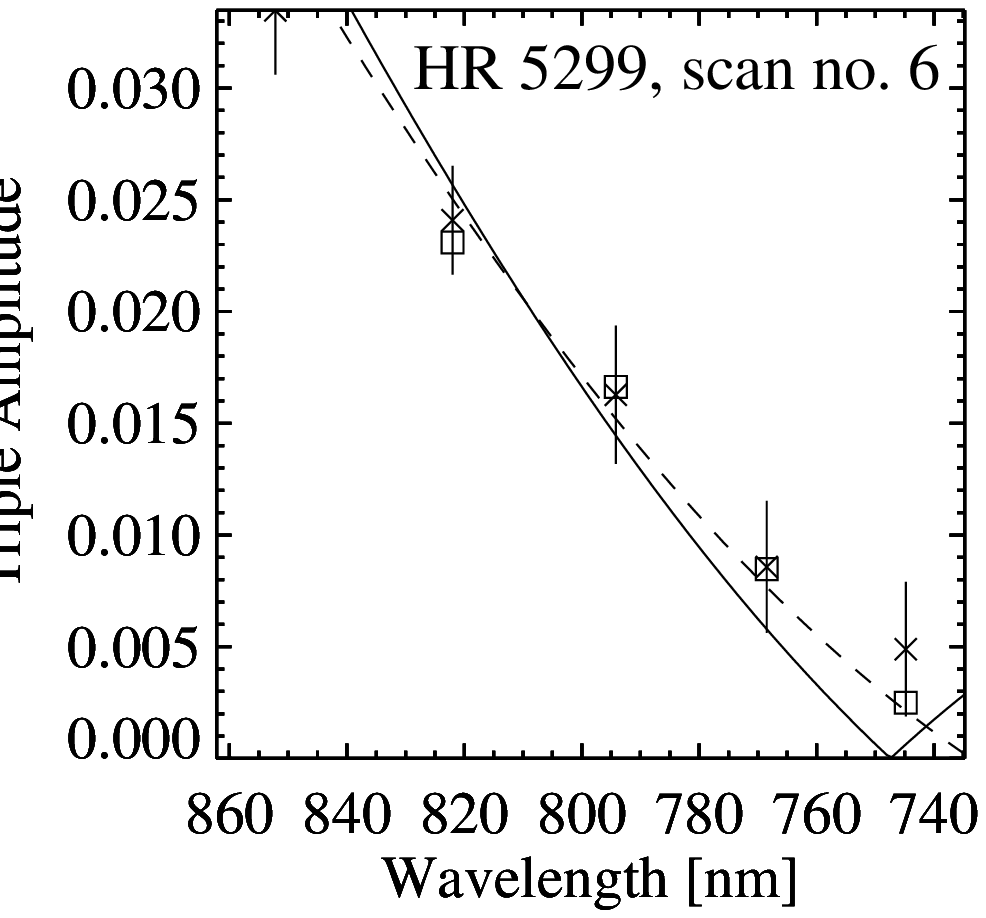}}
\hspace*{-1mm}%
\resizebox{0.32\hsize}{!}{\includegraphics{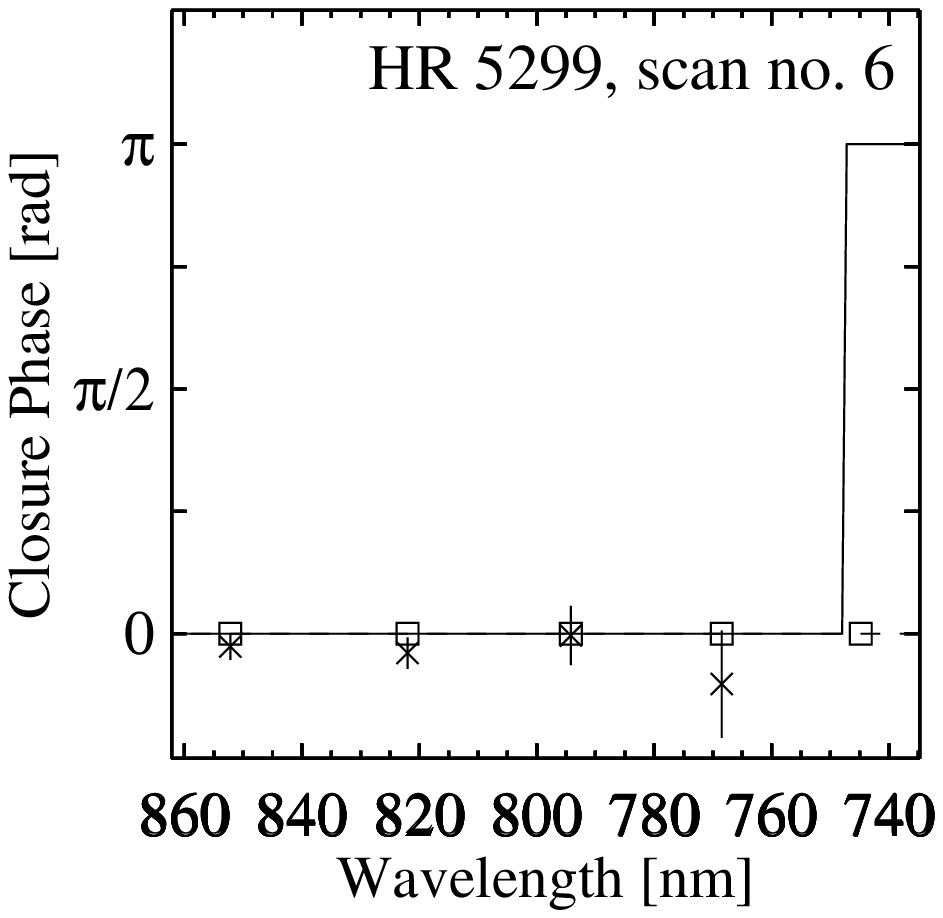}}
\end{minipage}
\parbox[b]{55mm}{
\caption{Squared visibility amplitudes (left), triple amplitudes (middle),
and closure phases (right) of HR\,5299 and of best fitting models, for
all obtained scans. The squared visibility amplitudes are plotted as
a function of the $uv$-radius, and the triple amplitudes and 
closure phases as a function of the spectral channels'
central wavelengths. The inset plots (left) provide an enlarged view
of the low squared visibility amplitude values around and beyond the 
first minimum.
The {\sf x}-symbols with error bars indicate the observations and their
errors, the solid line the model values based on a uniform disk, the dashed 
line those based on a fully darkened disk, and the squares those based on
the best fitting Kurucz stellar model atmosphere (Kurucz \cite{kurucz}). 
The error bars include the formal errors
as well as the calibration errors as described in Sect.~\ref{sec:reduction}.
\label{fig:fkv1368}}}
\end{figure*}
\begin{figure*}
\begin{minipage}[b]{12cm}
\resizebox{0.32\hsize}{!}{\includegraphics{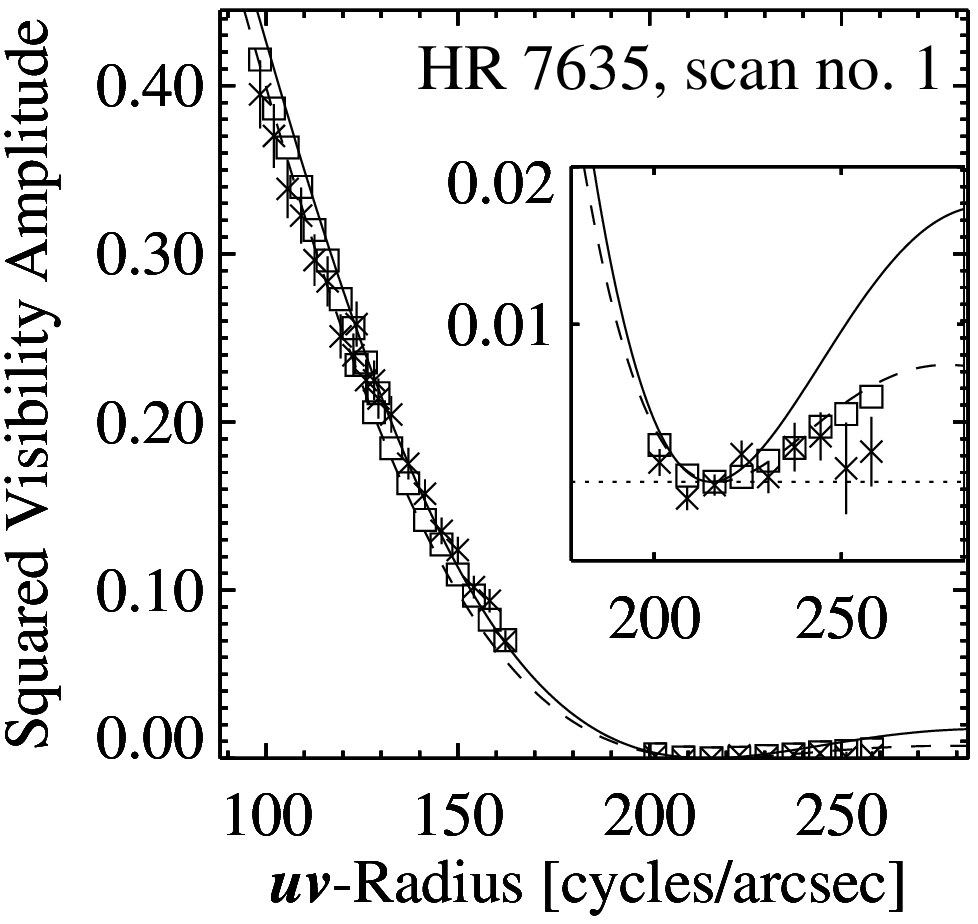}}
\hspace*{2mm}%
\resizebox{0.32\hsize}{!}{\includegraphics{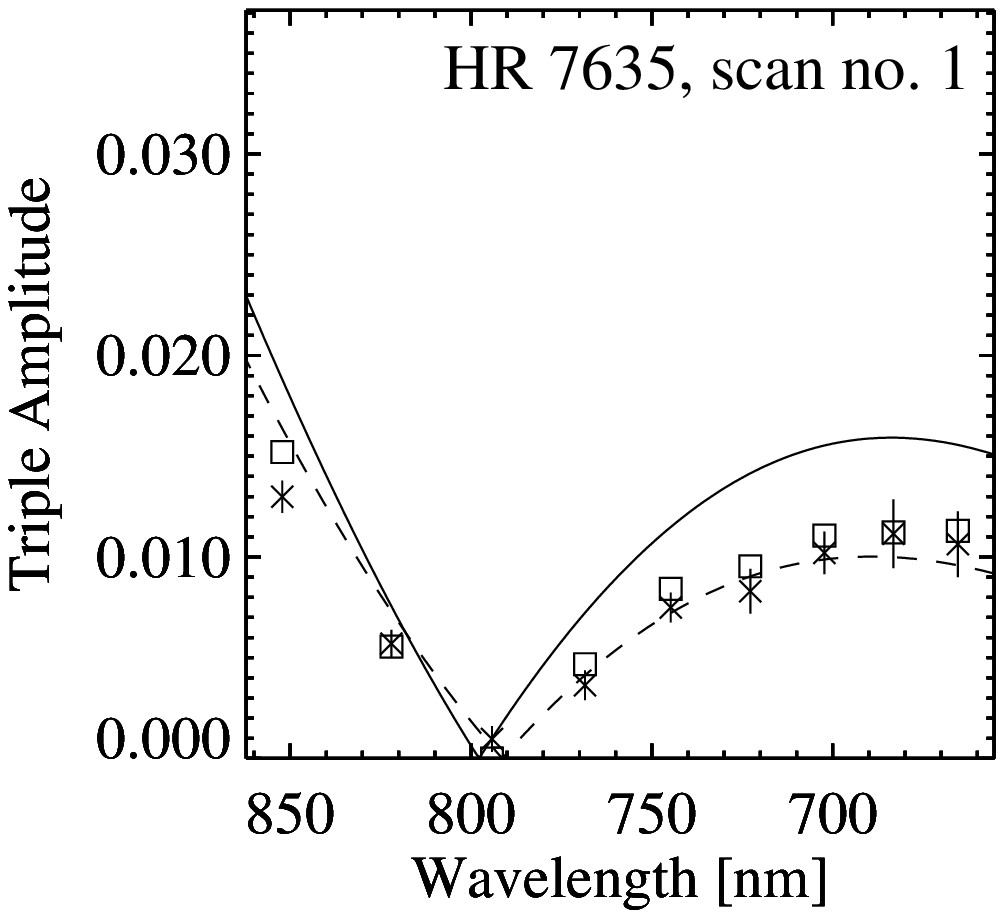}}
\hspace*{-1mm}%
\resizebox{0.32\hsize}{!}{\includegraphics{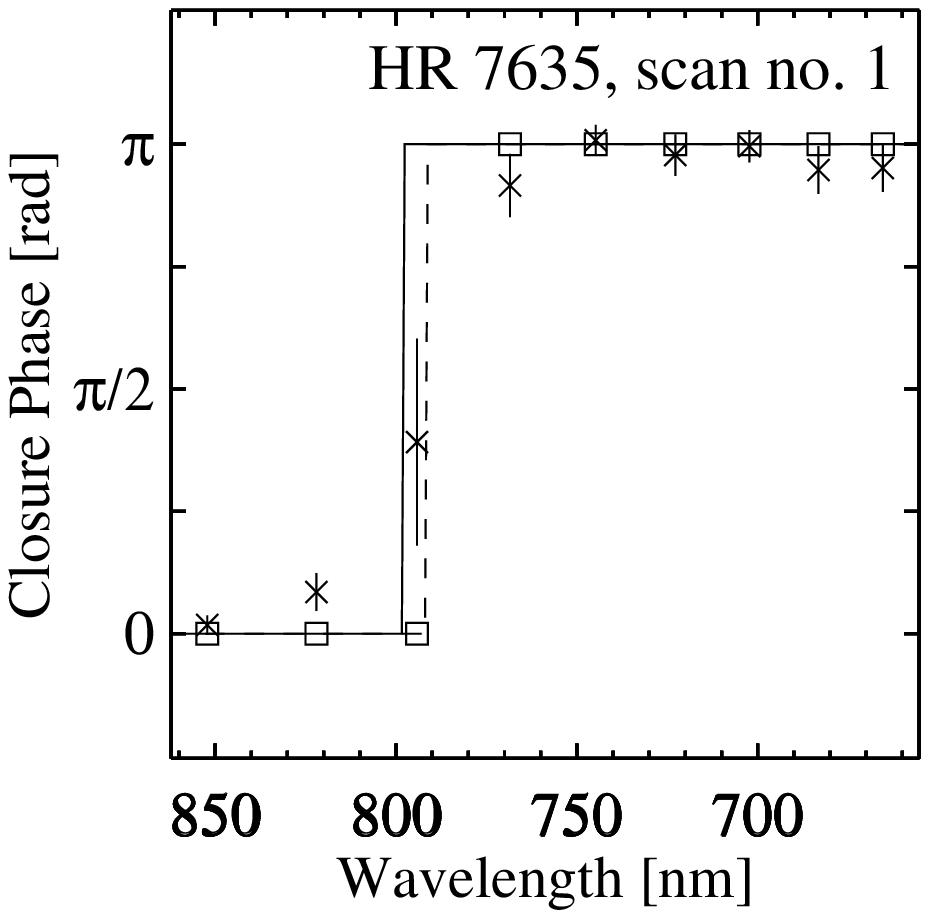}}

\resizebox{0.32\hsize}{!}{\includegraphics{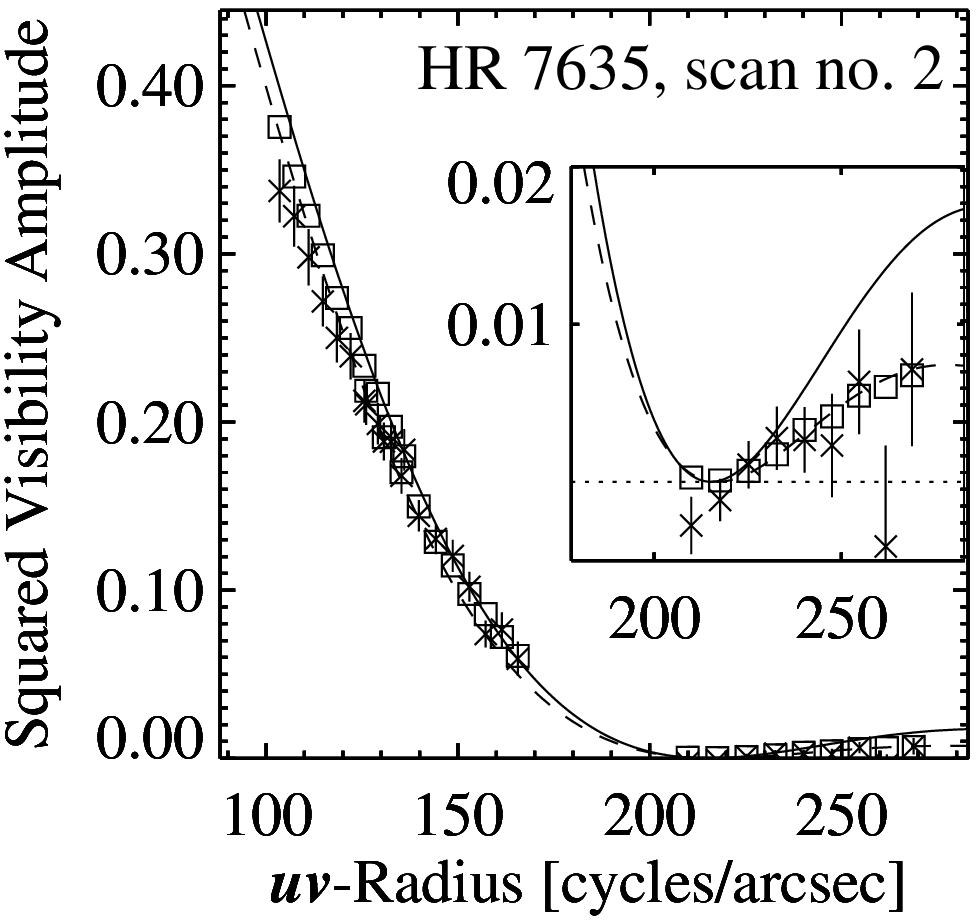}}
\hspace*{2mm}%
\resizebox{0.32\hsize}{!}{\includegraphics{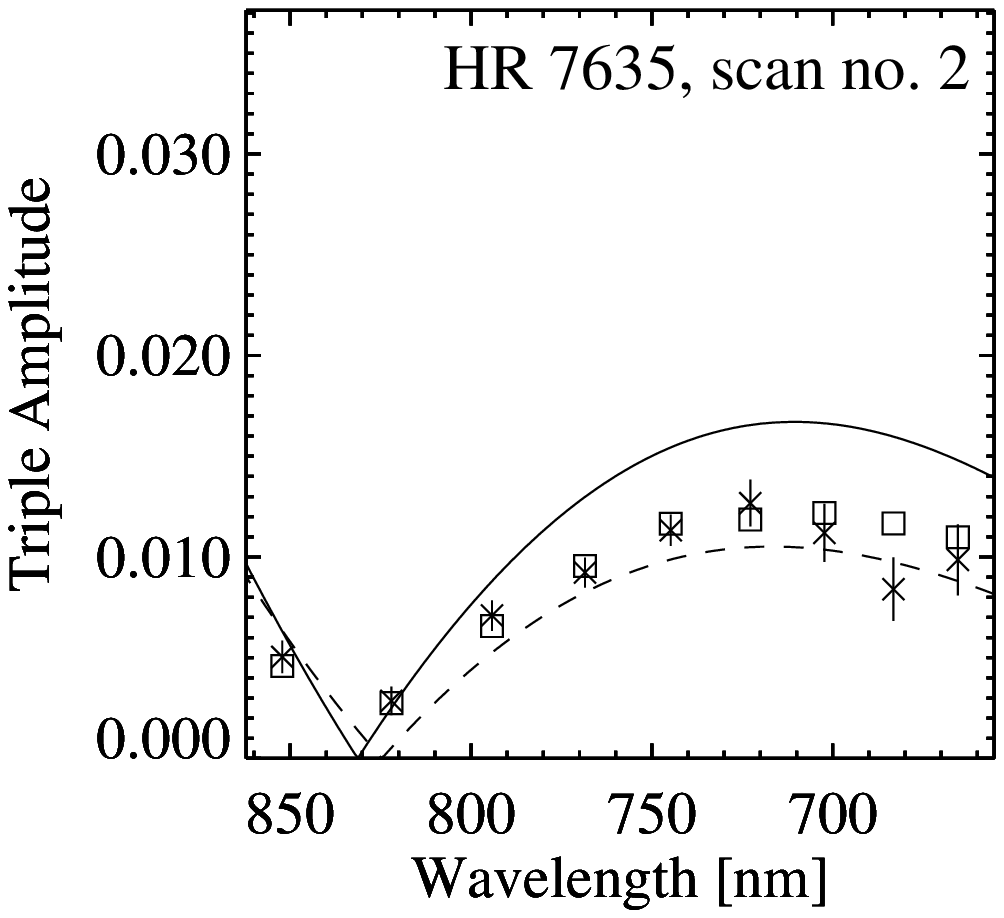}}
\hspace*{-1mm}%
\resizebox{0.32\hsize}{!}{\includegraphics{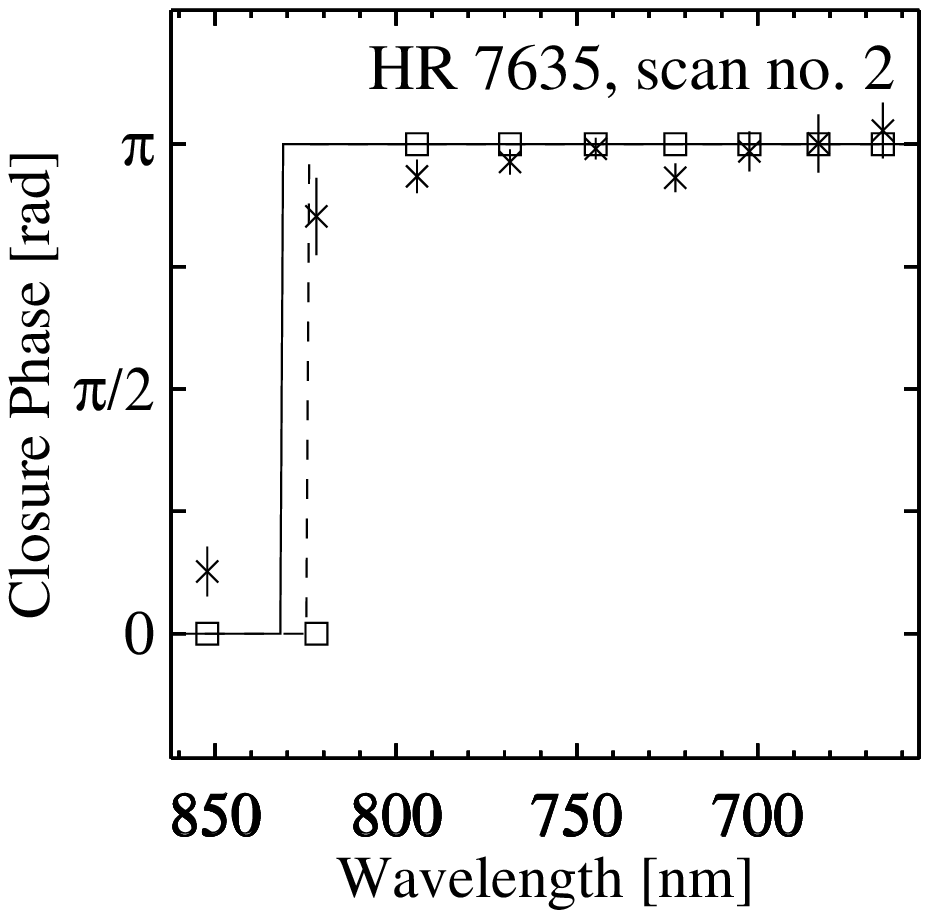}}

\resizebox{0.32\hsize}{!}{\includegraphics{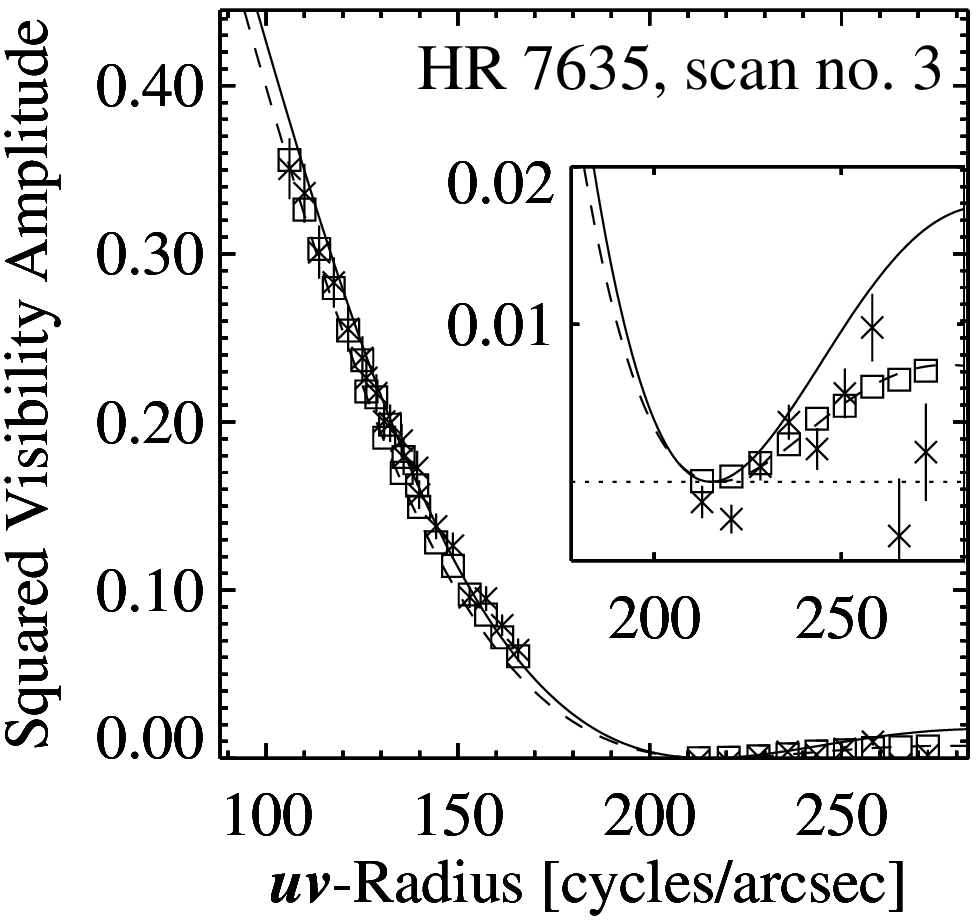}}
\hspace*{2mm}%
\resizebox{0.32\hsize}{!}{\includegraphics{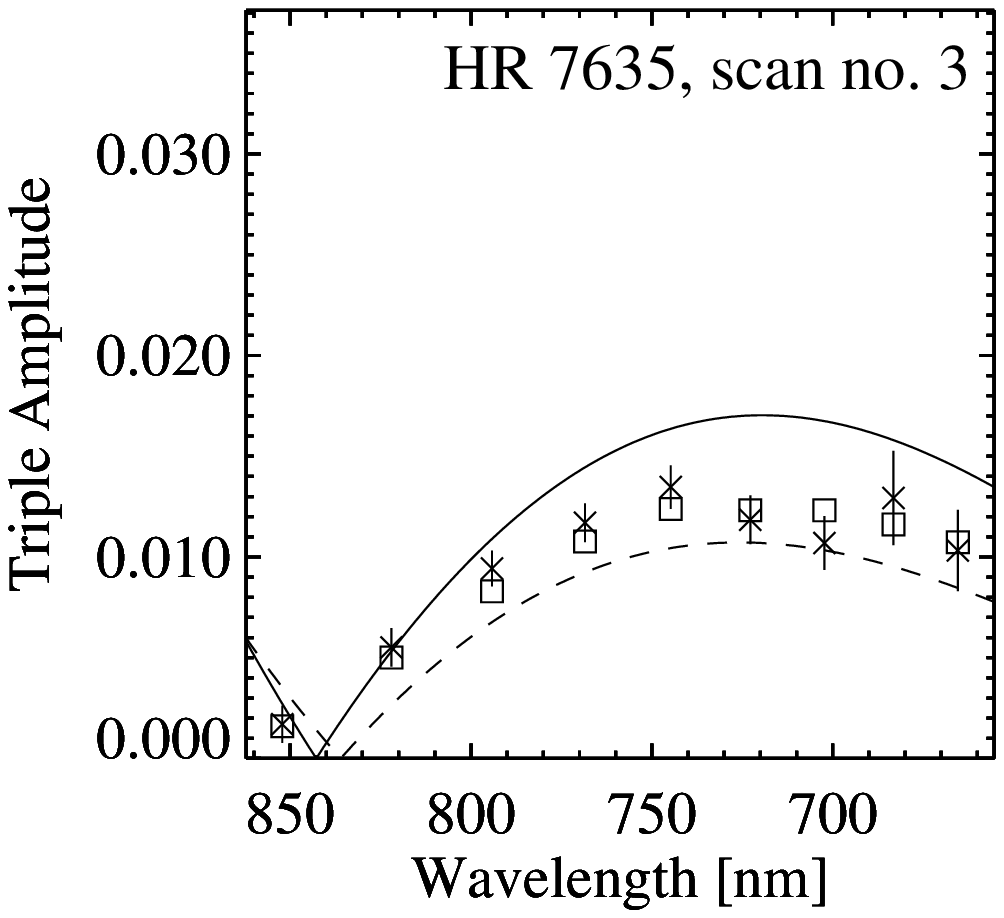}}
\hspace*{-1mm}%
\resizebox{0.32\hsize}{!}{\includegraphics{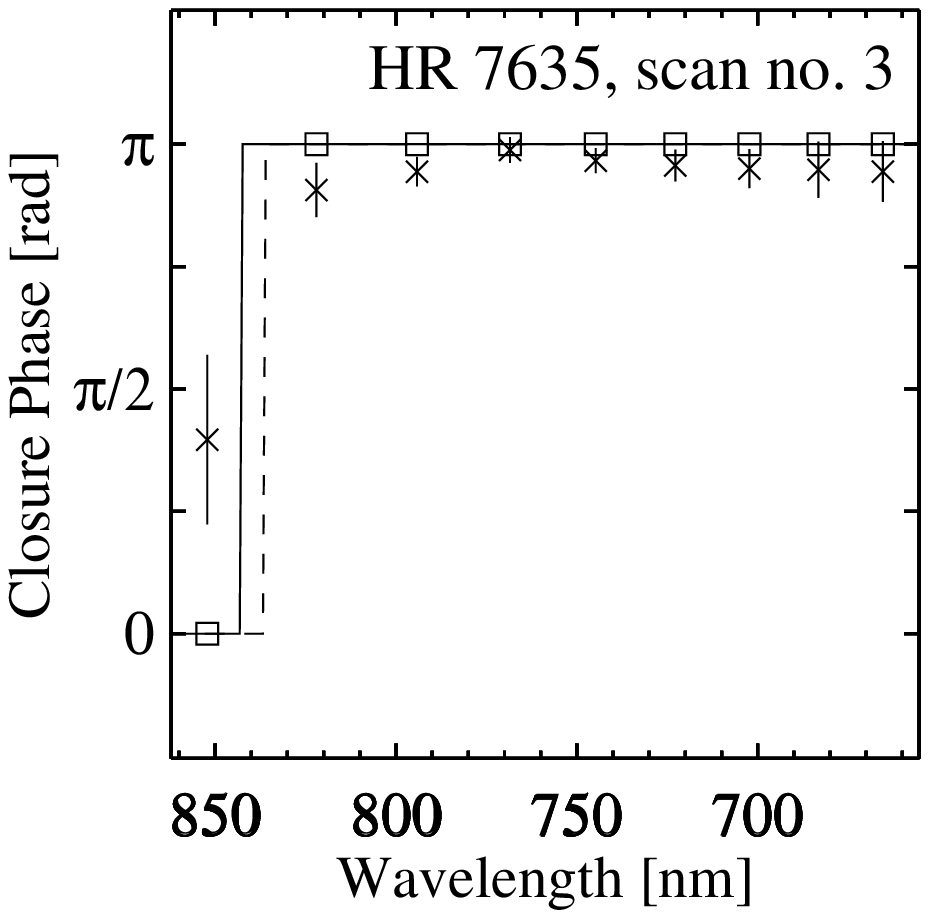}}

\resizebox{0.32\hsize}{!}{\includegraphics{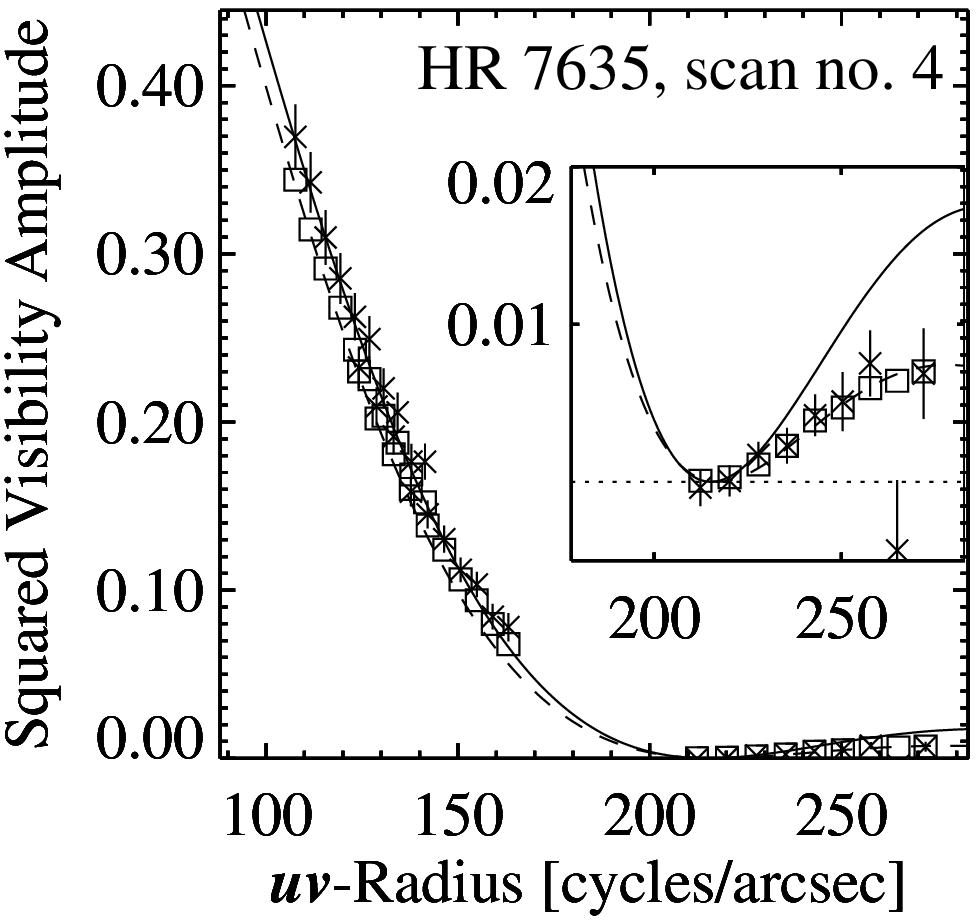}}
\hspace*{2mm}%
\resizebox{0.32\hsize}{!}{\includegraphics{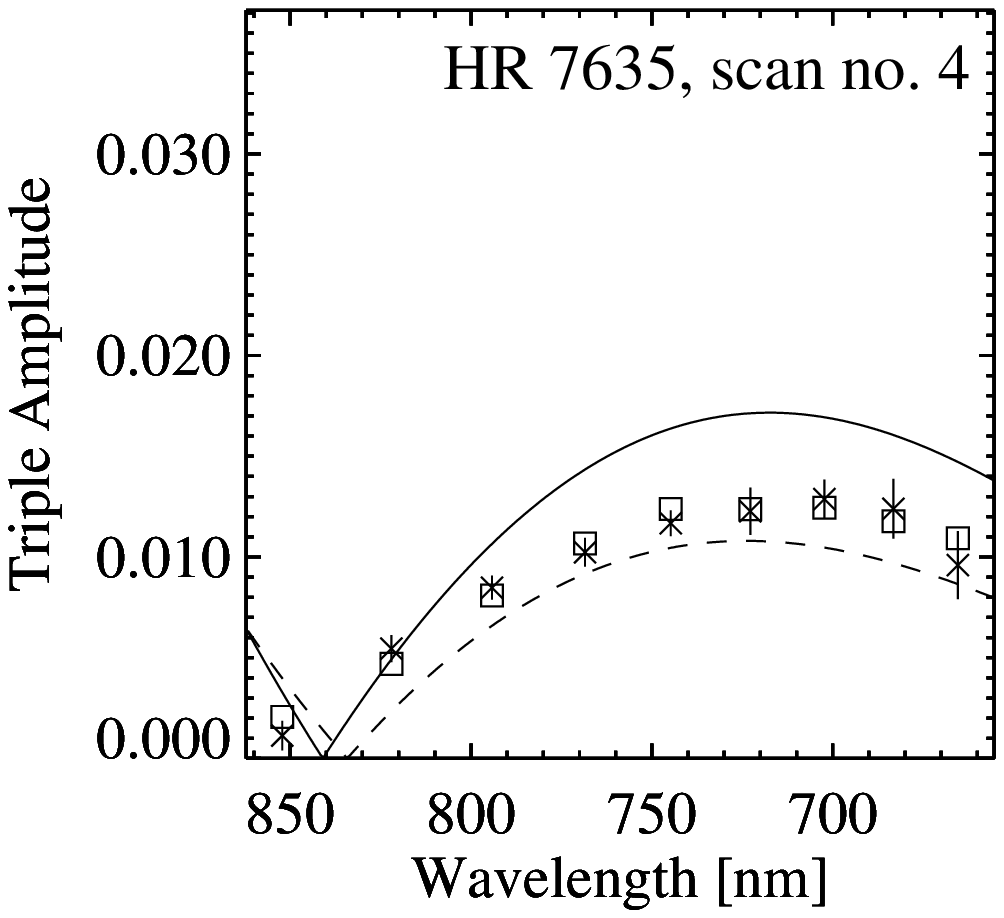}}
\hspace*{-1mm}%
\resizebox{0.32\hsize}{!}{\includegraphics{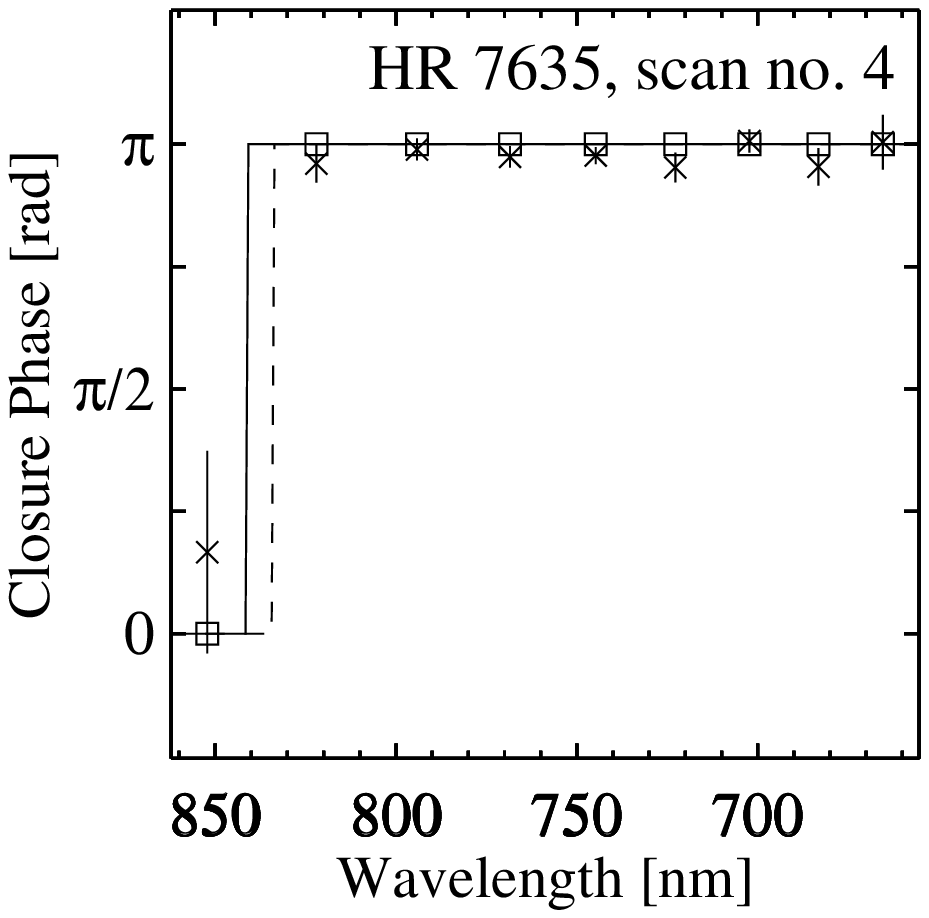}}

\resizebox{0.32\hsize}{!}{\includegraphics{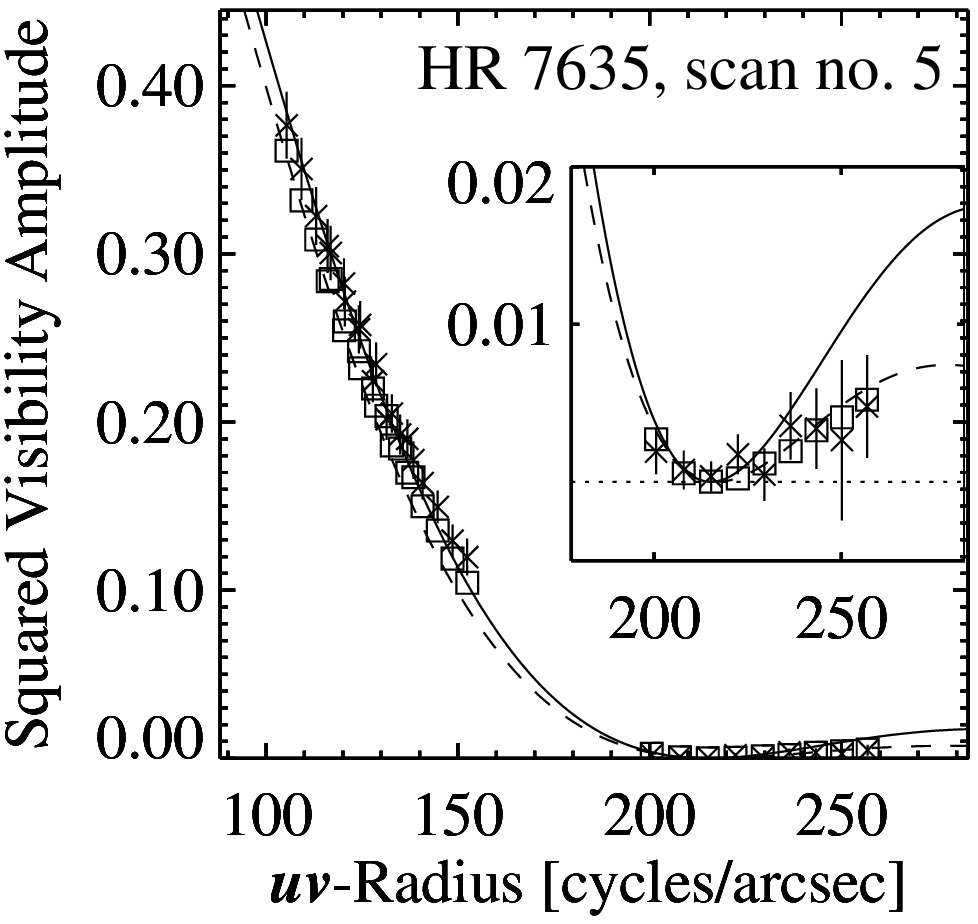}}
\hspace*{2mm}%
\resizebox{0.32\hsize}{!}{\includegraphics{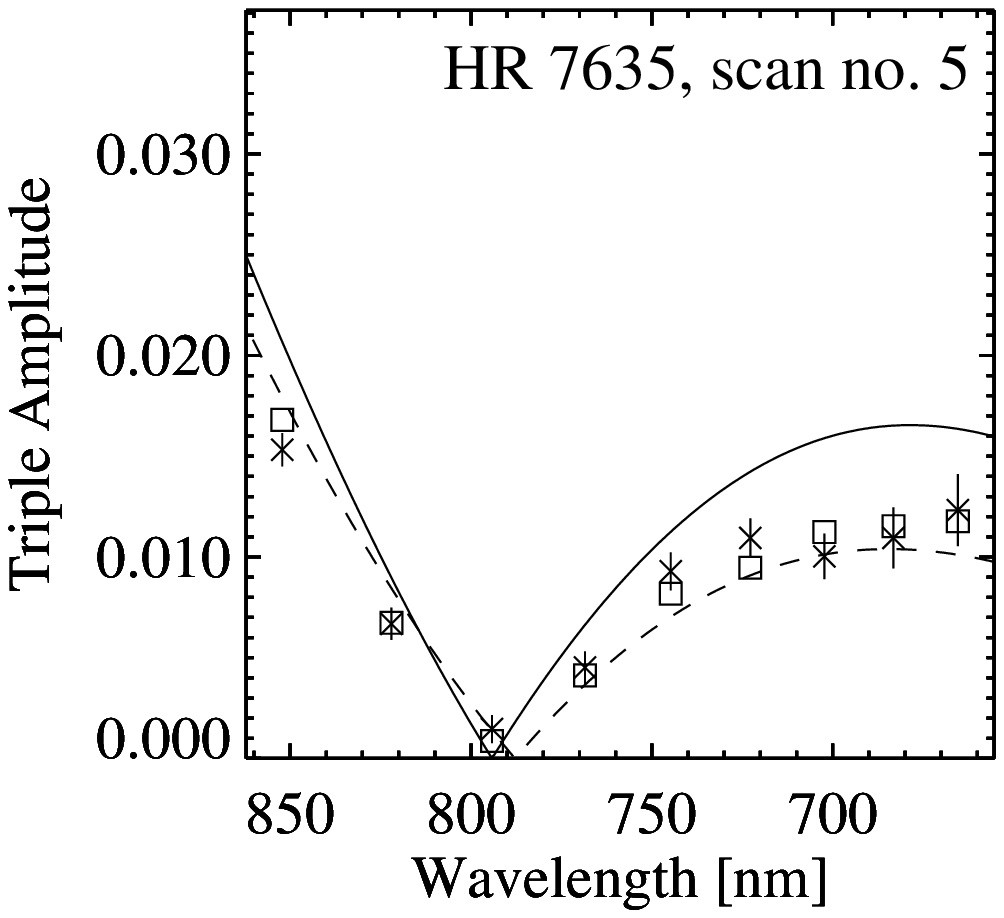}}
\hspace*{-1mm}%
\resizebox{0.32\hsize}{!}{\includegraphics{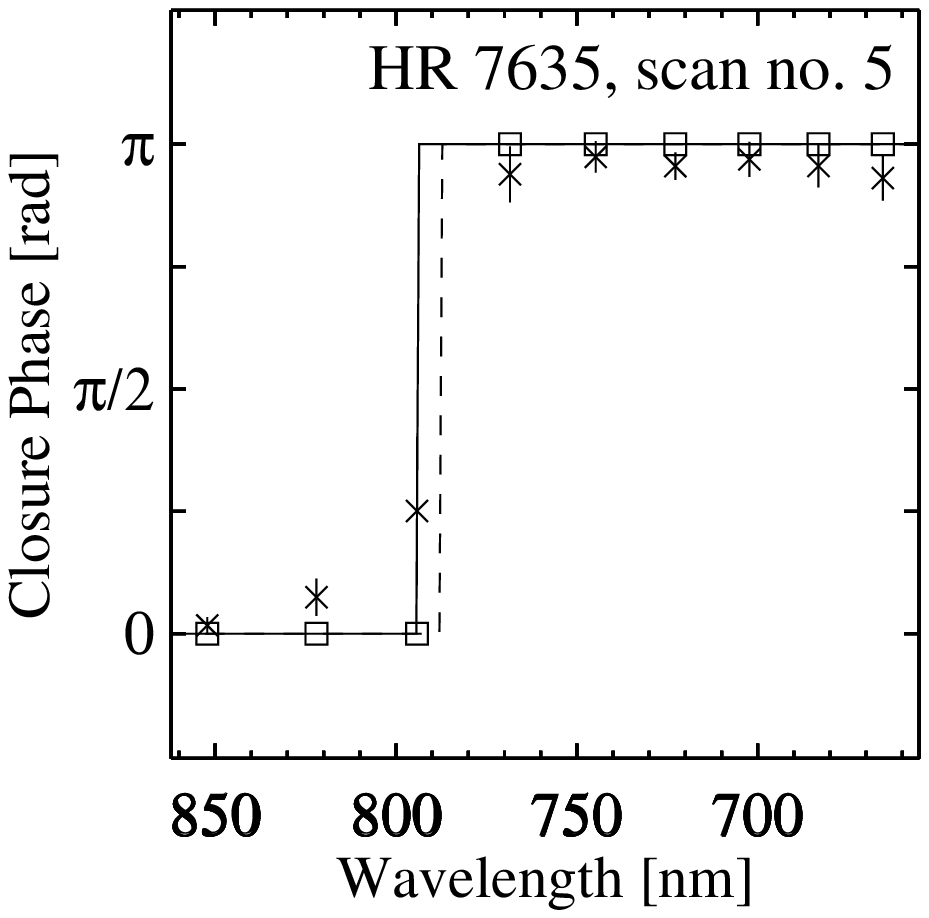}}

\resizebox{0.32\hsize}{!}{\includegraphics{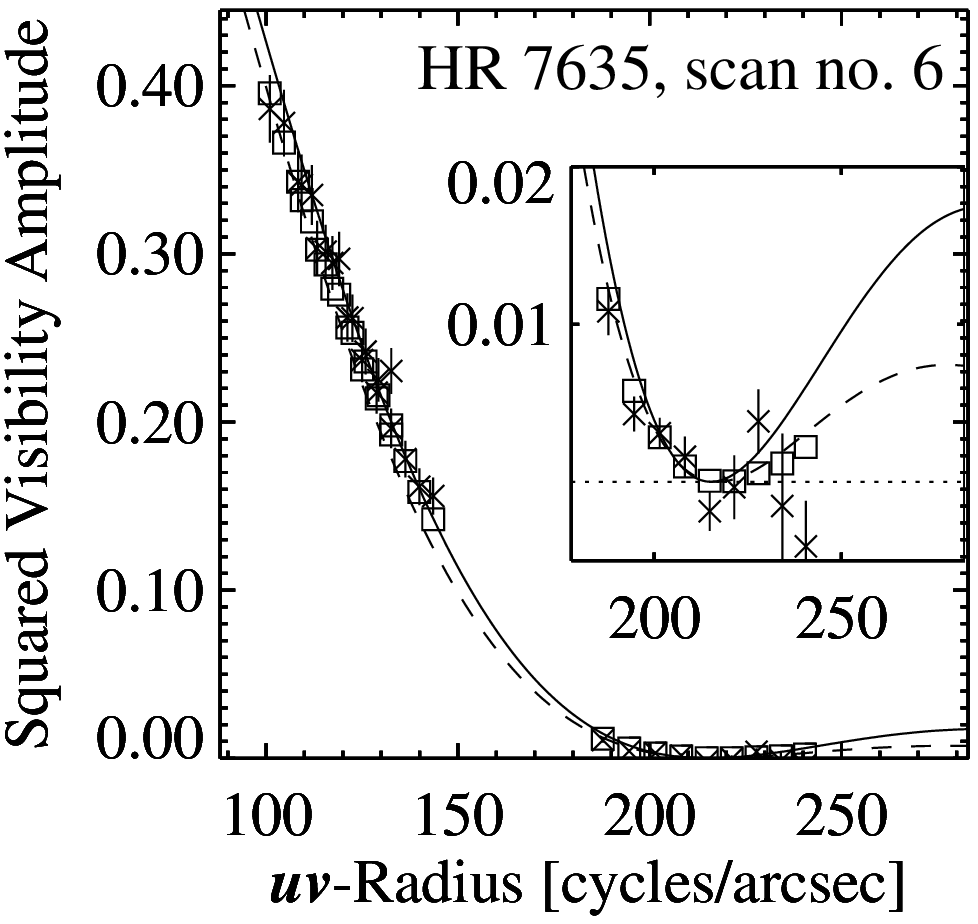}}
\hspace*{2mm}%
\resizebox{0.32\hsize}{!}{\includegraphics{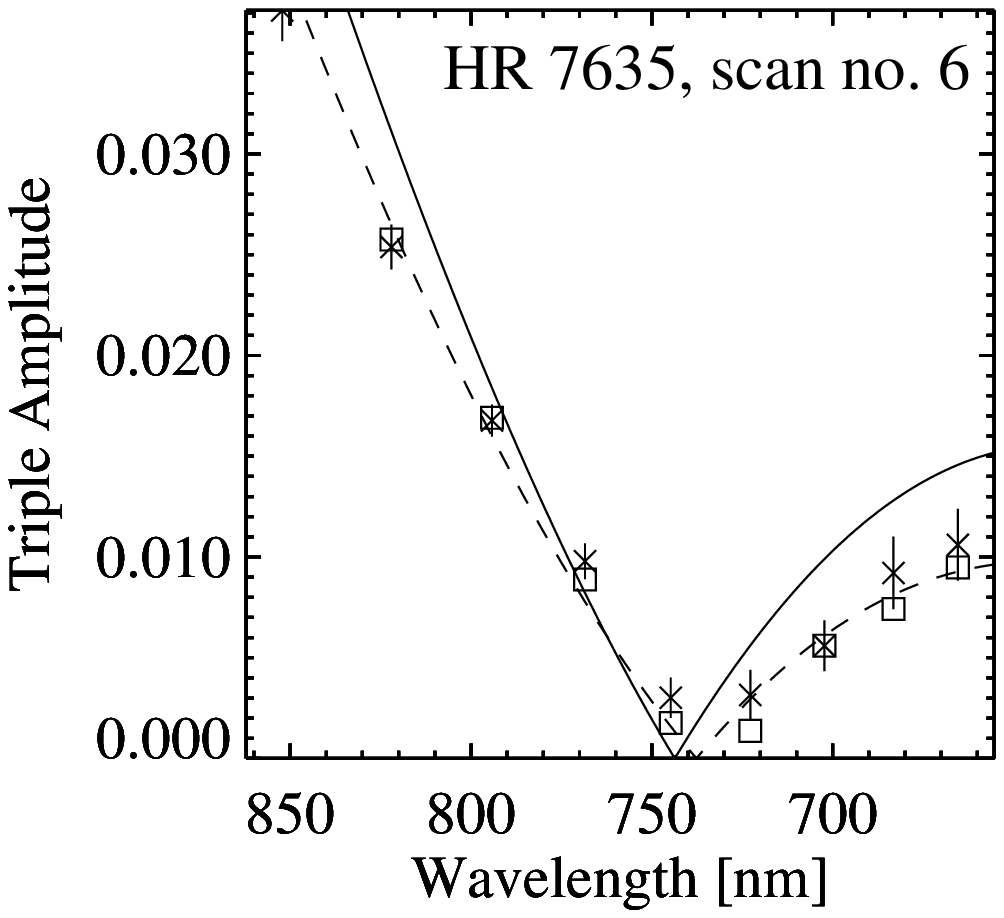}}
\hspace*{-1mm}%
\resizebox{0.32\hsize}{!}{\includegraphics{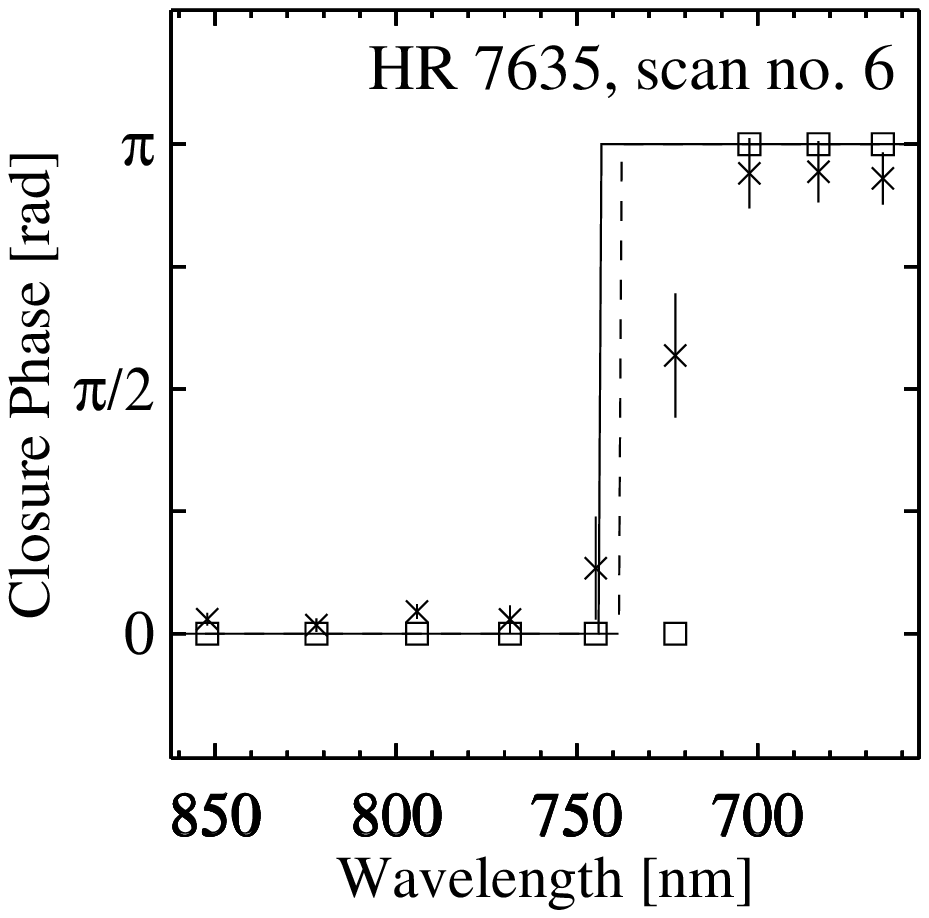}}
\end{minipage}
\parbox[b]{55mm}{
\caption{As Fig.~\ref{fig:fkv1368}, but for HR\,7635.}
\label{fig:fkv752}}
\end{figure*}
\begin{figure*}
\begin{minipage}[b]{12cm}
\resizebox{0.32\hsize}{!}{\includegraphics{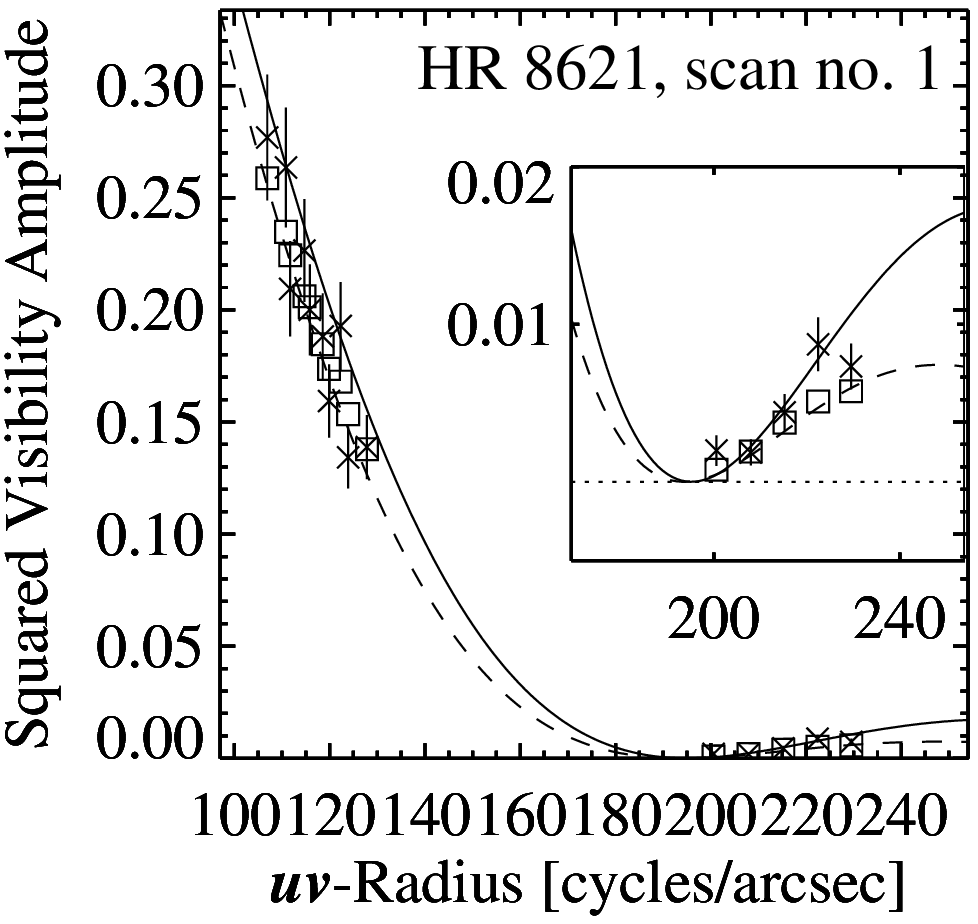}}
\hspace*{2mm}%
\resizebox{0.32\hsize}{!}{\includegraphics{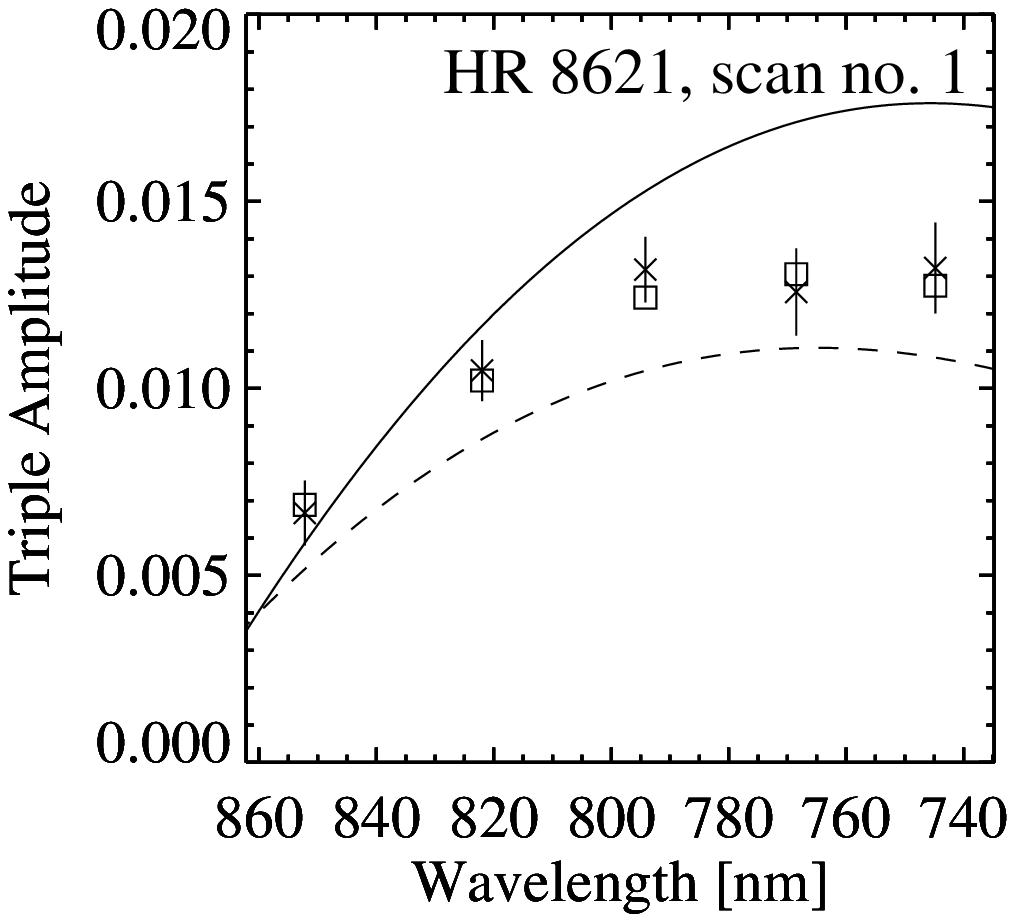}}
\hspace*{-1mm}%
\resizebox{0.32\hsize}{!}{\includegraphics{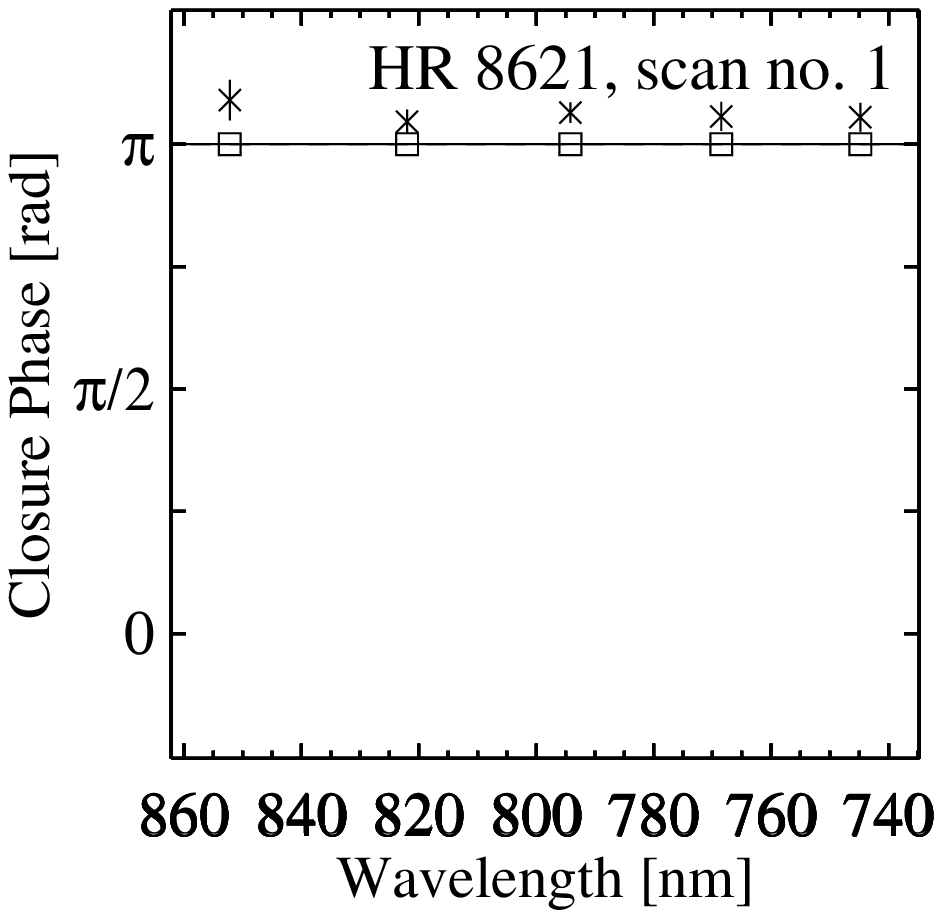}}

\resizebox{0.32\hsize}{!}{\includegraphics{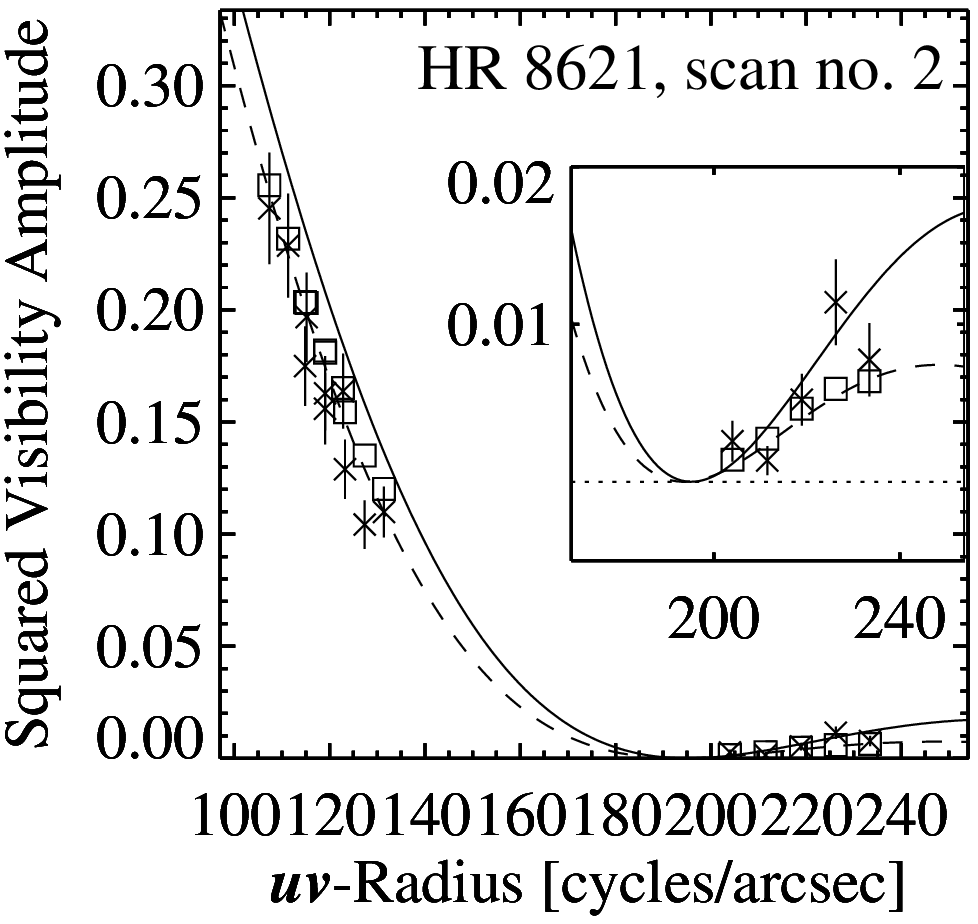}}
\hspace*{2mm}%
\resizebox{0.32\hsize}{!}{\includegraphics{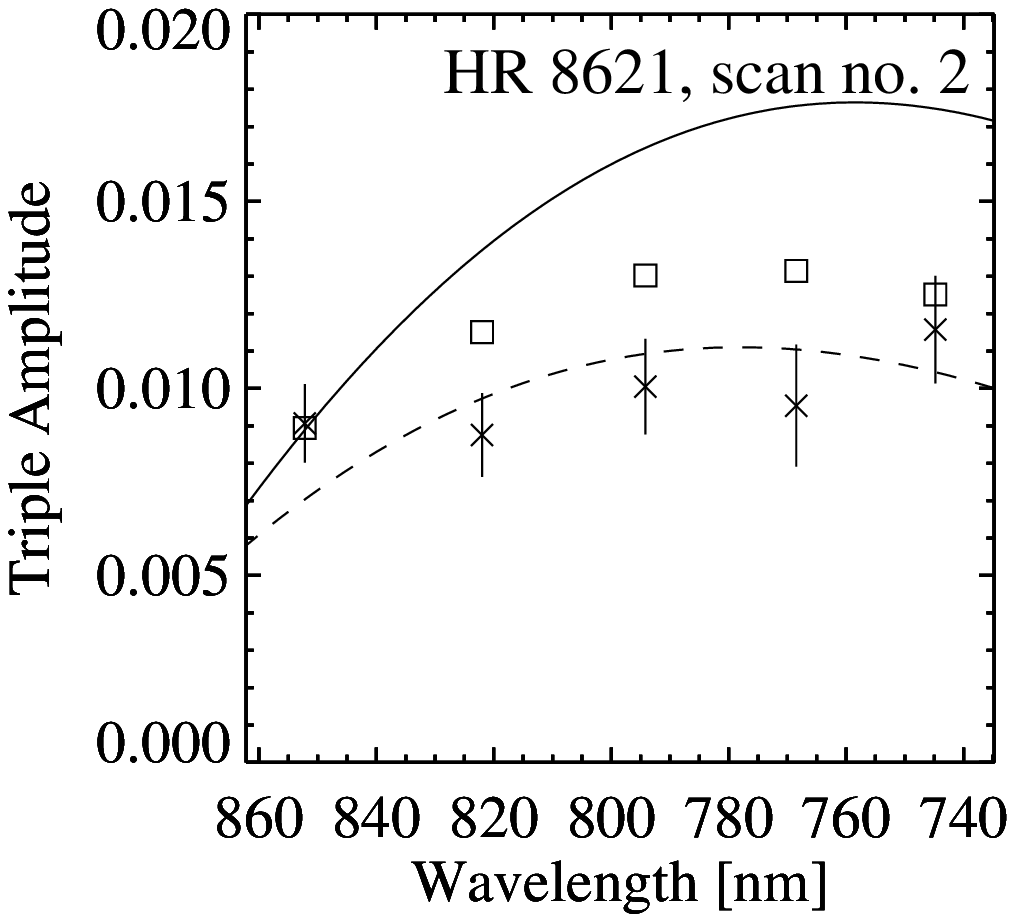}}
\hspace*{-1mm}%
\resizebox{0.32\hsize}{!}{\includegraphics{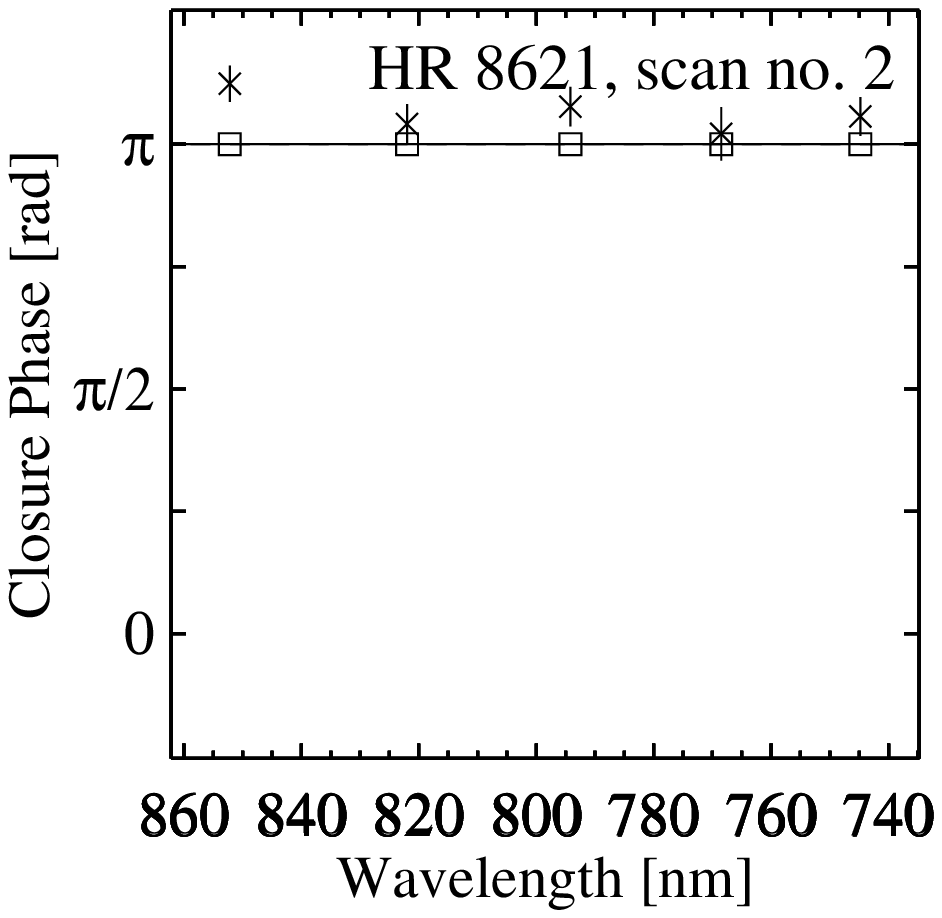}}

\resizebox{0.32\hsize}{!}{\includegraphics{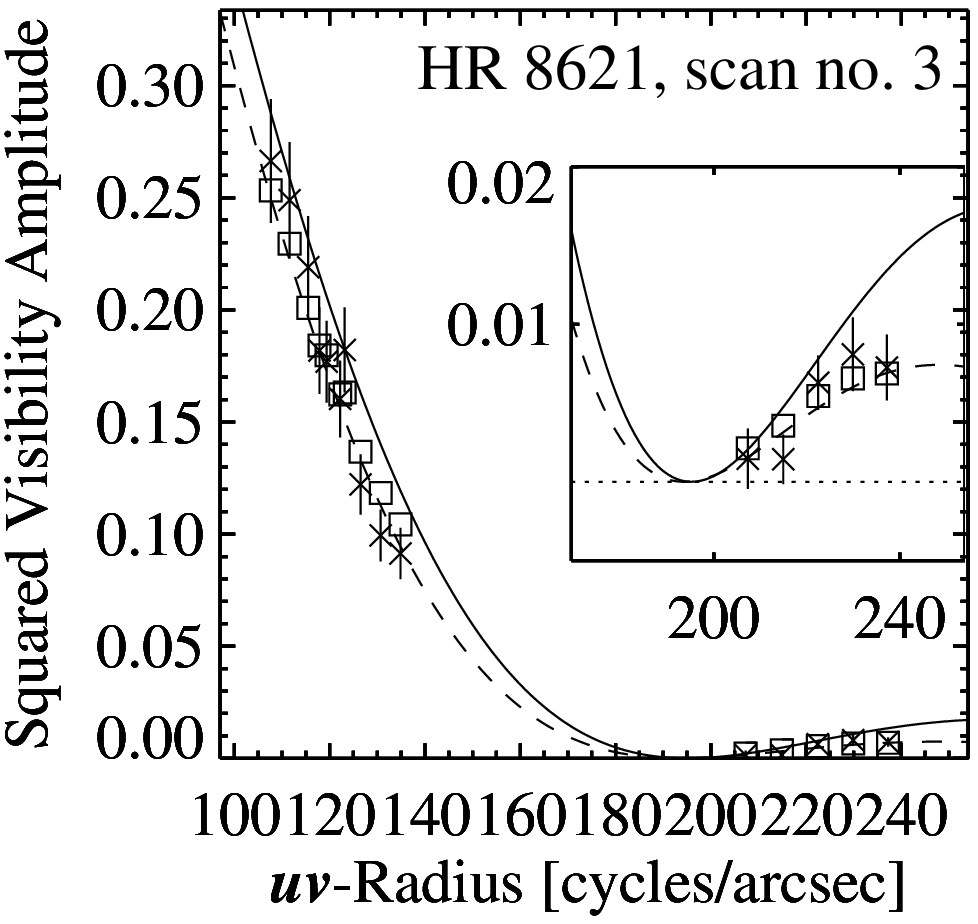}}
\hspace*{2mm}%
\resizebox{0.32\hsize}{!}{\includegraphics{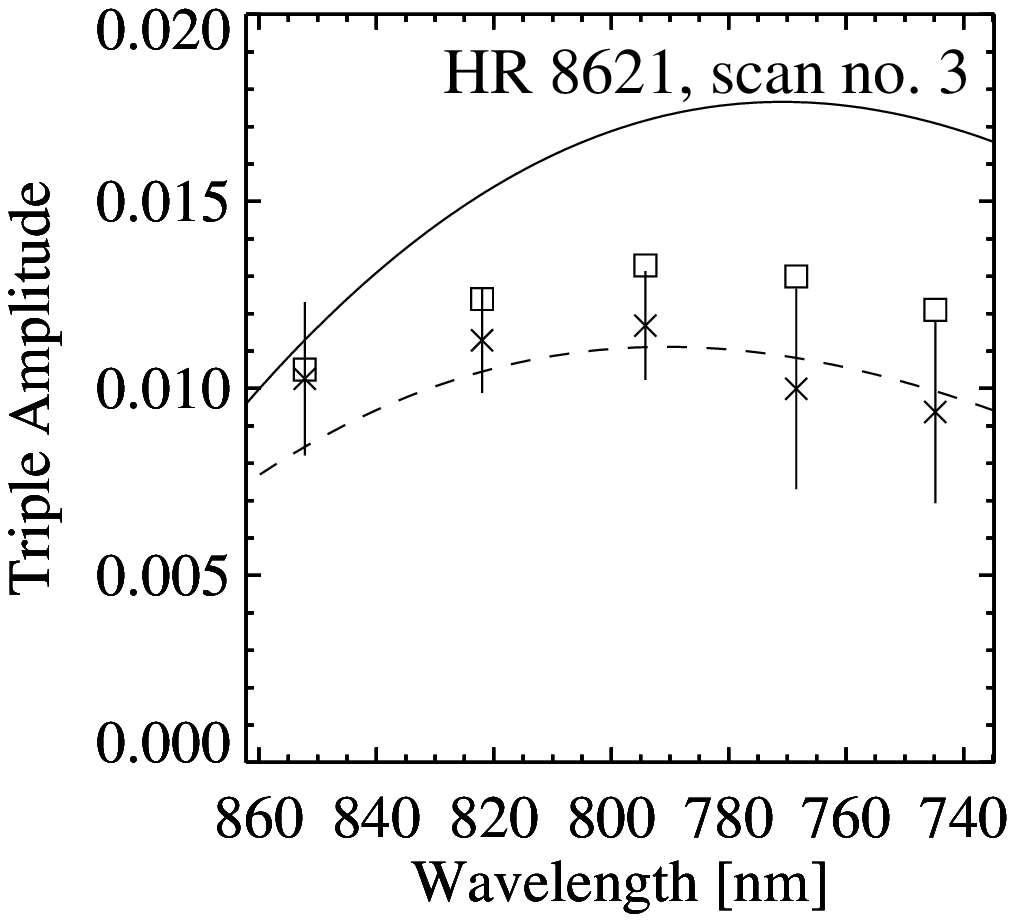}}
\hspace*{-1mm}%
\resizebox{0.32\hsize}{!}{\includegraphics{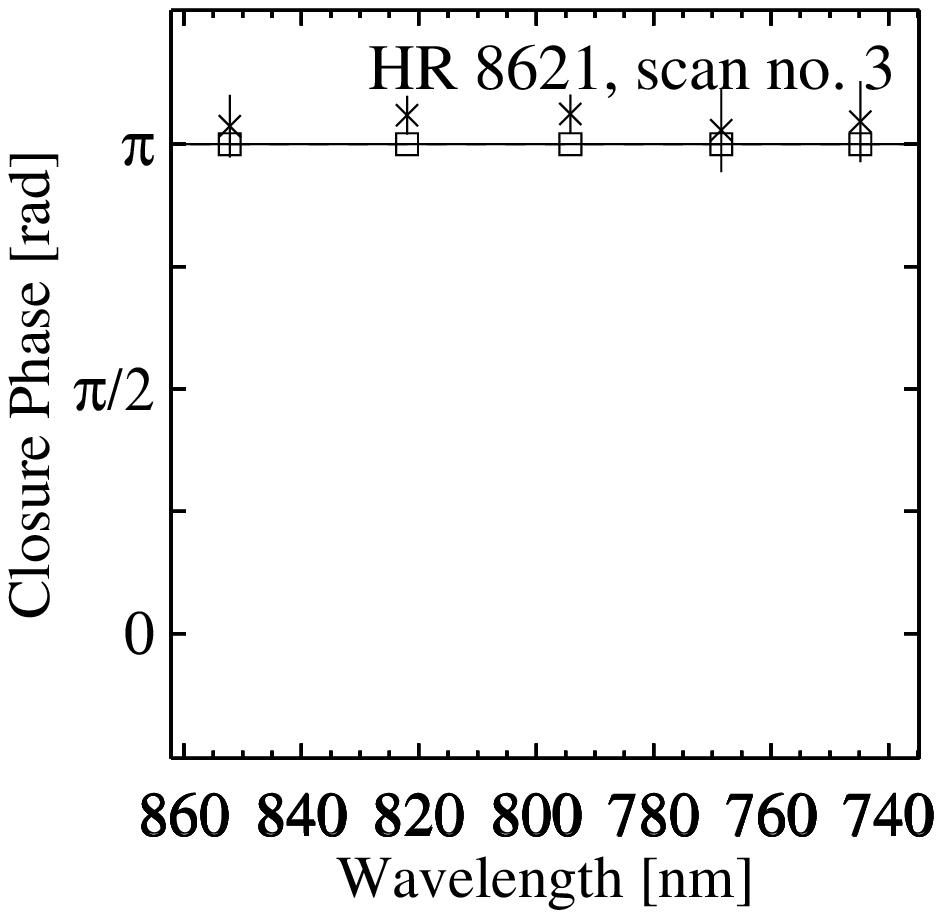}}

\resizebox{0.32\hsize}{!}{\includegraphics{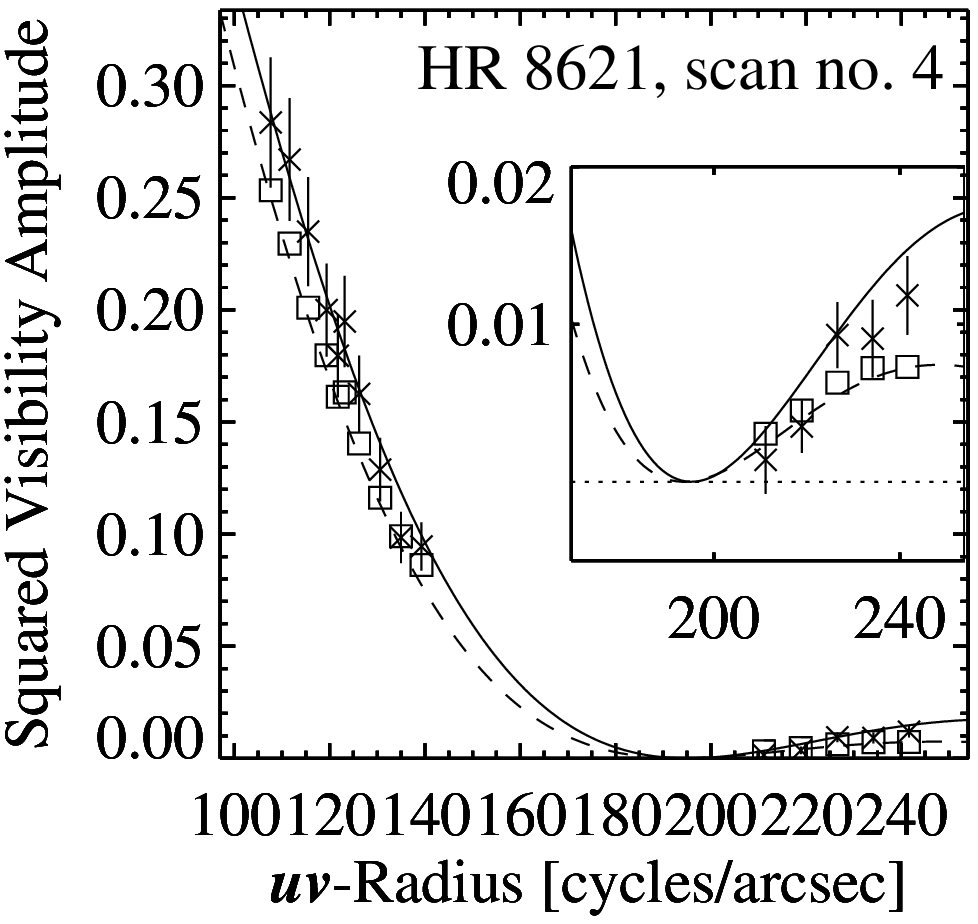}}
\hspace*{2mm}%
\resizebox{0.32\hsize}{!}{\includegraphics{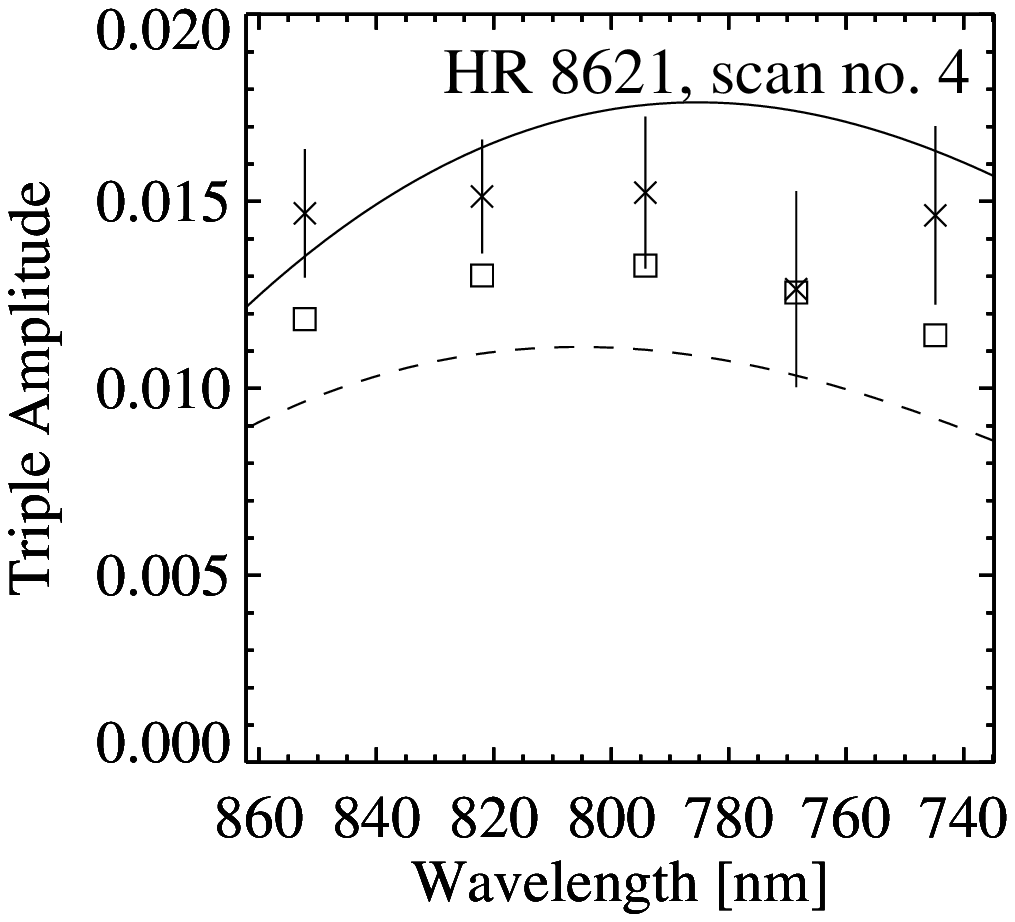}}
\hspace*{-1mm}%
\resizebox{0.32\hsize}{!}{\includegraphics{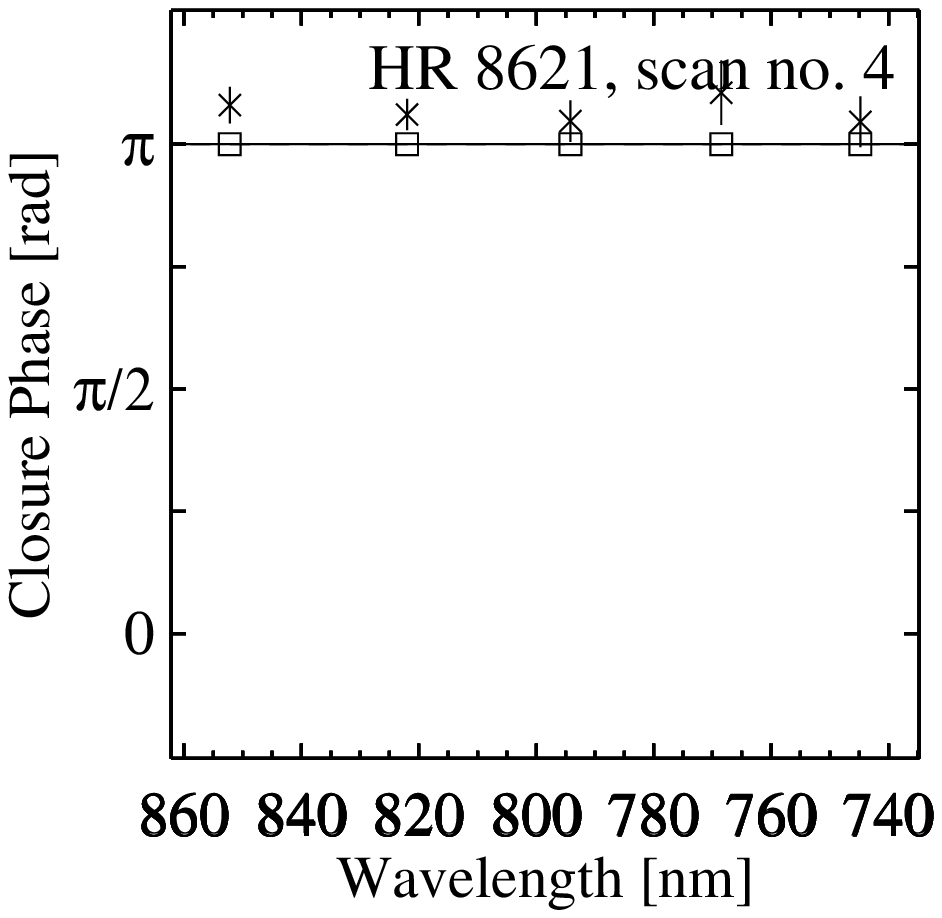}}

\resizebox{0.32\hsize}{!}{\includegraphics{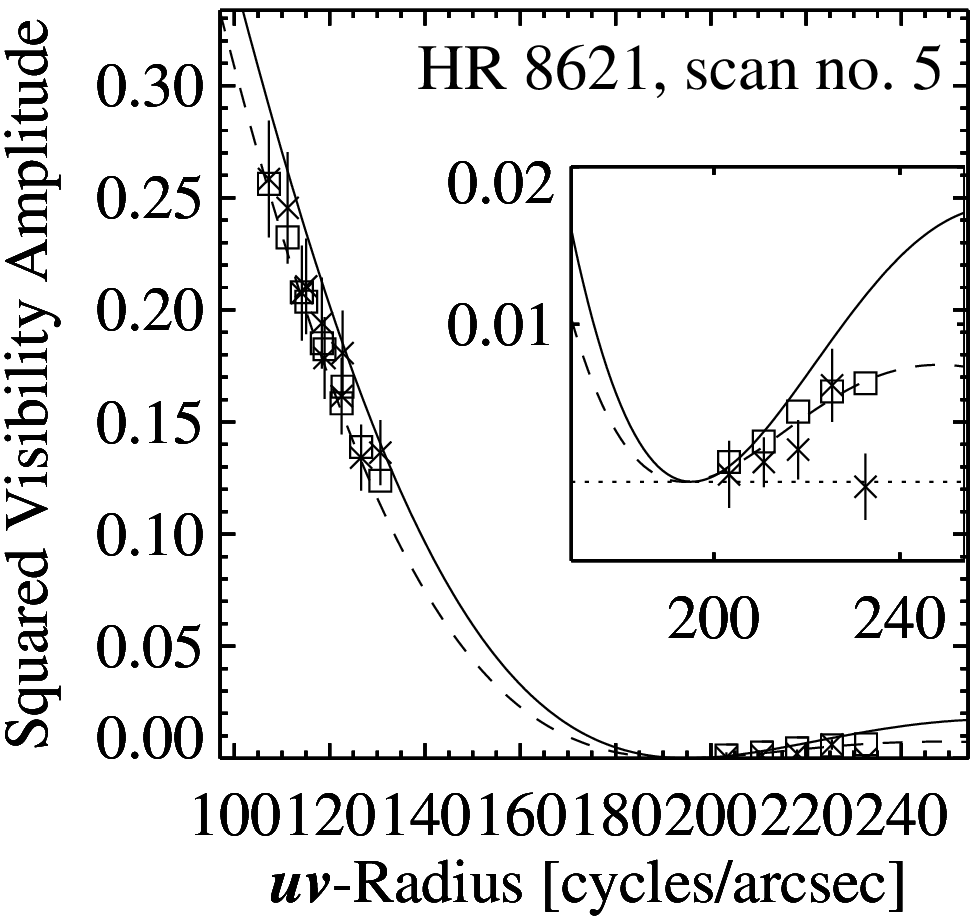}}
\hspace*{2mm}%
\resizebox{0.32\hsize}{!}{\includegraphics{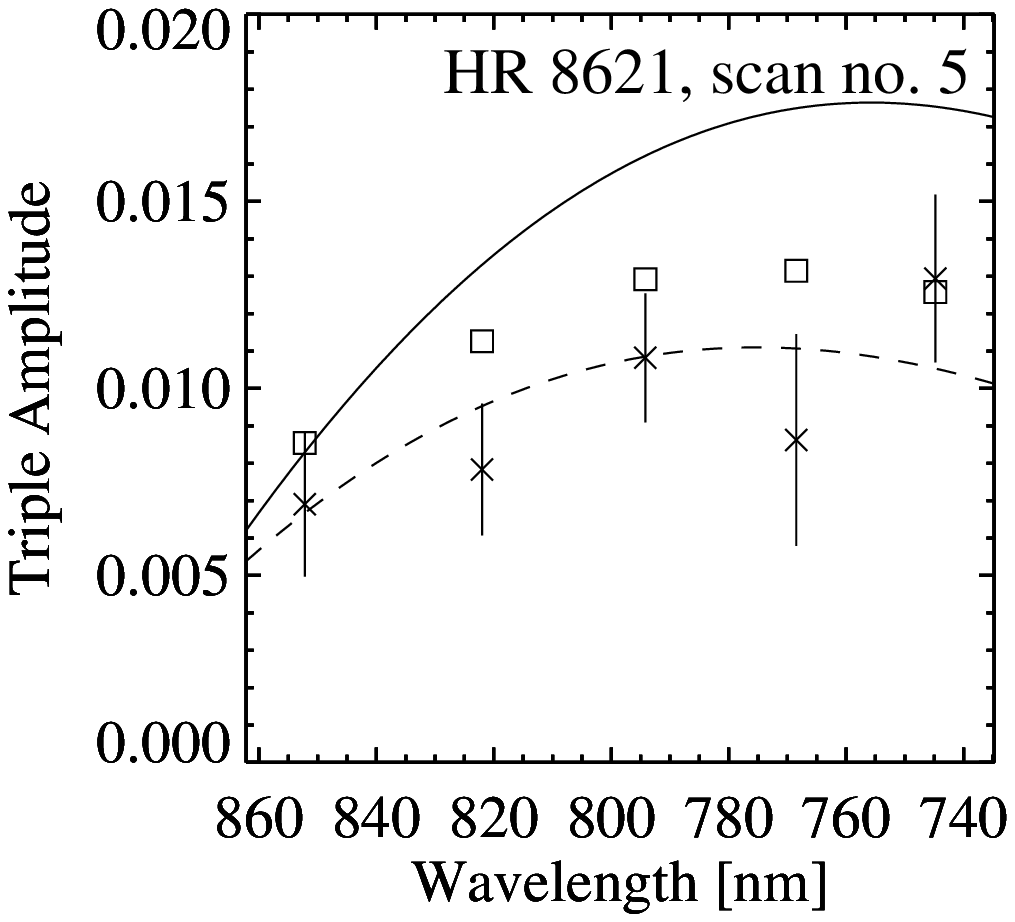}}
\hspace*{-1mm}%
\resizebox{0.32\hsize}{!}{\includegraphics{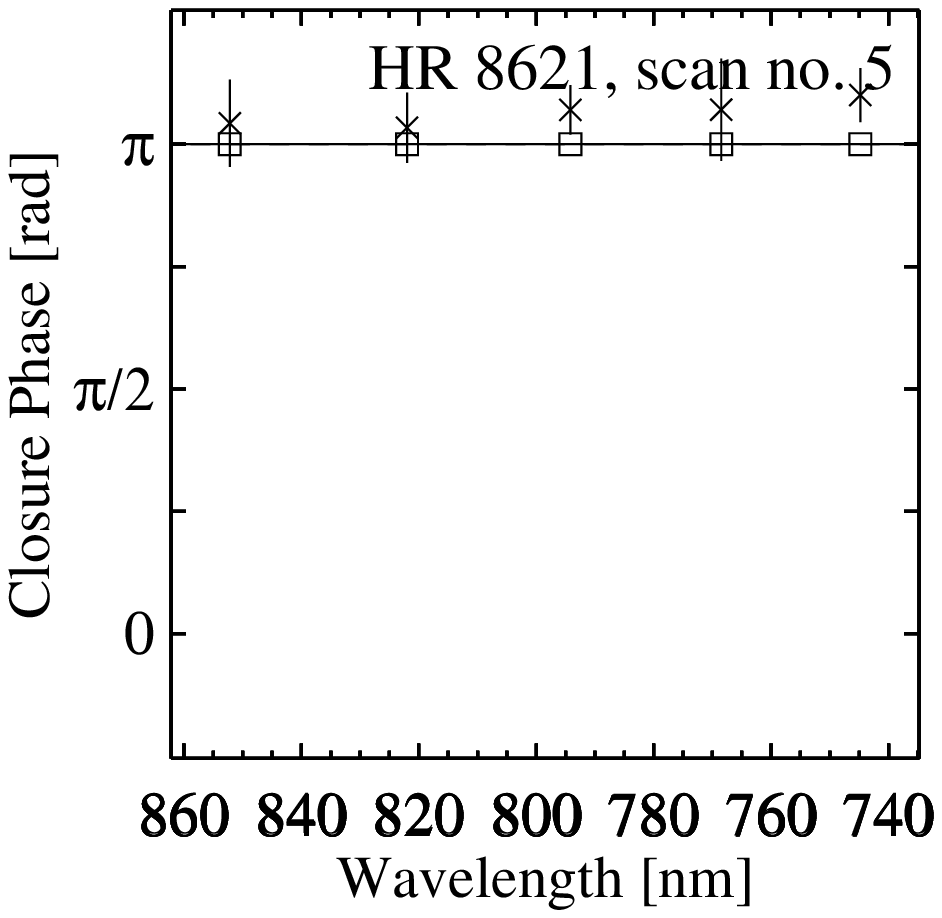}}

\resizebox{0.32\hsize}{!}{\includegraphics{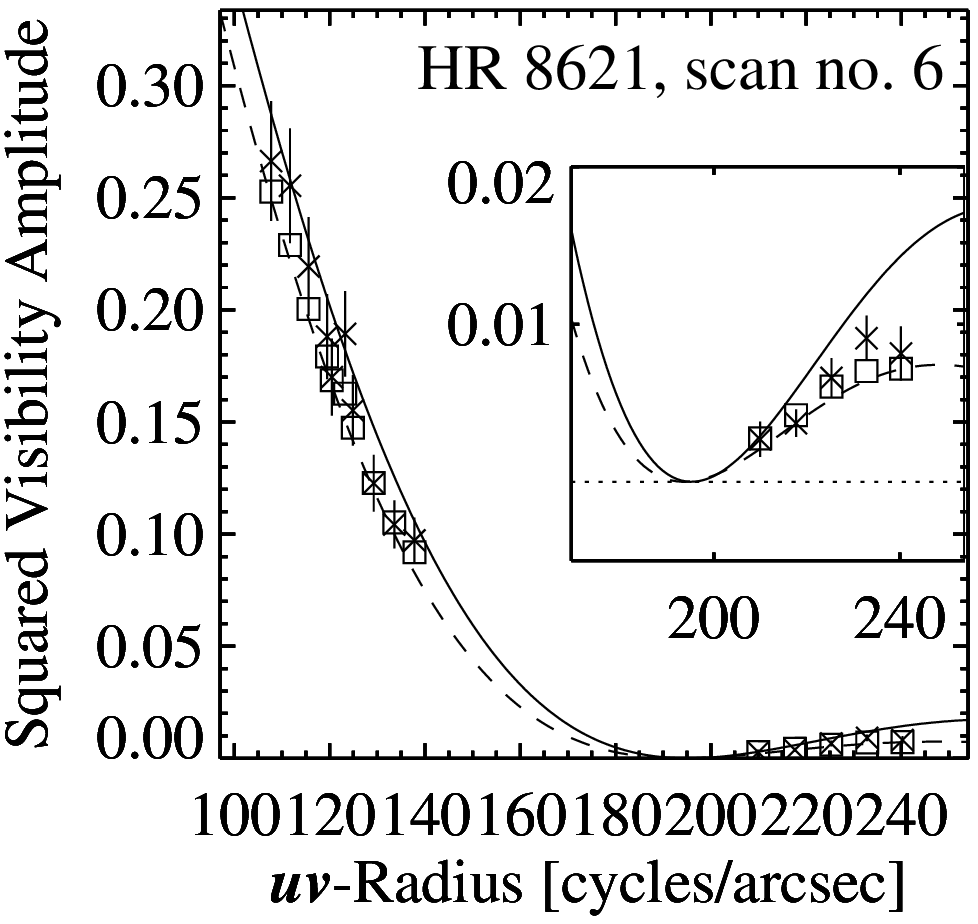}}
\hspace*{2mm}%
\resizebox{0.32\hsize}{!}{\includegraphics{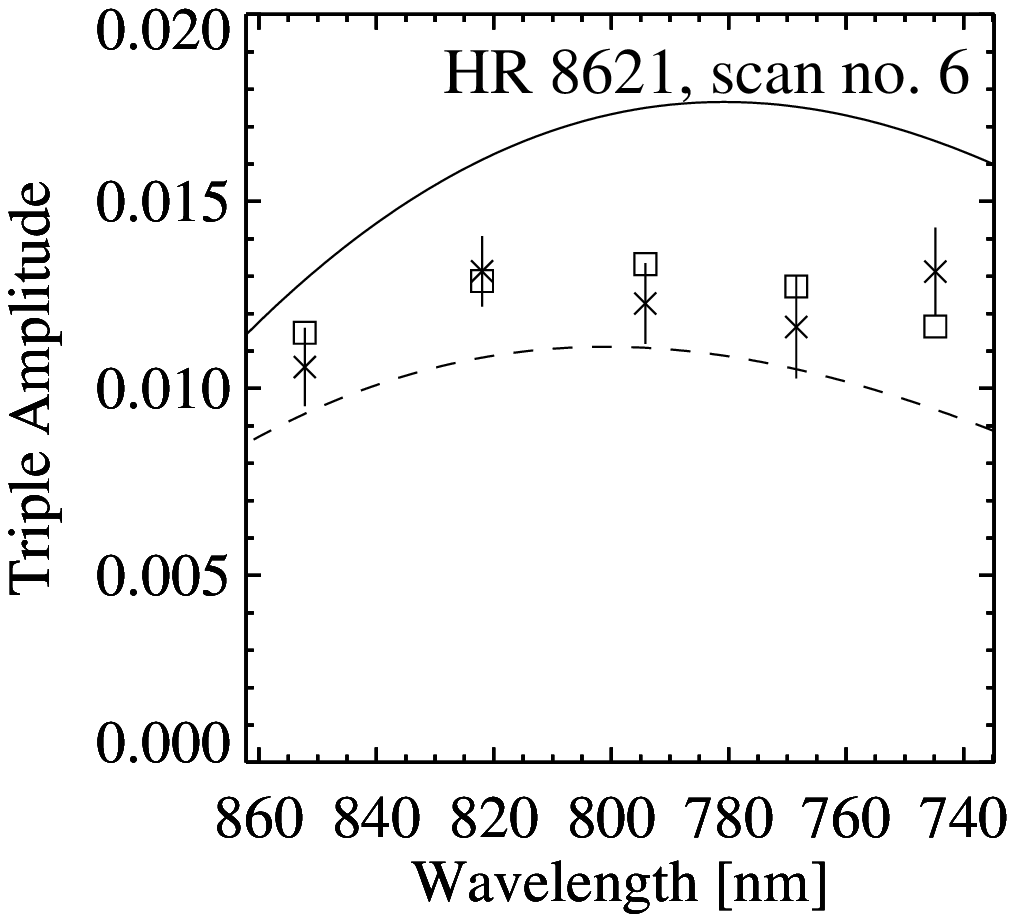}}
\hspace*{-1mm}%
\resizebox{0.32\hsize}{!}{\includegraphics{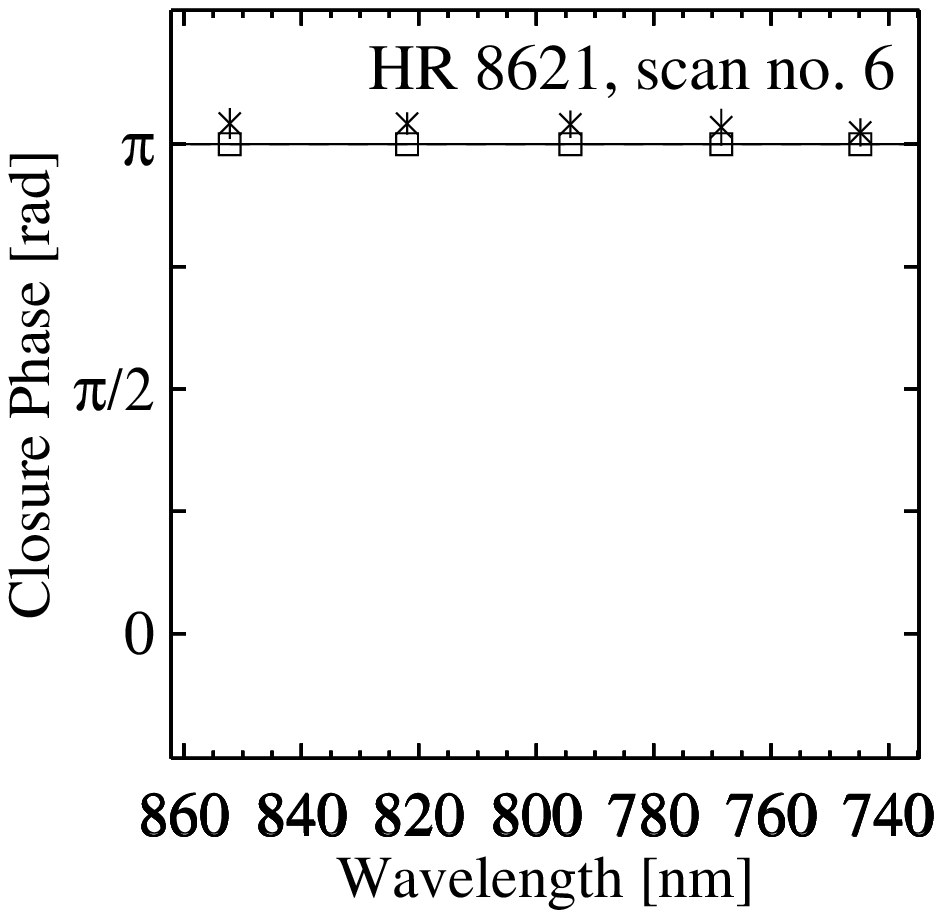}}
\end{minipage}
\parbox[b]{55mm}{
\hfill\resizebox{0.70\hsize}{!}{\includegraphics{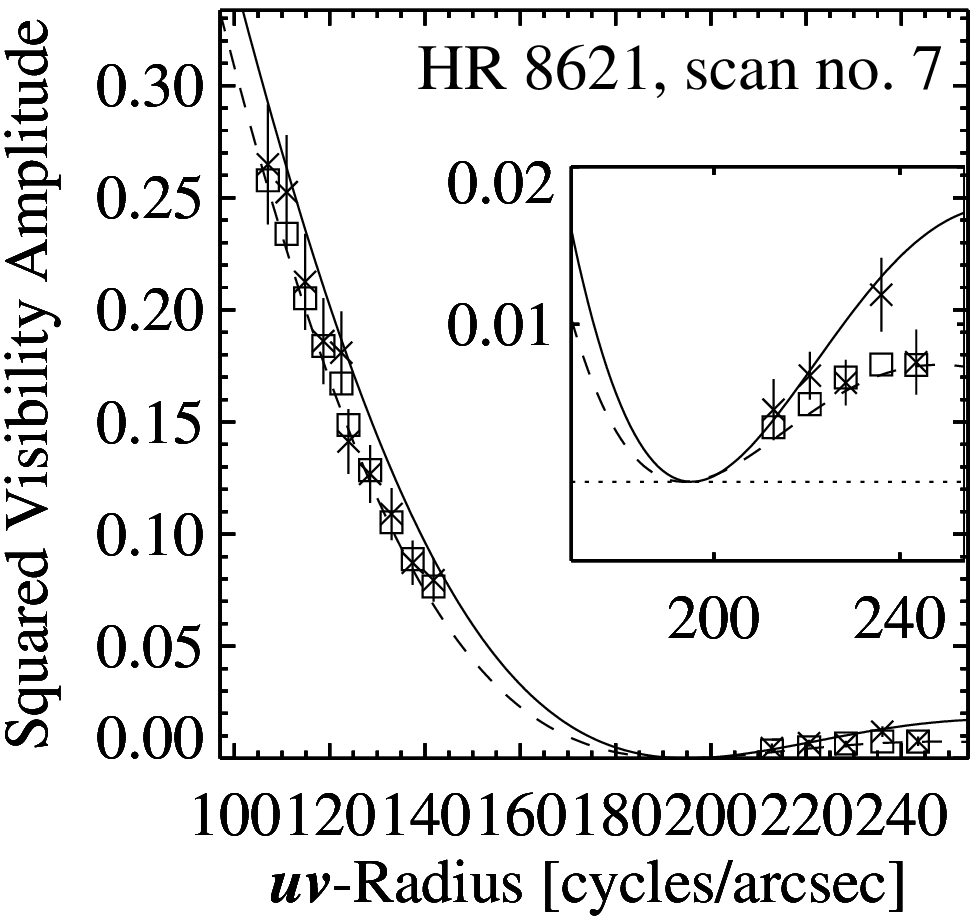}}

\hfill\resizebox{0.70\hsize}{!}{\includegraphics{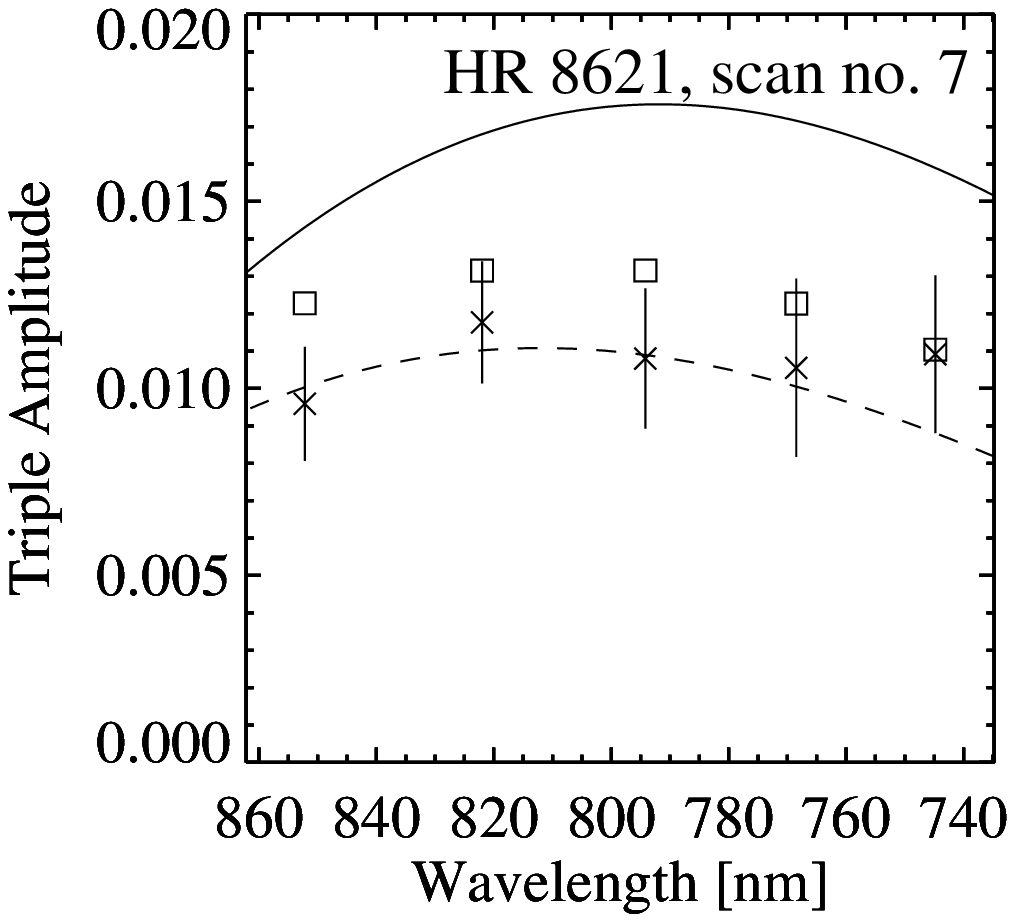}}

\hfill\resizebox{0.70\hsize}{!}{\includegraphics{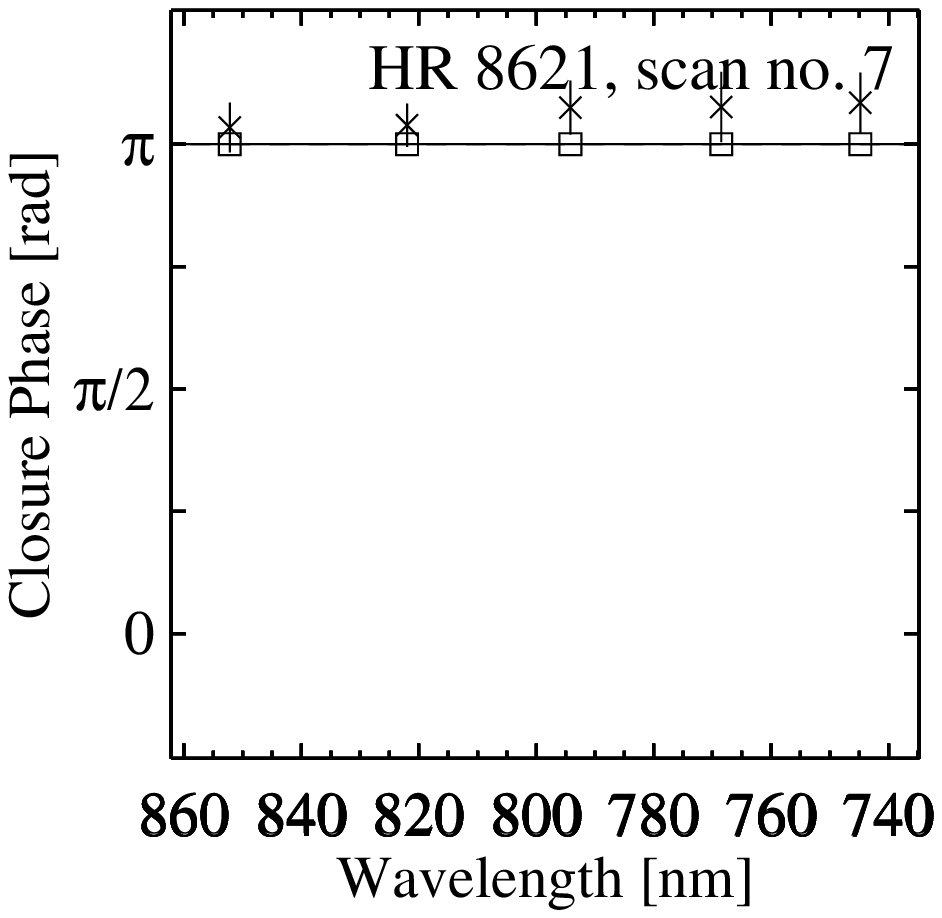}}

\vspace*{11.1cm}%

\caption{As Fig.~\ref{fig:fkv1368}, but for HR\,8621.}
\protect\label{fig:bsc8621}}
\end{figure*}
Figures~\ref{fig:fkv1368}, \ref{fig:fkv752} and \ref{fig:bsc8621} show
all obtained squared visibility amplitude, triple amplitude,
and closure phase data of HR\,5299, HR\,7635, and HR\,8621, respectively.
Also shown are best fitting uniform disk models, fully darkened disk
models, and Kurucz stellar model atmospheres as described below.

The measured squared visibility amplitude of all three program stars as a 
function of increasing $uv$-radius decreases monotonously towards 
a minimum, beyond which they increase.
The low values are difficult to measure since they correspond to vanishing 
fringe contrasts. The measurement of almost zero squared visibility amplitude 
values with acceptable error bars at the minima
confirms the feasibility of the bootstrapping technique as well as 
the correctness of the photon and detection noise bias compensation.
The observed functional form of the squared visibility amplitude is expected 
for a disklike object intensity distribution. The consistency of
visibility values with baselines of different orientations excludes large
deviations from circular symmetry. The absence of
systematic variations with smaller spatial frequencies excludes additional
large-scale structures like circumstellar material with considerable
intensity.

The triple amplitudes show a similar behaviour for decreasing wavelength,
i.e. increasing spatial resolution, with a minimum at a wavelength
where the closure phases clearly exhibit a flip from 0 to $\pi$.
The absence of intermediate closure phases indicates object intensity 
distributions which are symmetric through reflection.  Thus, for spectral 
channels with a bandwidth covering the location of the phase flip, 
intermediate values can occur. For HR\,8621, all recorded triple amplitudes 
and closure phases were beyond the minimum and the flip, due to the 
star's large diameter in relation to the effective baseline lengths.
The triple amplitudes and closure phases show a higher signal-to-noise
ratio than the squared visibility amplitudes on the long baseline,
as discussed in Sect.~\ref{sec:reduction}. 
 
The complex visibility of an astronomical object is related to the 
object intensity distribution through a Fourier transform. 
Consequently, the object intensity distribution can in principle be 
directly reconstructed from interferometric data using imaging techniques 
which effectively interpolate the limited coverage of the $uv$-plane.
This was performed with NPOI data e.g. by Benson et al. (\cite{benson}) for 
the double star Mizar\,A and by Hummel et al. (\cite{hummel}) for Matar. 
However, in order to obtain accurate estimates of physical parameters,
model fits are a better choice. This applies especially to 
stellar disks since their imaging would require more resolution elements. 
Here, our data provide, roughly, two resolution elements across the 
diameter of the stars.
The strength of the limb-darkening is related to the height of the
second maximum of the visibility function. The diameter
is then determined by the locations of the minima of the visibility function 
and of the flip of the closure phases.
\begin{table}
\caption{Best fitting diameters of HR\,5299, HR\,7635, and HR\,8621
based on the models of a uniform disk (UD), a fully darkened disk (FDD)
and the best fitting Kurucz model atmosphere (best K.)
together with the obtained $\chi^2_\nu$ values.
For HR\,5299, HR\,7635, and HR\,8621 the numbers of degrees of freedom
are 150, 300, and 175, respectively.
The quantity $\epsilon_\mathrm{final}$ denotes the final
error of the diameter based on the best Kurucz model atmosphere
due to the formal error ($\epsilon_\mathrm{formal}$) and calibration error 
($\epsilon_\mathrm{calibr.}$) of the underlying data
as well as on an error resulting from the choice of $T_\mathrm{eff}$ and 
$\log g$
($\epsilon_\mathrm{model}$). The last four rows give the mean values
and standard deviations of the UD and FDD diameters determined
for each spectral channel separately and
corrected with factors derived from the expected model atmosphere.}
\begin{tabular}{l|ll|ll|ll}
 & \multicolumn{2}{c}{HR\,5299} & \multicolumn{2}{c}{HR\,7635} &
\multicolumn{2}{c}{HR\,8621} \\
& $\Theta$  & $\chi^2_\nu$ & $\Theta$  & $\chi^2_\nu$ & $\Theta$  & 
$\chi^2_\nu$ \\
&[mas] &          &[mas] &          &[mas] &          \\\hline

UD                     & 6.82 & 3.15 & 5.67 & 4.36 & 6.25 & 4.23  \\
FDD                    & 7.85 & 1.38 & 6.61 & 2.26 & 7.39 & 1.67  \\[1ex]
best K.                & 7.44 & 1.15 & 6.18 & 1.17 & 6.94 & 1.31  \\
$\epsilon_\mathrm{formal}$    & 0.03 &      & 0.01 &      & 0.03 &   \\
$\epsilon_\mathrm{calibr.}$   & 0.10 &      & 0.07 &      & 0.10 &   \\
$\epsilon_\mathrm{model}$     & 0.02 &      & 0.02 &      & 0.05 &   \\
$\epsilon_\mathrm{final}$     & 0.11 &      & 0.07 &      & 0.12 &   \\[1ex]
corr. $d_\mathrm{UD}$     & 7.45 &      & 6.11 &      & 6.92 &       \\
$\sigma$($d_\mathrm{UD}$) & 0.07 &      & 0.02 &      & 0.08 &       \\
corr. $d_\mathrm{FDD}$    & 7.50 &      & 6.13 &      & 6.95 &       \\
$\sigma$($d_\mathrm{FDD}$)& 0.06 &      & 0.02 &      & 0.08 &       \\
\end{tabular}
\label{tab:fits}
\end{table}
\paragraph{Uniform disk and fully darkened disk models}
Since the data as described above suggest a circularly symmetric
disklike object intensity distribution $I$, models of a uniform disk
(UD, $I=1$ for $0\le\mu\le1$, $I=0$ otherwise, $\mu=\cos\Theta$ being
the cosine of the angle between the line of sight and the normal of the
surface element of the star) and a fully darkened disk (FDD, $I=\mu$) are
used as a first approach to describe the data.
In these cases, the visibility functions are given by
\begin{equation}
\hfill%
V_\mathrm{UD}=\frac{2\,J_1(x_\mathrm{UD})}{x_\mathrm{UD}}
\hfill%
V_\mathrm{FDD}=\frac{3\,\sqrt{\pi}\,J_{3/2}(x_\mathrm{FDD})}
{\sqrt{2}\,x_\mathrm{FDD}^{3/2}}
\hfill%
\label{eq:udfdd}
\end{equation}
with $x_\mathrm{UD,FDD}=\pi\,\Theta_\mathrm{UD,FDD}\,\sqrt{u^2+v^2}$
a dimensionless spatial frequency ($u$, $v$: spatial frequencies
$[\mathrm{cycles/arcsec}]$ as in
Figures~\ref{fig:uvcov},~\ref{fig:fkv1368},~\ref{fig:fkv752},~\ref{fig:bsc8621},
$\Theta$: angular diameter of the star) and $J_1$ and $J_{3/2}$ the
Bessel function of first kind and orders 1 and 3/2 (see e.g. Hestroffer
\cite{hestroffer}). With $V$ calculated for each baseline $i\in\{1,2,3\}$, 
the squared
visibility amplitudes $|V_i|^2$, the amplitude of the triple amplitude
$|V_1V_2V_3|$ and the closure phase $\phi_1+\phi_2-\phi_3$ 
($\phi_i=0$ where $V_i>0$ and $\phi_i=\pi$ where $V_i<0$) can be 
derived.

A $\chi^2$ minimization algorithm (simplex method) was applied in order to 
find the best fitting angular diameters $\Theta_\mathrm{UD}$ and 
$\Theta_\mathrm{FDD}$, using all
available data, i.e.  the squared visibility amplitudes, the triple 
amplitudes and the closure phases.
The derived diameters are shown in the first two rows of Table~\ref{tab:fits},
together with the corresponding reduced $\chi^2_\nu$ values. 
For HR\,5299, HR\,7635, and HR\,8621 the numbers of degrees of freedom 
are 150, 300, and 175, respectively.
These $\chi^2_\nu$ values show that the UD model can be rejected and that the
FDD model is a better description of our data.
The model functions are indicated by solid lines (UD) and dashed 
lines (FDD) in Figs.~\ref{fig:fkv1368}-\ref{fig:bsc8621}.
The minima of the visibility and triple amplitude functions are well 
defined by the data. Before and in particular beyond the minima, our measured
triple amplitudes
for all three stars are significantly lower than the UD model values 
and slightly higher than the FDD model values.
This indicates limb-darkened disks, less extreme than fully darkened disks,
as predicted for late-type giants. Therefore, the specific
limb-darkened profiles were investigated as described in the following 
paragraphs. 
\paragraph{Uniform disk and fully darkened disk models corrected for
the effect of limb-darkening}
\begin{figure*}
\resizebox{0.32\hsize}{!}{\includegraphics{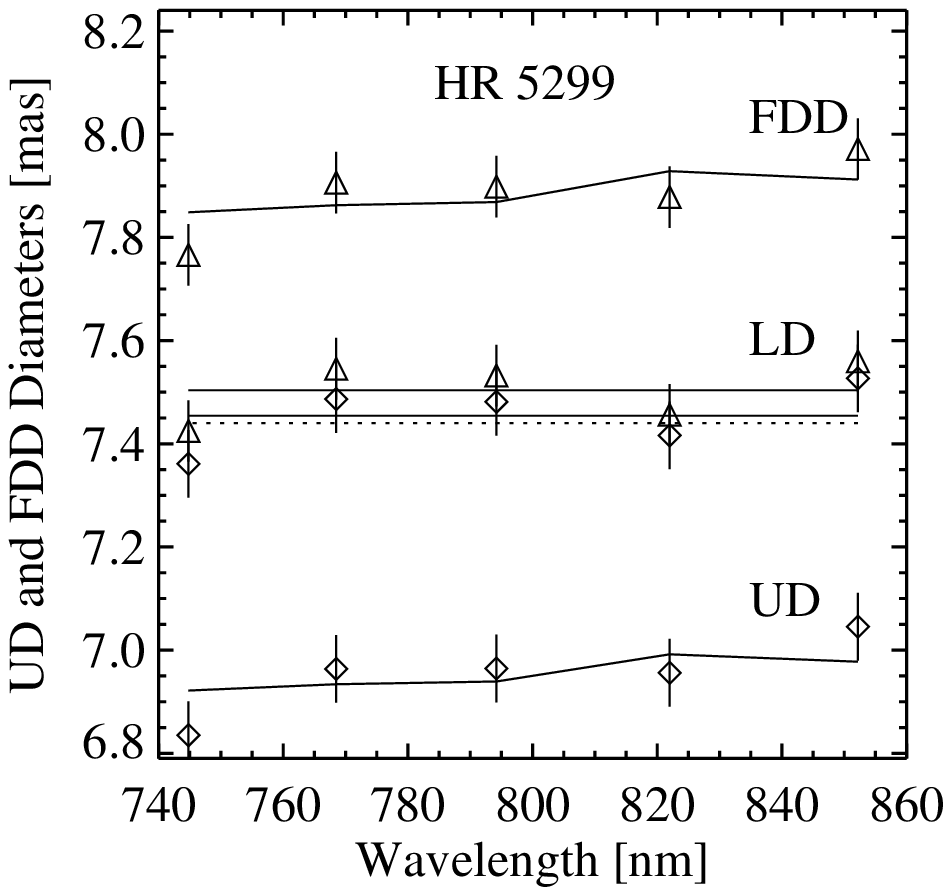}}
\resizebox{0.32\hsize}{!}{\includegraphics{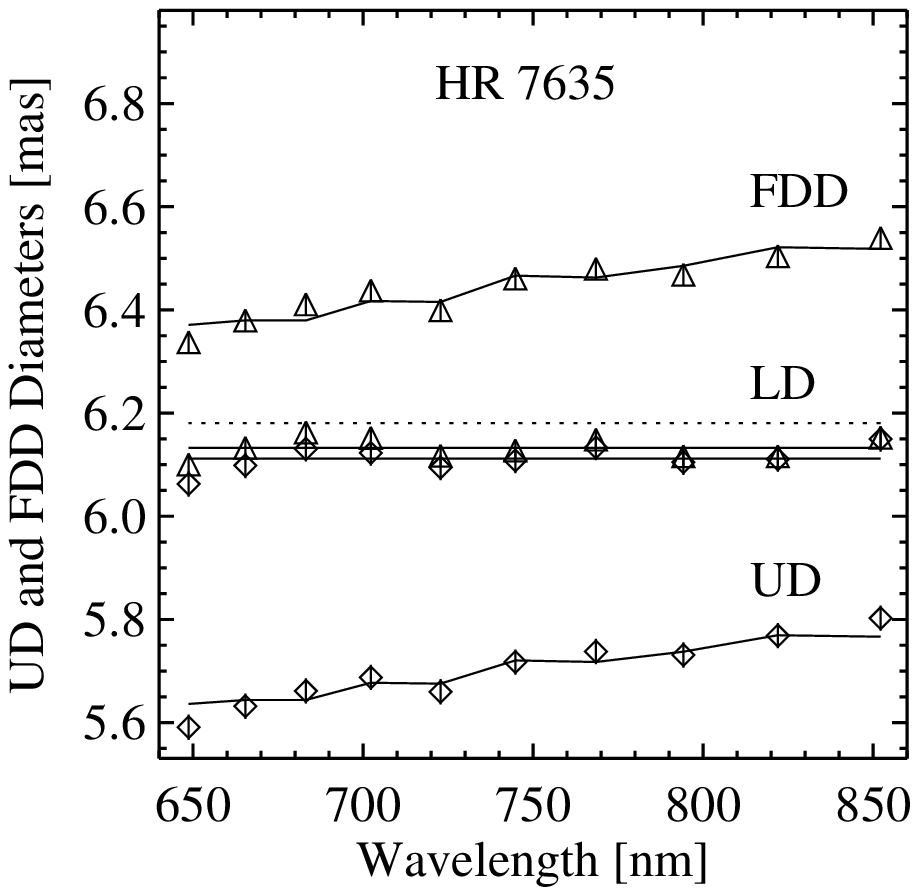}}
\resizebox{0.32\hsize}{!}{\includegraphics{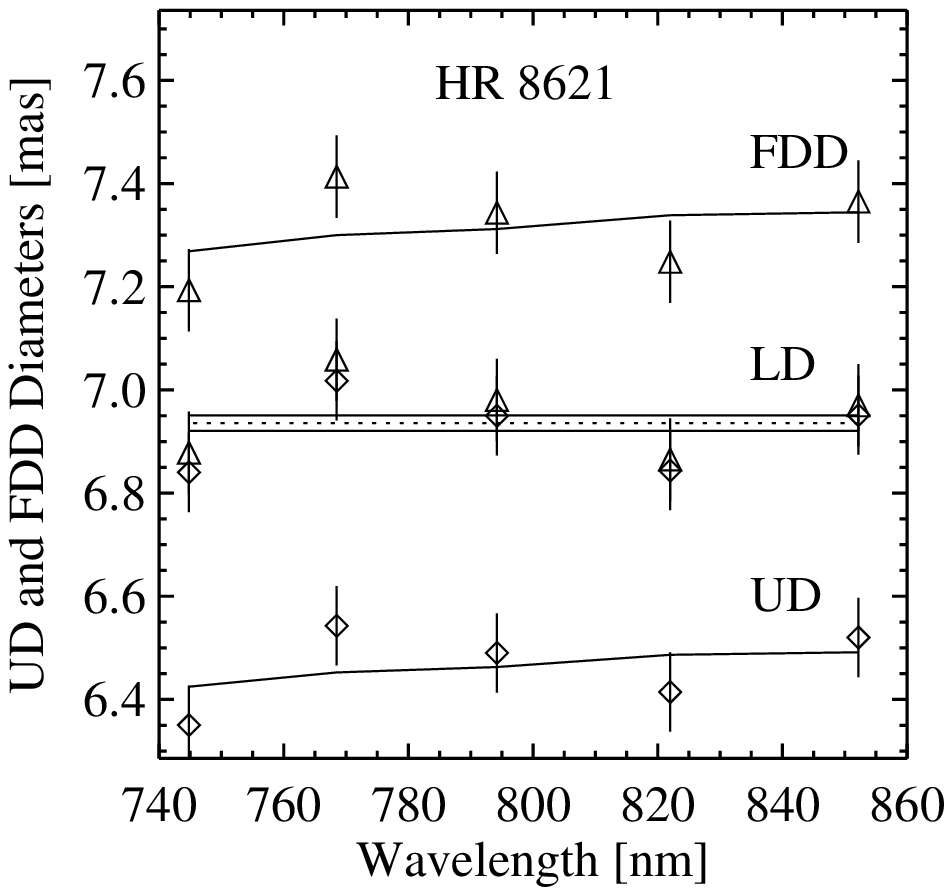}}
\hfill%
\vspace*{-4mm}%
\caption{Diameters based on uniform disk (UD, diamonds) and fully
darkened disk (FDD, triangles)
models for each spectral channel together with the Kurucz model atmosphere
predictions (connected by solid lines). In addition,
the limb-darkened (LD) diameters are shown which were derived by applying
correction factors to the UD and FDD
diameters. The mean LD diameters as quoted in Table \ref{tab:fits} are
indicated by the solid lines.
The error bars represent the standard deviation of the obtained LD diameters.
For comparison, the dotted lines indicate the limb-darkened diameters
which were derived by the direct fit to a grid of Kurucz model atmospheres.}
\label{fig:udfd-wavele}
\end{figure*}
Before we present a direct fit of all our data to theoretical model
atmospheres, we take an alternative approach that is based on
results which are independent of a particular model atmosphere.

Model atmospheres predict a decrease of the degree of limb-darkening 
with increasing wavelength, i.e. a transition from near-FDD to 
near-UD continuum shapes, while the ''true'' limb-darkened diameter
is wavelength-independent (Manduca et al. \cite{manduca},
Hofmann \& Scholz \cite{hofmann}). 
Since our data provide information at different spectral channels, 
they can be used to test stellar model atmospheres on their predicted 
wavelength dependence of UD and FDD diameters. 

Recently, Hestroffer (\cite{hestroffer}) discussed a 
limb-darkening law $I_\lambda=\mu^\alpha$, with $\alpha$ a positive real number,
as empirical brightness distribution function.
This representation of the center-to-limb variation which uses 
only one limb-darkening coefficient includes the UD and FDD models and 
is very well suited to describe a wide range of different realistic 
limb-darkening shapes (Hofmann \& Scholz \cite{hofmann}).
Thus, in principle our data could be used to simultaneously determine
both the apparent limb-darkened diameter of our program stars and their
limb-darkening parameter $\alpha_\lambda$ for each spectral channel.
However, by means of Monte-Carlo simulations based on the actual
coverage of the $uv$-plane of these observations and the claimed precision,
it was found that this determination is ambiguous.

The employed method was first used by Hanbury Brown
(\cite{hanbury}), then applied by Quirrenbach et al. (\cite{quirrenbach})
and recently used for theoretical studies by
Hofmann \& Scholz (\cite{hofmann}) and Davis et al. (\cite{davistangobooth}).
The diameters based on UD and FDD models were derived for each spectral
channel separately using only the squared visibility amplitudes up to the
first minimum, since these data can well be described by UD and FDD models.
The triple amplitudes and closure phases were not used since most of them
contain visibility values beyond the first minimum, which do not fit
UD and FDD models.
The resulting UD and FDD diameters were multiplied by limb-darkening
correction factors to obtain the limb-darkened diameter of the star.
Following the authors mentioned above, the correction factors were
derived as the ratios $x_\mathrm{LD,0.3}/x_\mathrm{UD,0.3}$ and
$x_\mathrm{LD,0.3}/x_\mathrm{FDD,0.3}$ with $x_\mathrm{LD,0.3}$,
$x_\mathrm{UD,0.3}=2.0818$, and $x_\mathrm{FDD,0.3}=2.3451$ being the
spatial frequencies where the squared
visibility amplitudes $|V_\mathrm{LD}|^2$, $|V_\mathrm{UD}|^2$ and
$|V_\mathrm{FDD}^2|$ (see Eq.~\ref{eq:udfdd} and \ref{eq:ld}) equal 0.3.
For the determination of $x_\mathrm{LD}$, the Kurucz model atmosphere
as described in detail in the following section
was used with values for $T_\mathrm{eff}$ and $\log g$
according to the spectral type of the star.

The obtained diameters are plotted in Fig.~\ref{fig:udfd-wavele}
as a function of wavelength. The mean values and standard deviations
of the corrected UD and FDD diameters are shown in the last four
rows of Table~\ref{tab:fits}.
Our observed wavelength dependence of the UD and FDD diameters corresponds 
well with the model predictions.
The small predicted deviations from a monotonous wavelength dependence
are caused by effects of molecular absorption bands. 
Especially in the case of the brightest of our program stars, HR\,7635,
our obtained diameters match the model predictions very well.
Here, a larger number of spectral channels could be used and the
effective temperature causes the transition from the UD to FDD model to
occur in the observed wavelength range.
For all our program stars, no systematic deviations between observations
and model predictions occur; which indicates the correctness of the
atmosphere models used.
Consequently, the derived limb-darkened diameters are, as required,
independent of wavelength. The results based on the UD fit and those
based on the FDD fit are consistent.

The uncorrected UD and FDD diameters can
be used for future comparisons with other model atmospheres and,
furthermore, for future analyses taking additional information at other 
wavelengths into account. However, this approach cannot make use of 
our data at long baselines and the corresponding triple amplitudes and closure
phases. In order to provide more accurate limb-darkened diameters
and to discriminate between different model assumptions a direct
fit of all our data to Kurucz model atmospheres is described in the
following paragraph.
\paragraph{Comparison of our interferometric data with a grid of Kurucz 
stellar model atmospheres}
\begin{table}
\caption{Comparison of minimum $\chi^2_\nu$ values obtained 
by fitting our HR\,5299 and HR\,7635 data to a grid of
Kurucz model atmospheres with solar chemical abundances
(Kurucz \cite{kurucz}) 
based on effective temperatures ranging from 3500\,K to 4500\,K and 
$\log g$ ranging from 0.0 to 2.5. For each program star, the best 
$\chi^2_\nu$ values are marked by a box.
For HR\,5299 and HR\,7635 the numbers of degrees of freedom
are 150 and 300, respectively.}
\begin{tabular}{lr|llllll}
HR      & $\log g$/   & 0.0 & 0.5 & 1.0 & 1.5 & 2.0 & 2.5 \\
        & $T_\mathrm{eff}$ &    &      &      &          \\
        & [K] &    &      &      &          \\\hline
5299    & 3500 & 1.25 & 1.19 & \fbox{1.15} & 1.17 & 1.19 & 1.23 \\
        & 3750 & 1.19 & 1.19 & 1.19 & 1.18 & 1.18 & 1.17 \\
        & 4000 & 1.21 & 1.21 & 1.21 & 1.20 & 1.20 & 1.20 \\
        & 4250 & 1.27 & 1.27 & 1.26 & 1.26 & 1.26 & 1.26 \\
        & 4500 & 1.42 & 1.40 & 1.38 & 1.37 & 1.36 & 1.35 \\
\hline
7635    & 3500 & 1.50 & 1.40 & 1.35 & 1.38 & 1.47 & 1.58 \\
        & 3750 & 1.29 & 1.29 & 1.29 & 1.29 & 1.27 & 1.26 \\
        & 4000 & 1.18 & 1.17 & \fbox{1.17} & 1.17 & 1.18 & 1.18 \\
        & 4250 & 1.20 & 1.20 & 1.19 & 1.19 & 1.19 & 1.19 \\
        & 4500 & 1.26 & 1.25 & 1.24 & 1.23 & 1.23 & 1.23 \\
\end{tabular}
\label{tab:chicomp}
\end{table}
In an effort to compare different model atmosphere predictions based on 
a grid of effective temperatures and surface gravities with our
interferometric data, rather than assuming a particular model profile 
a priori, the best $\chi^2_\nu$ values based on different models were 
determined. In this way, theoretically predicted differences of the
strength of the limb-darkening effect for different effective temperatures 
and surface gravities can be compared to direct measurements.

Kurucz (\cite{kurucz}) tabulates monochromatic intensities $I(1)$
and limb-darkening ratios $I(\mu)/I(1)$ for 17 values of $\mu$ in 
1221 frequency intervals ranging from 9.09\,nm to 160.0\,$\mu$m, 
based on grids of model atmospheres for different chemical abundances. 
Here, his grid for solar chemical abundances and a microturbulent
velocity of $v_\mathrm{turb}=2\,\mathrm{km\,s^{-1}}$ was used 
(file ''cfaku5.harvard.edu/grids/gridP00/ip00k2.pck19'').  
The data is available for effective temperatures $T_\mathrm{eff}$ ranging from
3500\,K to 50000\,K in steps of 250\,K (for low $T_\mathrm{eff}$)
and surface gravities $\log g$ (cgs) from 0 to 5 in steps of 0.5.
For all our program stars, $T_\mathrm{eff}\leq 4000$\,K and $\log g\leq 2$
are predicted, based on their spectral types (see Table \ref{tab:endres}).
Thus, only models with $T_\mathrm{eff}\in[3500\,\mathrm{K},4500\,\mathrm{K}]$
and $\log g\in[0,2.5]$ were considered. 

NPOI passband-specific limb-darkening ratios $I_c(\mu)/I_c(1)$ were 
calculated by integrating the Kurucz data over each of the NPOI spectral
channel's $c\in[1,10]$ sensitivity functions. The NPOI spectral channels may
be affected by molecular absorption bands, which is taken into
account by the calculation of passband-specific limb-darkening model profiles.
Compact photospheres were assumed, i.e. $I_c(\mu=0)=0$. $I_c(1)$
was set to 1 since our measured visibility values are scaled to $V(0)=1$ for
each spectral channel separately.

Following Davis et al. (\cite{davistangobooth}), the model visibility values 
were derived by numerical evaluation of the
Hankel transform of the obtained tabulated intensity profiles 
\begin{equation}
V_\mathrm{LD}=
\frac{\int_0^1\,I_c(\mu)J_0[x_\mathrm{LD}(1-\mu^2)^{1/2}]\mu\,d\mu}
{\int_0^1\,I_c(\mu)\mu\,d\mu},
\label{eq:ld}
\end{equation}
with $x_\mathrm{LD}=\pi\,\Theta_\mathrm{LD}\,\sqrt{u^2+v^2}$ the 
dimensionless spatial frequency as described above, but now based on the 
star's limb-darkened angular diameter $\Theta_\mathrm{LD}$. 
No approximation of the tabulated model limb-darkened profiles by any 
limb-darkening law was used. 
The model squared visibility amplitudes, model triple amplitudes, and model 
closure phases were derived as described above.
 
Based on each of the 30 considered Kurucz models (five values for 
$T_\mathrm{eff}$ and six values for $\log g$), the wavelength-independent
limb-darkened diameter, treated as the only free parameter, and
the corresponding $\chi^2_\nu$ value were derived as described above
using all our available data, i.e. the squared visibility amplitudes, the
triple amplitudes, and the closure phases. 

For our three program stars, the lowest obtained $\chi^2_\nu$ values 
are listed in Table~\ref{tab:fits}, together with the obtained best fitting
limb-darkened diameters. Table~\ref{tab:chicomp} shows all
resulting $\chi^2_\nu$ values for HR\,5299 and HR\,8621, where
the lowest $\chi^2_\nu$ values are marked by a box.

The occurrence of minimum $\chi^2_\nu$ values larger than 1.0 might,
in principle, be caused by optimum parameters lying in between our
grid points, a wrong model assumption, an underestimation
of the calibration errors, or systematic calibration errors.
An underestimation of the calibration errors leading to total errors
underestimated by only ~7\% and ~8\% is most likely to be
the main cause for the deviations from unity in the cases of HR\,5299 and
HR 7635 since these errors can only be roughly
estimated (see Sect.~\ref{sec:reduction}).
A considerable part of the larger deviation from unity, a value of 1.31,
in the case of HR\,8621 might also be caused by an effective temperature
lower than 3500\,K, i.e. an incorrect model assumption, or systematic
effects during the calibration process.
The stellar atmosphere model was adopted as the best fit to our
data for HR 5299 and HR 7635. The $\chi^2_\nu$ values were analyzed
as a function of $T_\mathrm{eff}$ and $\log g$ for these program stars only, 
as follows.

For these program stars, significantly different $\chi^2_\nu$ values are 
obtained for different model parameters, for example, in the case of 
HR\,5299 we obtain values between 1.15 and 1.42. At higher 
temperatures, differences for varying $T_\mathrm{eff}$ are larger than for 
varying $\log g$, because of the latter's lesser effect on the 
limb-darkened profile (see e.g. Manduca et al. \cite{manduca}).

To take the deviations of the $\chi^2_\nu$ values from unity into account, 
the values in Tab.~\ref{tab:chicomp} were normalized to unity at the minimum
for the following analysis. Using this method, the  total assumed errors
in our data are increased by common mean factors of 1.07 and 1.08,
neglecting that calibration errors depend on the value of the visibility. 
It was verified that due to the smallness of this correction the results 
obtained are still valid and that it is insignificant whether the total
data errors are rescaled or just the calibration errors.

Near the minimum, the $\chi^2$ function is expected to be a quadratic 
function of each of the varied parameters. Therefore, for each star,
a parabola was fitted to the $\chi^2$ values as a function of 
$T_\mathrm{eff}$ with fixed best-fitting $\log g$ and as a function
of $\log g$ with fixed best-fitting $T_\mathrm{eff}$. Here,
more digits were used than shown in Table~\ref{tab:chicomp}.
The most likely values for the parameters $T_\mathrm{eff}$ and $\log g$ 
can be estimated by the locations of the minima of the parabola.
Assuming purely Gaussian noise, the corresponding $1\,\sigma$ errors can 
be estimated as the variation in the parameters which will increase the 
normalized total $\chi^2$ values by 1 from its value at the 
minimum of the fitted parabola (see e.g.  
Bevington \& Robinson \cite{bevington}).
For HR\,5299 and HR\,7635, the $\chi^2$ values as a function of 
$T_\mathrm{eff}$ as well as of $\log g$ match a parabola very
well. This confirms that effects due to systematic calibration errors
or an incorrect model assumption are not of considerable size. However, 
small additional errors due to these effects cannot be ruled out and
are not included in the error analysis presented here.
For HR\,5299, the $\chi_\nu^2$ values in Table \ref{tab:chicomp} 
as a function of $T_\mathrm{eff}$ extend to the minimum
but not beyond. However, the one-sided $\chi_\nu^2$ values fit a parabola 
with a minimum at 3520\,K very well, confirming that our grid point
at 3500\,K is in fact close to the minimum.
The derived most likely values and the errors are shown 
in Table~\ref{tab:endres} and are
compared to independent estimates in Sect.~\ref{sec:concl}.

For all our program stars, the best $\chi^2_\nu$ values derived here are 
significantly better than those based on the FDD model 
(see Table~\ref{tab:fits}). This effect is most noticeable in the case
of HR\,7635, due to the higher signal-to-noise ratio of the data and
the higher $T_\mathrm{eff}$ resulting in limb-darkening that is not as 
close to the FDD case as for the cooler stars.

The limb-darkened diameters corresponding to the best fitting
models are shown in Table~\ref{tab:fits}. Errors $\epsilon_\mathrm{final}$
were derived based on the formal errors and calibration
errors mentioned in Sect.~\ref{sec:reduction} and on those due to
the choice of $T_\mathrm{eff}$ and $\log g$. For the latter error,
the standard deviations of diameters based on models
with $T_\mathrm{eff}$ and $\log g$ values within their error bars
were taken. For HR\,8621 all 30 considered models were included.

The squared model visibility amplitudes, model triple amplitudes, and
model closure phases obtained with the best fitting Kurucz model atmosphere
are indicated by the squares in 
Figs.~\ref{fig:fkv1368}-\ref{fig:bsc8621}. They coincide well with our
measured data and describe them considerably better
than the UD and FDD models.
The observed data of HR\,8621 differ slightly from the model values
which might be explained by calibration errors or a wrong model assumption
as mentioned above.
The absence of further systematic deviations between the model predictions 
and our data confirms that extended photospheres with $I(\mu=0)>0$ need not 
to be considered and that the width of the NPOI spectral channels does not
noticeably affect our analysis.
\begin{figure*}
\begin{minipage}[b]{12cm}
\resizebox{0.32\hsize}{!}{\includegraphics{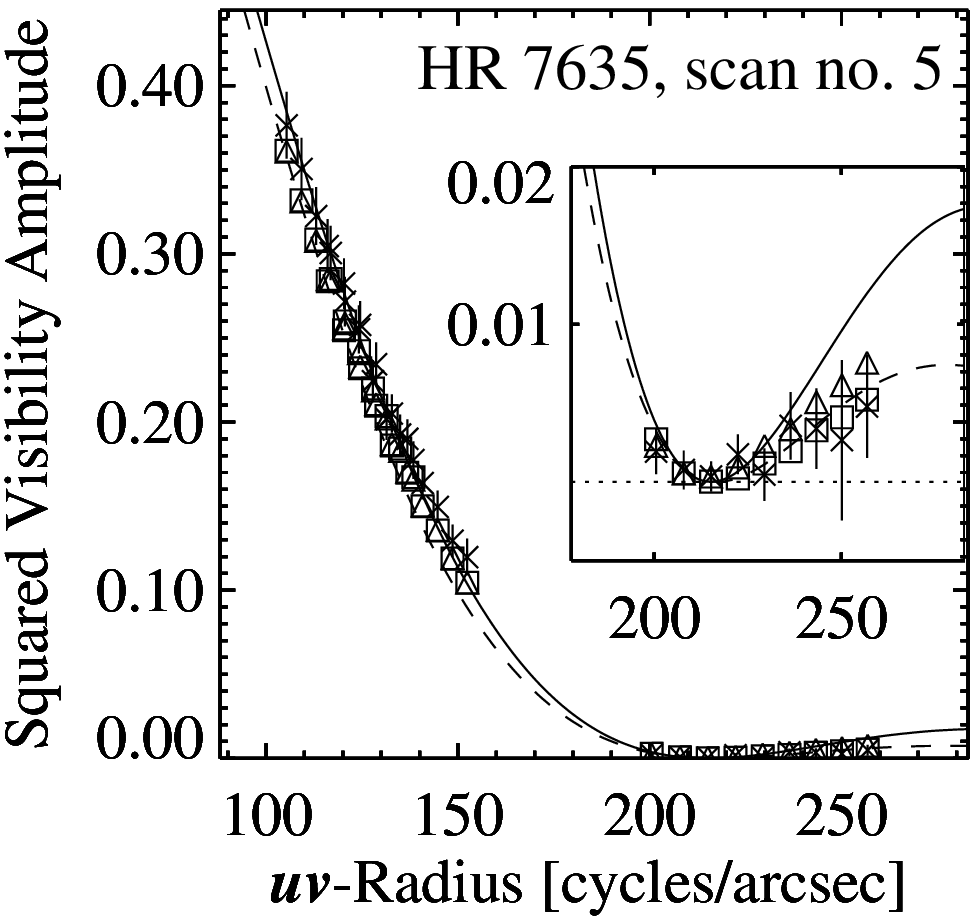}}
\hspace*{2mm}%
\resizebox{0.32\hsize}{!}{\includegraphics{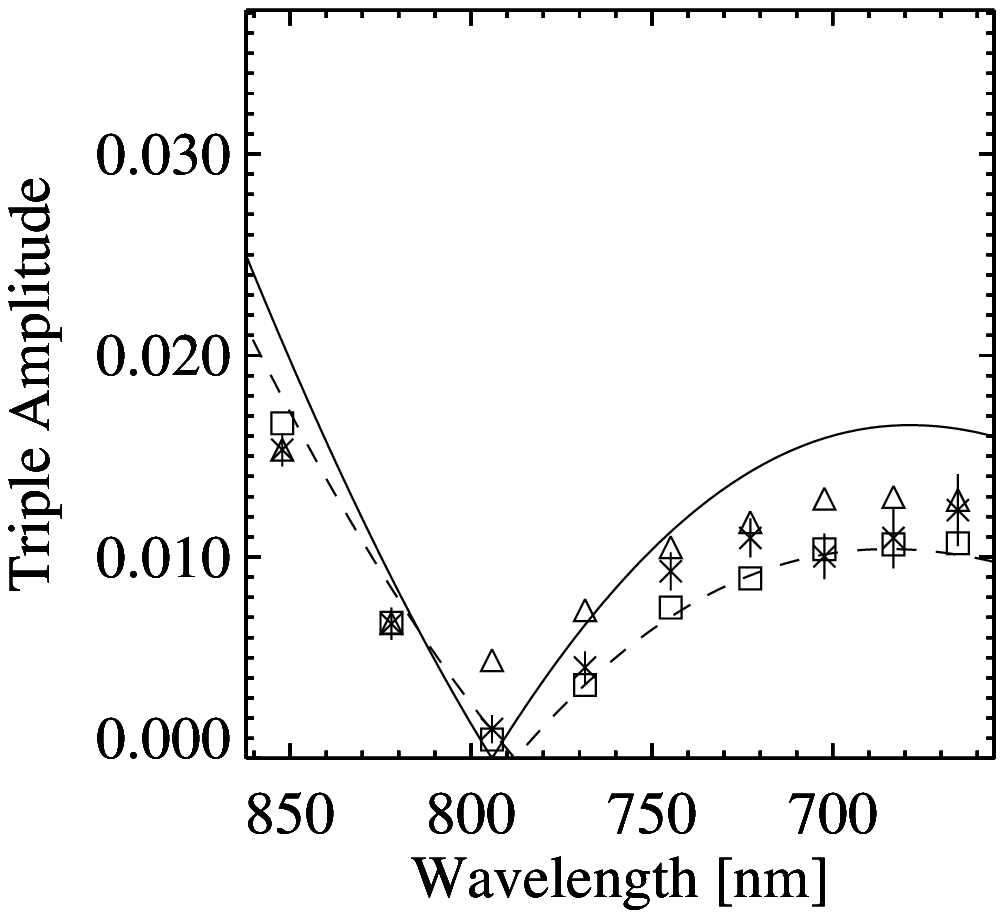}}
\hspace*{-1mm}%
\resizebox{0.32\hsize}{!}{\includegraphics{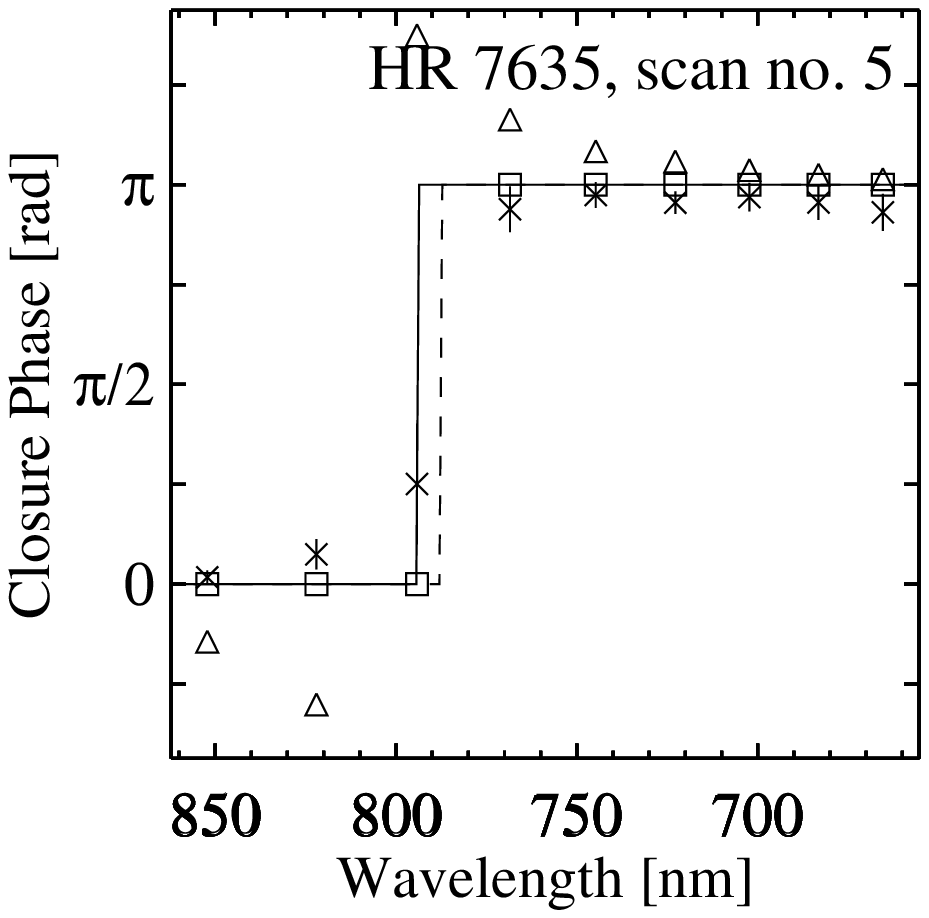}}

\resizebox{0.32\hsize}{!}{\includegraphics{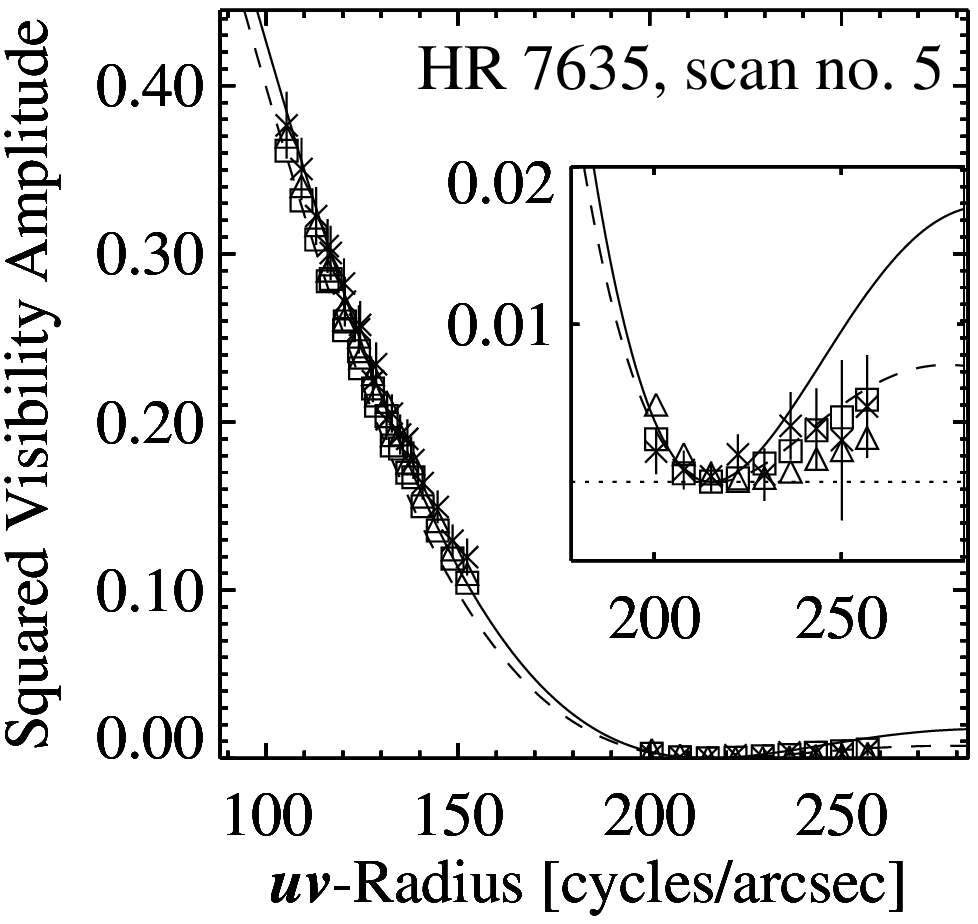}}
\hspace*{2mm}%
\resizebox{0.32\hsize}{!}{\includegraphics{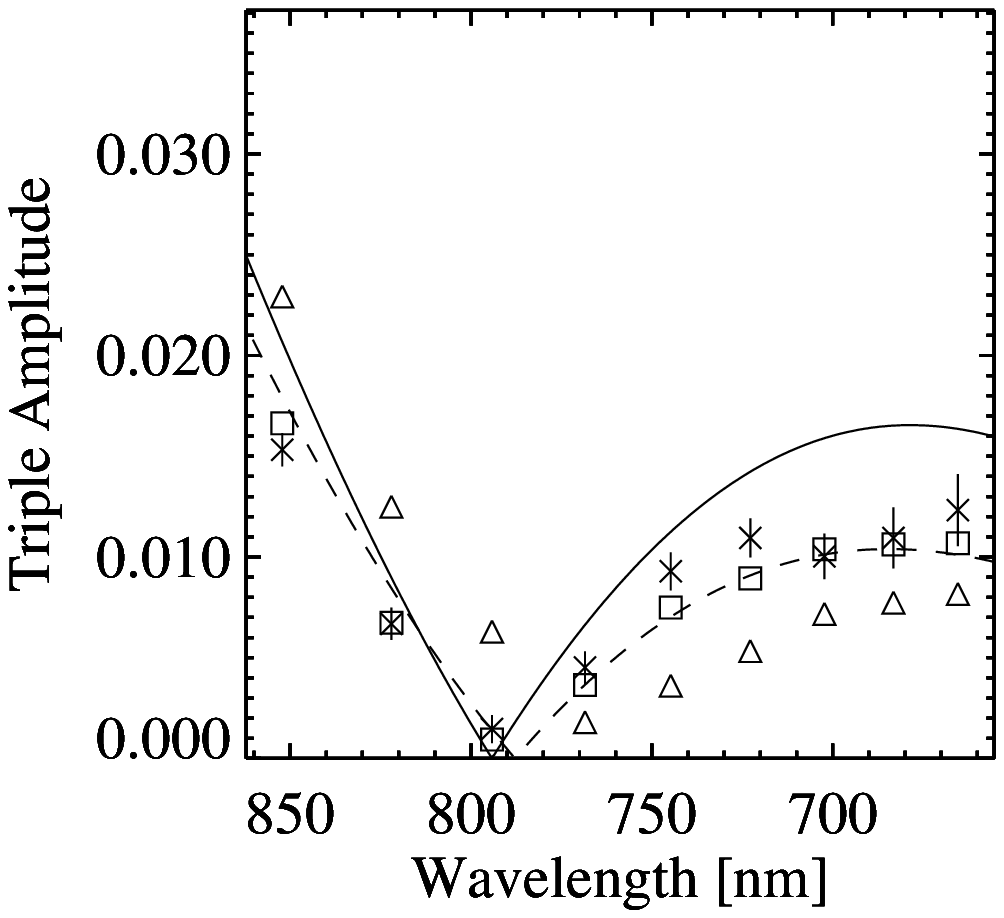}}
\hspace*{-1mm}%
\resizebox{0.32\hsize}{!}{\includegraphics{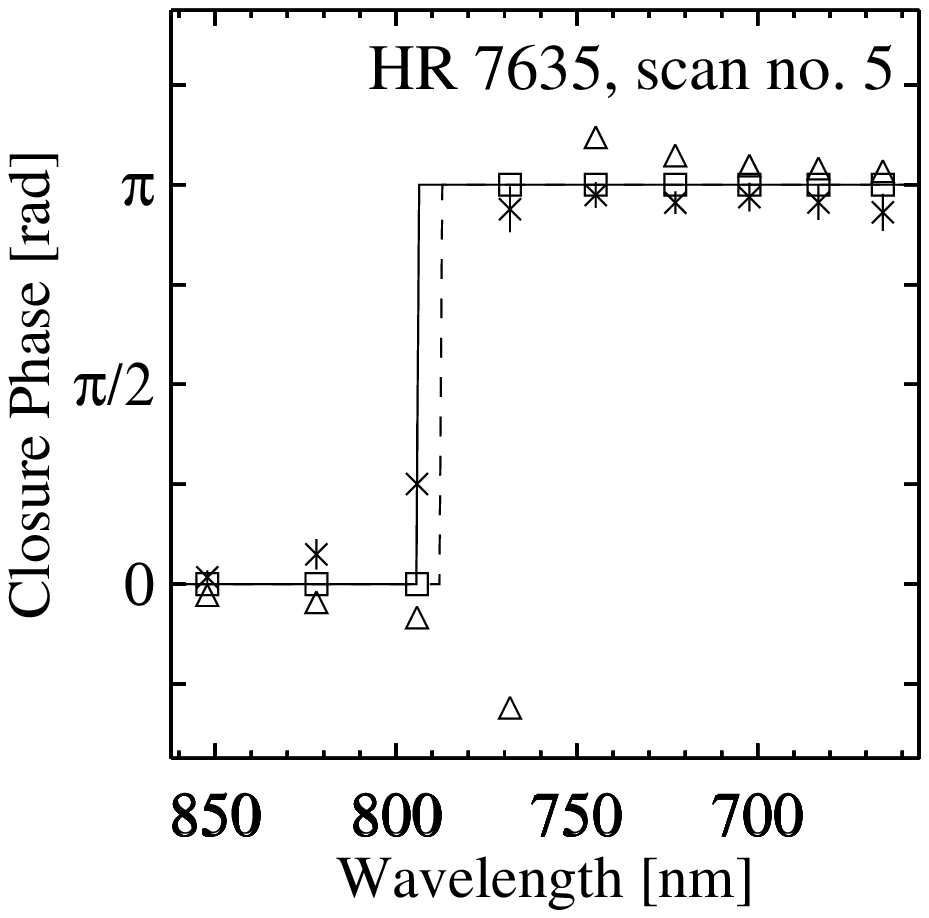}}
\end{minipage}
\parbox[b]{55mm}{
\caption{Impact of hot spots on model data for the example of HR\,7635,
scan no. 5. The triangles denote the best fitting limb-darkening model
with one additional hot spot. The
spot's intensity relative to that of the star is 2\%, its separation
is $\theta_\mathrm{LD}/4$, its position angle is (top) along the preferred
direction
of the $uv$ plane coverage (3\degr) and (bottom) perpendicular to it (93\degr).
For comparison, the {\sf x}-symbols with error bars and the squares
denote the measurements and the best fitting limb-darkening model, 
respectively, as in Fig.~\protect\ref{fig:fkv752}.}\label{fig:spots}}
\end{figure*}
\paragraph{Deviations from featureless symmetric disks}
Although, as discussed above, our data indicate a featureless symmetric disk,
a formal search for solutions with an asymmetric object
intensity distribution was performed in order to validate this interpretation.
All our data were fitted
to a limb-darkened profile as derived before but with an elliptical shape.
Starting values for the position angle of the major axis were chosen from
0\degr to 180\degr in steps of 10\degr and for the axis ratios
(major axis/minor axis) two starting values of 1.05 and 1.1 were considered. 
For the stars HR\,5299, HR\,7635,
and HR\,8621, slightly better $\chi^2_\nu$ values than for the
symmetric intensity profile were derived with axes ratios of 1.02, 1.01,
and 1.01, respectively. These small deviations from circular symmetry
are not significant for an asymmetric object intensity distribution, but
can be caused, for instance, by systematic calibration errors.
However, this study does
exclude elliptical object intensity distributions with larger axis ratios.

In order to estimate whether additional hot spots, as they were found
on the surfaces of $\alpha$\,Ori, $\alpha$\,Sco, and $\alpha$\,Her
(see Sec.~\ref{sec:introduction})
could be detected in our data, model squared visibility amplitudes,
triple amplitudes, and closure phases were calculated for
one example based on our best fitting limb-darkening model with one
additional hot spot. The spot's intensity was chosen to be only 2\% of
the star's intensity, which is clearly less than that of the spots on
$\alpha$\,Ori, $\alpha$\,Sco, and $\alpha$\,Her.
The spots were assumed to be unresolved and to have a separation of half of 
the star's limb-darkened radius. Two position angles were considered,
namely that of the preferred direction of the $uv$-plane coverage
and that perpendicular to it.
Figure \ref{fig:spots} shows the model predictions,
indicating that the existence of such a spot would
significantly affect the closure phases and the triple
and visibility amplitudes around their minima. 
The different triple
and visibility amplitudes might be modeled by another stellar diameter
and limb-darkening profile, but the occurrence of closure phases significantly
different from values of 0 and $\pi$, however, could only be explained by
an asymmetric intensity distribution. Our data is not consistent with
such a noticeable asymmetry.
Consequently, it can be concluded that the 
existence of an unresolved single spot on the surfaces of our program stars 
with an intensity at least as high as studied above is highly unlikely.
A resolved spot is unlikely, too, since its intensity contribution
would be higher.
However, by an analysis of the closure phases we cannot rule out a centered
spot.
\begin{table*}
\caption{Best fitting limb-darkened diameters $\Theta_\mathrm{LD}$,
effective temperatures, and surface gravities as derived by the
direct fit of our interferometric data to a grid of Kurucz model atmospheres 
as described
above (Tables~\ref{tab:fits} and \ref{tab:chicomp}). 
Using $\Theta_\mathrm{LD}$ together with the HIPPARCOS
parallax $\pi_\mathrm{trig}$ and the bolometric flux $F_\mathrm{Bol}$
(see Table \ref{tab:obs}) the linear stellar diameter
$D$ ($\Theta_\mathrm{LD}$, $\pi_\mathrm{trig}$)  and the
effective temperature
$T_\mathrm{eff}$ ($\Theta_\mathrm{LD}$, $F_\mathrm{Bol}$) are derived.
For further comparison,
the effective temperatures $T_\mathrm{eff}$ (Sp.T.) and surface gravities 
$\log g$ (Sp.T.) derived from empirical calibrations of the spectral type 
(Schmidt-Kaler \cite{schmidt-kaler})
are shown.}
\begin{tabular}{l|lll|ll|ll}
HR & $\Theta_\mathrm{LD}$ & $T_\mathrm{eff}$ [K] & $\log g$ &
$D$ ($\Theta_\mathrm{LD}$, $\pi_\mathrm{trig}$) &
$T_\mathrm{eff}$ ($\Theta_\mathrm{LD}$, $F_\mathrm{Bol}$) &
$T_\mathrm{eff}$ (Sp.T.) & $\log g$ (Sp.T.)  \\
   & [mas]  &  &  & [R$_\odot$]  &  [K] &  [K]  &  \\[2ex] \hline 
5299 & 7.44$\pm$0.11 & 3520$\pm$190 & 1.3$\pm$0.4 &
       228$^\mathrm{+28}_\mathrm{-23}$ & 3420$\pm$160 &
       3410 & $<$1.3 \\[1ex]
7635 & 6.18$\pm$0.07 & 4160$\pm$100 & 0.9$\pm$1.0 &
       112$^\mathrm{+8}_\mathrm{-8}$   & 3850$\pm$ 70 &
       3950 & 1.7 \\[1ex]
8621 & 6.94$\pm$0.12 & -- & -- &
       197$^\mathrm{+19}_\mathrm{-17}$   & &
       3430 & $<$1.3 \\
\end{tabular}
\label{tab:endres}
\end{table*}
\section{Discussion}
\label{sec:concl}
Based on the data analysis described in Sect.~\ref{sec:results}
it follows that all our interferometric data are consistent
with featureless circularly symmetric limb-darkened disks.
The method of comparing all our interferometric data to a grid
of Kurucz model atmospheres allows the determination of the effective 
temperature and surface gravity without additional information 
in the cases of HR\,5299 and HR\,7635 and provides 
an accurate estimate of the limb-darkened diameter $\Theta_\mathrm{LD}$.

Independent estimates for $T_\mathrm{eff}$ and $\log g$ can be obtained
by empirical calibrations of the spectral type. 
Additionally, $T_\mathrm{eff}$ and the linear limb-darkened diameter $D$
can be derived by using the bolometric flux $F_\mathrm{Bol}$ 
(see Table~\ref{tab:obs}) and the HIPPARCOS parallax $\pi_\mathrm{trig}$,
together with our value for $\Theta_\mathrm{LD}$.
The errors are dominated by the uncertainties of $F_\mathrm{Bol}$
and $\pi_\mathrm{trig}$. As a result, variations of $\Theta_\mathrm{LD}$ have 
little impact on $T_\mathrm{eff}$ and this estimate of
$T_\mathrm{eff}$ can be regarded as sufficiently independent of our 
determination by the direct fit to Kurucz model atmospheres as well.

Table~\ref{tab:endres} lists the results obtained by the direct fit to 
Kurucz model atmospheres together with the independent estimates.
The error estimates for $T_\mathrm{eff}$ and $\log g$ are based
on the analysis of Table~\ref{tab:chicomp} as described above.

\begin{figure}
\resizebox{1\hsize}{!}{\includegraphics{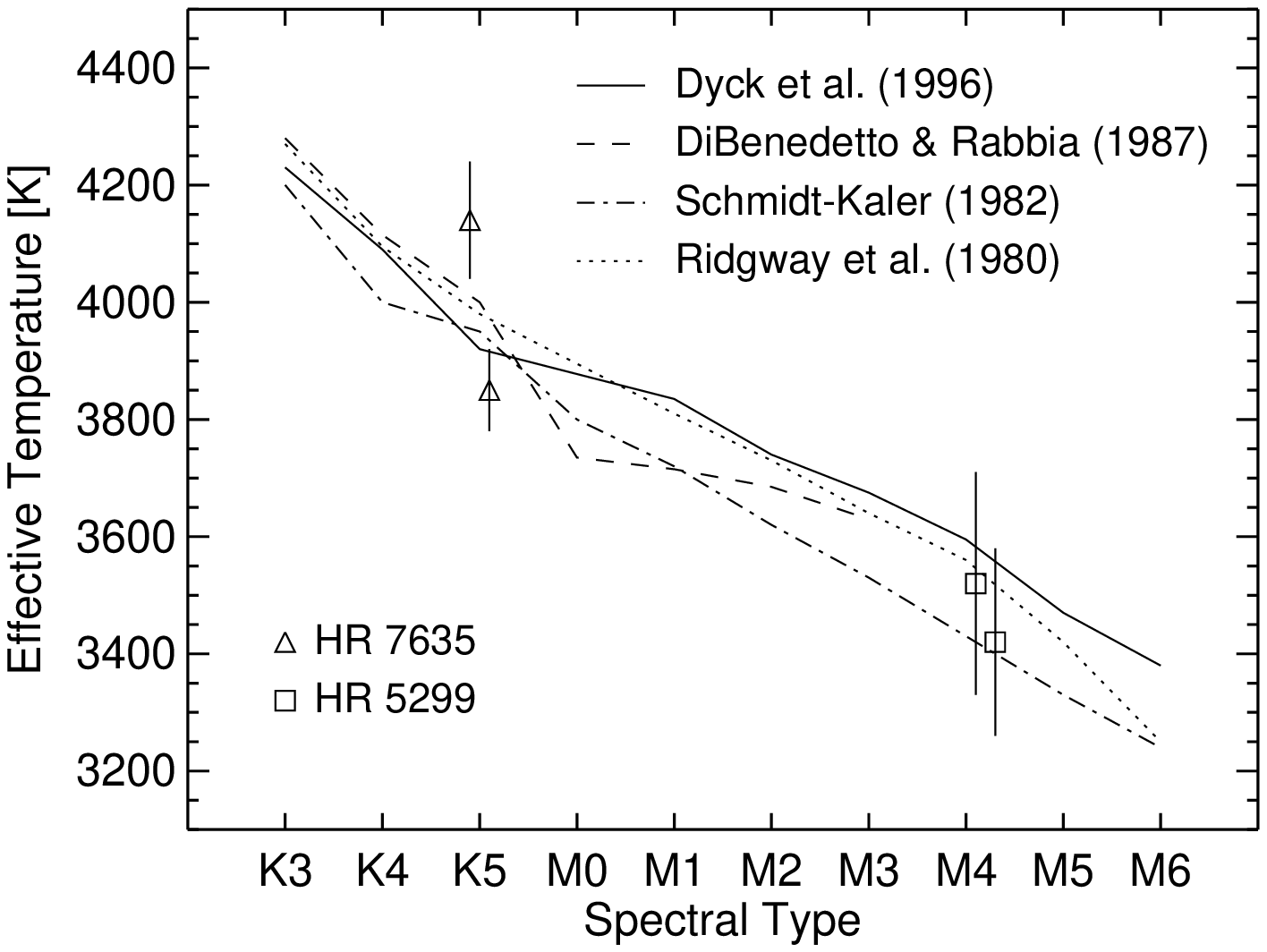}}
\vspace*{-0.5cm}%
\caption{Comparison of our derived HR\,5299 and HR\,7635 $T_\mathrm{eff}$ 
values with different effective temperature scales  
(Ridgway et al. \cite{ridgway}, Schmidt-Kaler \cite{schmidt-kaler}, 
DiBenedetto \& Rabbia \cite{dibenedetto}, Dyck et al. \cite{dyck1}). 
The symbols shifted to the left of the nominal spectral type position
indicate the values derived by the direct fit to a grid of Kurucz model
atmospheres. The symbols shifted to the right indicate the values
derived from the limb-darkened angular diameter $\Theta_\mathrm{LD}$ 
and the bolometric flux $F_\mathrm{bol}$.}
\label{fig:spt}
\end{figure}
Figure~\ref{fig:spt} compares the two derived $T_\mathrm{eff}$ values,
obtained by the direct fit to a grid of Kurucz model atmospheres, and by
$\Theta_\mathrm{LD}$ and $F_\mathrm{bol}$, with different empirical
calibrations.
For HR\,8621 the data quality is, as mentioned above, not sufficiently 
high to obtain $T_\mathrm{eff}$ and $\log g$ by means 
of the direct fit to Kurucz model atmospheres. For HR\,5299,
both the $T_\mathrm{eff}$ value and $\log g$ value derived by this 
fit to Kurucz model atmospheres are well consistent with the independent
estimates. For HR\,7635, the obtained $T_\mathrm{eff}$ value
is higher than that of the independent
estimates and consistent only within $\sim\,2\,\sigma$. 
The value obtained for $\log g$ is consistent with
the empirical calibration.

Our derived values for $\Theta_\mathrm{LD}$ are generally
consistent with earlier determinations of uniform disk diameters
corrected for limb-darkening. They are available
for HR\,5299 (7.0\,mas\,$\pm$\,0.3\,mas by Dyck et al. \cite{dyck1}, 
\cite{dyck2}) and for
HR\,7635 (7.4\,mas\,$\pm$\,0.2\,mas by Hutter et al. \cite{hutter}; 
5.5\,mas\,$\pm$\,0.5\,mas by Dyck et al. \cite{dyck1}, \cite{dyck2}).
For HR\,7635,
Alonso et al. \cite{alonso2} derived a limb-darkened diameter
of 6.12\,mas\,$\pm$\,0.2\,mas by means of the infrared flux method
(IRFM), wich is in very good agreement with our value.

Our derived linear limb-darkened radii for the three stars of
114\,R$_\odot\,\pm\,$13\,R$_\odot$, 56\,R$_\odot\,\pm\,$4\,R$_\odot$,
and 98\,R$_\odot\,\pm\,$9\,R$_\odot$ are in good agreement with
those obtained with the empirical calibration for M giants by 
Dumm \& Schild (\cite{dumm}) based on the HIPPARCOS parallaxes,
$V$ magnitudes, and $V-I$ color indexes, which are 119\,R$_\odot$,
60\,R$_\odot$, and 97\,R$_\odot$, respectively.

The circular symmetry of our observed object intensity distributions
is expected because at optical wavelengths only the surfaces of the stars
themselves are observed rather than additional circumstellar envelopes where
asymmetric morphologies were discovered. These asymmetries in the envelopes
can be caused by, e.g., rotations so slow that they do not observably affect 
the star's shape.
The absence of additional surface features as observed on the surfaces of 
the apparently largest supergiants might be explained by the higher surface 
gravities of our program giant stars.
\section{Summary}
\label{sec:disc}
Featureless symmetric limb-darkened stellar disks provide good
fits to our NPOI interferometric data of HR\,5299, HR\,7635, and HR\,8621.
We are able to discriminate between model atmospheres with different
effective temperatures and surface gravities. We find that our
interferometric measurements and stellar model atmosphere
predictions by Kurucz (\cite{kurucz}) of the limb-darkening effect are
consistent.
We obtain high-precision (1\%-2\% accuracy) limb-darkened angular disk
diameters and derive linear radii and effective temperatures using the
HIPPARCOS parallaxes and bolometric fluxes reported in the literature.
With reduced noise terms and the upcoming simultaneous combination of six
beams at NPOI we will be able to obtain even more precise limb-darkened
diameters of a much larger number of stars. Furthermore, interferometric
data will soon allow the discrimination between model atmospheres with
different effective temperatures, surface gravities and even chemical
abundances with higher precision than in this first attempt. Thus, further
observational constraints for model atmospheres will become available,
in addition to observations of stellar spectra.
\begin{acknowledgements}
We thank the observers B.\ O'Neill and C.\ Denison for their careful 
operation of the NPOI array.
M.\ Wittkowski acknowledges support from the Alexander von Humboldt
Foundation through a Feodor Lynen Fellowship. 
We thank Robert L. Kurucz for making his model atmosphere data 
easily available.
The Navy Prototype Optical Interferometer is a joint project of the 
Naval Research Lab and the US Naval Observatory in cooperation with
Lowell Observatory, and is funded by the Office of Naval Research and 
the Oceanographer of the Navy. 
This research has made use of the SIMBAD database, 
operated at CDS, Strasbourg, France. 
\end{acknowledgements}

\begin{thebibliography}{}
%
\bibitem[1999]{alonso} 
Alonso A., Arribas S., Mart\'{\i}nez-Roger C., 1999, A\&AS 139, 335
%
\bibitem[2000]{alonso2}
Alonso A., Salaris M., Arribas S., Mart\'{\i}nez-Roger C., Asensio Ramos A.,
2000, A\&A 355, 1060
%
\bibitem[1998]{armstrong} 
Armstrong J.T., Mozurkewich D., Rickard L.J., et al., 1998, ApJ 496, 550
%
\bibitem[1997]{benson} 
Benson J.A., Hutter D.J., Elias II N.M., et al., 1997, AJ 114(3), 1221
%
\bibitem[1991]{bevington}
Bevington P.R., Robinson D.K., 1991, Data Reduction and Error Analysis
for The Physical Sciences, McGraw-Hill, ISBN 0079112439
%
\bibitem[1997]{burns}
Burns D., Baldwin J.E., Boysen R.C., et al., 1997, MNRAS 290, L11
%
\bibitem[1990]{buscher}
Buscher D.F., Haniff C.A., Baldwin J.E., Warner P.J., 1990, MNRAS 245, 7p
%
\bibitem[1999]{colavita}
Colavita M.M., 1999, PASP 111, 111
%
\bibitem[1999]{davis}
Davis J., Tango W.J., Booth A.J., Thorvaldson E.D., Giovannis J., 1999,
MNRAS 303, 783
%
\bibitem[2000]{davistangobooth}
Davis J., Tango W.J., Booth A.J., 2000, MNRAS 318, 387
%
\bibitem[1987]{dibenedetto}
DiBenedetto G.P., Rabbia Y., 1987, A\&A 188, 114
%
\bibitem[1998]{dumm}
Dumm T., Schild H., 1998, New Astronomy 3, 137
%
\bibitem[1996]{dyck1} 
Dyck H.M., Benson J.A., van Belle G.T., Ridgway S.T., 1996 AJ 111(4), 1705
%
\bibitem[1998]{dyck2} 
Dyck H.M., van Belle G.T., Thompson R.R., 1998, AJ 116, 981
%
\bibitem[1997]{esa}
Perryman M.A.C. and ESA, 1997, The HIPPARCOS and TYCHO catalogues, 
ESA SP Series vol. 1200, Noordwijk, Netherlands: ESA Publications Division
%
\bibitem[1996]{gilliland}
Gilliland R.L., Dupree A.K., 1996, ApJ 463, L29
%
\bibitem[1998]{hajian} 
Hajian A.R., Armstrong J.T., Hummel C.A., et al., 1998, ApJ 496, 484
%
\bibitem[1974]{hanbury} 
Hanbury Brown R., Davis J., Lake R.J.W., Thompson R.J., 1974, MNRAS 167, 475
%
\bibitem[1995]{haniff}
Haniff C.A., Scholz M., Tuthill P.G., 1995, MNRAS 276, 640
%
\bibitem[1997]{hestroffer} 
Hestroffer D., 1997, A\&A 327, 199
%
\bibitem[1998]{hofmann}
Hofmann K.-H., Scholz M., 1998, A\&A 335, 637
%
\bibitem[1998]{hummel} 
Hummel C.A., Mozurkewich D., Armstrong J.T., et al., 1998, AJ 116, 2536
%
\bibitem[1989]{hutter}
Hutter D.J., Johnston K.J., Mozurkewich D., et al., 1989, AJ 340, 1103
%
\bibitem[2000]{jacob}
Jacob A.P., Bedding T.R., Robertson J.G., Scholz M., 2000, MNRAS 312, 733
%
\bibitem[1958]{jennison}
Jennison R.C., 1958, MNRAS 118, 276
%
\bibitem[1993]{kurucz}
Kurucz R., 1993, Limbdarkening for 2 km/s grid (No.\ 13): [+0.0] to [-5.0].\ 
Kurucz CD-ROM No.\ 17.\  Cambridge, Mass.: Smithsonian Astrophysical 
Observatory, 1993
%
\bibitem[1977]{manduca}
Manduca A., Bell R.A., Gustafsson B., 1977, A\&A 61, 809
%
\bibitem[1973]{morgan}
Morgan W.W., Keenan P.C., 1973, ARA\&A 11, 29
%
\bibitem[1991]{mozurkewich} 
Mozurkewich D., Johnston K.J., Simon R.S., et al., 1991, AJ 101, 2207
%
\bibitem[1996]{quirrenbach} 
Quirrenbach A., Mozurkewich D., Buscher D.F., Hummel C.A., Armstrong J.T.,
1996, A\&A 312, 160
%
\bibitem[1980]{ridgway}
Ridgway S.T., Joyce R.R., White N.M., Wing R.F., 1980, ApJ 235, 126
%
\bibitem[1988]{roddier} 
Roddier F., 1988, in ESO Conf. and Workshop 29, High Resolution Imaging by
Interferometry II ed. F. Merkle (Garching:ESO), 565
%
\bibitem[1987]{scholztakeda}
Scholz M., Takeda Y., 1987, A\&A 186, 200
%
\bibitem[1982]{schmidt-kaler}
Schmidt-Kaler T., 1982, in Landolt-B\"ornstein, New Series VI/2b,
ed. Schaifers K. \& Voigt H.H. (Springer Berlin, Heidelberg, New York),p. 451
%
\bibitem[1988]{shao}
Shao M., Colavita M.M., Hines B.E., et al., 1988, A\&A 193, 357
%
\bibitem[1981]{tsuji} 
Tsuji T., 1981, A\&A 99, 48
%
\bibitem[1997]{tuthill}
Tuthill P.G., Haniff C.A., Baldwin J.E., 1997, MNRAS 285, 529
%
\bibitem[1998]{weigelt}
Weigelt G., Balega Y., Bl\"ocker T., et al., 1998, A\&A 333, L51
%
\bibitem[1992]{wilson}
Wilson R.W., Baldwin J.E., Buscher D.F., Warner P.J., 1992, MNRAS 257, 369
%
\bibitem[1998]{wittkowski}
Wittkowski M., Langer N., Weigelt G., 1998, A\&A 340, L39
%
\bibitem[2000]{young} 
Young J.S., Baldwin J.E., Boysen R.C., et al., 2000, MNRAS 315, 625
%
\end{thebibliography}
\end{document}